\documentclass[12pt]{article}
\usepackage{mathrsfs}
\makeatletter \@addtoreset{equation}{section} \makeatother
\renewcommand{\theequation}{\thesection.\arabic{equation}}
\newcommand{\ba}{\begin{array}}
\newcommand{\ea}{\end{array}}
\newcommand{\beq}{\begin{equation}}
\newcommand{\eeq}{\end{equation}}
\newcommand{\bea}{\begin{eqnarray}}
\newcommand{\eea}{\end{eqnarray}}

\def\bce{\begin{center}}
\def\ece{\end{center}}

\def\nonu{\nonumber}

\def\pa{\partial}
\def\al{\alpha}
\def\be{\beta}
\def\ga{\gamma}

\def\de{\delta}

\def\la{\lambda}
\def\La{\Lambda}

\newcommand{\p}{\hat{p}}

\def\eps6{{\displaystyle \mathop{\epsilon}^{6}}{}}
\def\g6{{\displaystyle \mathop{g}^{6}}{}}

\def\nab6{{\displaystyle \mathop{\nabla}^{6}}{}}

\def\0{{\sst{(0)}}}
\def\1{{\sst{(1)}}}
\def\2{{\sst{(2)}}}
\def\3{{\sst{(3)}}}
\def\4{{\sst{(4)}}}
\def\5{{\sst{(5)}}}
\def\6{{\sst{(6)}}}
\def\7{{\sst{(7)}}}
\def\8{{\sst{(8)}}}

\def\p{\partial}

\def\ba{\begin{array}}
\def\ea{\end{array}}
\def\beq{\begin{equation}}
\def\eeq{\end{equation}}
\def\be{\begin{equation}}
\def\ee{\end{equation}}

\def\la{\lambda}
\def\eps{\epsilon}

\def\p{\partial}

\def\ba{\begin{array}}
\def\ea{\end{array}}
\def\beq{\begin{equation}}
\def\eeq{\end{equation}}
\def\be{\begin{equation}}
\def\ee{\end{equation}}

\def\la{\lambda}
\def\eps{\epsilon}

\def\eps6{{\displaystyle \mathop{\epsilon}^{6}}{}}

\def\nab6{{\displaystyle \mathop{\nabla}^{6}}{}}

\newcommand{\bean}{\begin{eqnarray*}}
\newcommand{\eean}{\end{eqnarray*}}

\begin{document}
\thispagestyle{empty} \addtocounter{page}{-1}
   \begin{flushright}
\end{flushright}

\vspace*{1.3cm}
  
\centerline{ \Large \bf
The ${\cal N}=2$ Supersymmetric $w_{1+\infty}$ Symmetry
}
\vspace*{0.3cm}
\centerline{ \Large \bf
in the Two-Dimensional  SYK Models} 
\vspace*{1.5cm}
\centerline{ {\bf  Changhyun Ahn}
} 
\vspace*{1.0cm} 
\centerline{\it 
 Department of Physics, Kyungpook National University, Taegu
41566, Korea} 
\vspace*{0.5cm}
\centerline{\tt ahn@knu.ac.kr
} 
\vskip2cm

\centerline{\bf Abstract}
\vspace*{0.5cm}

We identify the rank $(q_{syk}+1)$ of the interaction of the
two-dimensional ${\cal N}=(2,2)$ SYK model
with the deformation parameter
$\la$
in the Bergshoeff, de Wit and Vasiliev(in 1991)'s
linear $W_{\infty}[\la]$ algebra via $\la =\frac{1}{2(q_{syk}+1)}$
by using a matrix generalization.
At the vanishing $\la$ (or the infinity limit of $q_{syk}$),
the ${\cal N}=2$ supersymmetric linear $W_{\infty}^{N,N}[\la=0]$
algebra  contains the matrix version of known
${\cal N}=2$  $W_{\infty}$ algebra, as a subalgebra,
by realizing that the $N$-chiral multiplets and the $N$-Fermi multiplets
in the above SYK models play the role of the same number of
$\beta\, \ga$ and
$b\, c$ ghost systems in the linear $W_{\infty}^{N,N}[\la=0]$ algebra.
For the nonzero $\la$, we determine the complete
${\cal N}=2$ supersymmetric linear $W_{\infty}^{N,N}[\la]$ algebra
where the structure constants are given by the linear combinations of
two different generalized hypergeometric functions having the
$\la$ dependence.
The weight-$1, \frac{1}{2}$ currents occur
in the right hand sides of this algebra and
their structure constants have the $\la$ factors.
We also describe the $\la =\frac{1}{4}$ (or $q_{syk}=1$) case
in the truncated subalgebras by calculating the vanishing
structure constants.


\baselineskip=18pt
\newpage
\renewcommand{\theequation}
{\arabic{section}\mbox{.}\arabic{equation}}

\tableofcontents

\section{ Introduction}

The celestial holography \cite{PPR}
connects  the gravitational scattering in asymptotically
flat spacetimes  with  the conformal field theory which lives
on the celestial sphere.
By using the low energy scattering problems, the symmetry
algebra of the conformal field
theory for flat space   
has been found in \cite{GHPS}.
Furthermore, in \cite{Strominger},
the group of symmetries on the celestial sphere plays the role of
the wedge subalgebra of $w_{1+\infty}$ algebra \cite{Bakas}.
We should understand
the unknown structures behind these findings 
in order to convince the above duality.
In \cite{Ahn2111,Ahn2202},
the supersymmetric $w_{1+\infty}$ algebra has been identified with
the corresponding soft current algebra in the supersymmetric
Einstein-Yang-Mills theory.
Recently, in \cite{PV}, the holographic map from
two-dimensional SYK models to the
conformally soft sector of gravity in four-dimensional
asymptotically flat spacetimes is studied.
One of the motivations in this paper is to consider other types of
SYK models and to check whether we have similar $w_{1+\infty}$ symmetry
or not.
See the review papers
\cite{PPR,Aneeshetal,Pasterski,Raclariu} on the
celestial holography \footnote{
  The relevant works on the celestial holography in
  the connection with the $w_{1+\infty}$ symmetry 
  can be found in \cite{
Costello:2022wso,
Freidel:2021ytz,
Bu:2021avc,MRSV,Ball:2021tmb,Adamo:2021lrv,
Jiang:2021csc,Jiang:2021ovh}.}. 

In ${\cal N}=(2,2)$ SYK models \cite{MSW,Bulycheva,AP},
the two $U(1)$ symmetries of the ${\cal N}=(0,2)$ SYK models
can be combined with $U(1)$ $R$ symmetry and the
chiral and Fermi multiplets are also combined into
${\cal N}=(2,2)$ chiral multiplet. This implies
that their charges are related to each other.
It turns out that the stress energy tensor takes simple form
and the coefficients of the stress energy tensor are related to 
the rank of the interaction of the SYK models.
The standard ${\cal N}=2$ superconformal algebra
is realized by the chiral multiplets and Fermi multiplets
in quadratic form with various powers of (antiholomorphic) derivatives.

In the ${\cal N}=2$ supersymmetric linear
$W_{\infty}[\la]$ algebra \cite{Bergetalnpb,Bergetalplb},
by so-called $\beta \, \ga$ and $b \, c$
ghost systems, the higher spin currents with one-parameter
are determined by the quadratic forms of these bosonic and fermionic
operators \footnote{This is not the asymptotic symmetry algebra
introduced in \cite{GH}.}.
In this case, the standard ${\cal N}=2$ superconformal algebra
can be written in terms of the currents of low weights.
Moreover, the so-called ${\cal N}=2$ scalar multiplet 
can be described by the lowest bosonic and fermionic currents.
As a subalgebra, the bosonic algebra contains
$W_{\infty}[\la]$ algebra and $W_{\infty}[\la +\frac{1}{2}]$ algebra.
They claim that  the ${\cal N}=2$ supersymmetric linear
$W_{\infty}[\la]$ algebra is isomorphic to
 the ${\cal N}=2$ supersymmetric linear
 $W_{\infty}[\frac{1}{2}-\la]$ algebra because there exist some transformations
 between the above  $\beta \, \ga$ and $b \, c$
 ghost systems by introducing two real anticommuting parameters
 \footnote{The notations for the subscript $\infty$
   will be clearer when we
discuss about the algebra itself in section $3$.
 }.

In this paper,
by realizing that the above two models have their own one parameter,
i) the rank $(q_{syk}+1)$ of the interaction of the SYK models, and ii)
the $\la$ parameter and the fundamental building blocks are characterized
by chiral and Fermi multiplets on the one hand and by
$\beta \, \ga$ and $ b \, c $ ghost systems
on the other hand,  we would like to study
the precise relation between the ${\cal N}=(2,2)$ SYK models and 
the ${\cal N}=2$ $\beta \, \ga$ and $b \, c$ ghost systems.

\begin{itemize}
\item[]
  At first, we
  make a generalization of \cite{Bergetalnpb,Bergetalplb}
  by introducing the multiple $\beta \, \ga$ and $b \, c$ ghost
  systems.
  Then we can compare with each stress energy tensor (or
  the generators of ${\cal N}=2$ superconformal algebra) described above. 
  This will provide the exact correspondence between
  the two parameters mentioned before.
  At $\la =0$, we identify the free field realization in \cite{Ahn2202}
  with the ones from $\beta \, \ga$ and $b \, c$ ghost
  systems. This implies that
  the realization of
  ${\cal N}=2$ supersymmetric linear $W^{N,N}_{\infty}[\la=0]$ algebra
  is described by the above ${\cal N}=(2,2)$ SYK models together with
  the infinity limit of the rank of the interaction.

  At nonzero $\la$, by using the higher spin currents of
   the matrix generalized $\beta \, \ga$ and $b \, c$ ghost
  systems \cite{Bergetalnpb,Bergetalplb}, we determine the complete 
  ${\cal N}=2$ supersymmetric linear $W^{N,N}_{\infty}[\la]$ algebra
  in terms of various (anti)commutator relations.
  The structure constants originate from the oscillator construction
  in the $AdS_3$ Vasiliev higher spin theory  \cite{AK2009}.
  At $\la =\frac{1}{4}$ (corresponding to the rank $(q_{syk}+1)=2$ of the interaction
  of the ${\cal N}=(2,2)$ SYK models), we show how the truncated
  subalgebra arises by calculating the vanishing structure constants.
  Finally, we also describe the relation with celestial holography
  briefly.
  
\end{itemize}

In section $2$,
we review both
${\cal N}=(2,2)$ SYK models and 
the ${\cal N}=2$ supersymmetric linear
$W_{\infty}[\la]$ algebra.

In section $3$,
by matrix generalization of
the ${\cal N}=2$ supersymmetric linear
$W_{\infty}[\la]$ algebra,
 the realization of
${\cal N}=2$ supersymmetric linear $W^{N,N}_{\infty}[\la=0]$ algebra
 in the ${\cal N}=(2,2)$ SYK models
 is described.
 For nonzero $\la$, starting from the
$\la$ dependent higher spin currents, we construct
 the (anti)commutator relations by checking the structure constants
 explicitly.
The realization of
${\cal N}=2$ supersymmetric linear
$W^{N,N}_{\infty}[\la=\frac{1}{4}]$ algebra
 in the ${\cal N}=(2,2)$ SYK models
 is studied.
 The relation with celestial holography
 is obtained.

In section $4$,
we summarize what we have obtained in this paper
and further directions are also described.

In Appendices,
some detailed calculations in section $3$
are explained. 

We are using the Thielemans package \cite{Thielemans}
with a mathematica \cite{mathematica}.

\section{Review}

\subsection{ Two-dimensional ${\cal N}=(2,2)$ SYK models}

In the two-dimensional SYK model \cite{MSW},
there are $N$ chiral multiplets
$\Phi^a (a=1,2, \cdots, N)$ and
$M$ Fermi multiplets $\La^i (i=1,2, \cdots, M)$ with a random
coupling.
This random coupling of the interaction of the SYK model
has a rank of $(q_{syk}+1)$.
The model with $N=M$ has an enhanced
${\cal N}=(2,2)$ supersymmetry and reduces to the one
studied in \cite{MSW,Bulycheva} \footnote{From now on, we
use the terminology of ${\cal N}=2$ rather than ${\cal N}=(2,2)$ for
simplicity. See also relevant works in
\cite{SW,Witten1,Dedushenko,KW,KM}. More literatures can be found in
\cite{AP}.}.
The lowest components of these superfields satisfy
the following operator product expansions (OPEs)
\bea
 \bar{\pa} \, \bar{\phi}^a(\bar{z}) \, \phi^b(\bar{w}) \, =
\frac{\de^{a b}}{(\bar{z} -\bar{w})}
+ \cdots, \qquad
\frac{i \,}{\sqrt{2}}\,  \bar{\la}^a(\bar{z}) \,
\frac{i }{\sqrt{2}} \, \la^b(\bar{w})=
\frac{\de^{a b}}{(\bar{z} -\bar{w})}
+ \cdots.
\label{fieldOPE}
\eea
For the fermions in the second equation of
(\ref{fieldOPE}), the proper normalization is
performed, compared to the one in \cite{AP}.
The conformal weights for
$\phi^a$, $\bar{\pa} \, \bar{\phi}^a$, $\la^a$ and $ \bar{\la}^a$
in the antiholomorphic sector 
are given by $\frac{1}{2(q_{syk}+1)}$, $1-\frac{1}{2(q_{syk}+1)}$,
$\frac{1}{2} +\frac{1}{2(q_{syk}+1)}$ and
$\frac{1}{2} -\frac{1}{2(q_{syk}+1)}$.

The lowest supermultiplet contains the weight-$1$ operator,
two supercharges and the stress energy tensor.
Then the standard ${\cal N}=2$ superconformal algebra
is realized by \cite{Witten,AP}
\bea
J &=& \frac{ q_{syk}}{2(q_{syk}+1)}
\bar{\la}^a \, {\la}^a
-
\frac{1}{(q_{syk}+1)}
\phi^a \, \bar{\pa} \, \bar{\phi}^a,
\nonu \\
G^{+} & = &
\frac{1}{\sqrt{2}}
\, \bar{\pa} \, \bar{\phi}^a \, \la^a,
\nonu
 \\
G^{-} & = & -\frac{ q_{syk}}{\sqrt{2}\, (1+q_{syk})}
\, \Bigg[
  \bar{\pa} \, \phi^a \,
  \bar{\la}^a -\frac{1}{q_{syk}} \,
  \phi^a \, \bar{\pa} \,
\bar{\la}^a \Bigg],
\nonu
\\
T &=& \frac{1}{(2q_{syk}+2)}
\Bigg[ {(2q_{syk}+1)} \, \bar{\pa} \, \phi^a \, \bar{\pa}
\, \bar{\phi}^a -\, \phi^a \, \bar{\pa}^2 \, \bar{\phi}^a
- \frac{(q_{syk}+2)}{2}   \, \bar{\pa} \,
\bar{\la}^a\, \la^a +\frac{q_{syk}}{2}   \, \bar{\la}^a\,
\bar{\pa} \, \la^a
\Bigg].
\nonu \\
&& \label{SCA}
\eea
The central charge in (\ref{SCA}), where the fourth order pole
in the OPE $T(\bar{z})\, T(\bar{w})$
is $\frac{c}{2}$, is given by
\bea
c = 3N\, \frac{(q_{syk}-1)}{(q_{syk}+1)}.
\label{cen}
\eea
Each independent term in the stress energy tensor
contributes to its own central term and the overall factor $N$ appears in
(\ref{cen}).
Some typo in \cite{AP} is corrected in (\ref{SCA}).
Note that we can multiply any (pure imaginary) numerical number
into the $G^{+}$ and its inverse into the $G^{-}$ without changing
the definition of the ${\cal N}=2$ superconformal algebra.
The central terms of the OPEs, $J(\bar{z}) \, J(\bar{w})$
and $G^{+}(\bar{z})\, G^{-}(\bar{w})$, are given by $\frac{c}{3}$
and $\frac{c}{3}$ respectively
\footnote{Note that in the ${\cal N}=(0,2)$ SYK model,
  the stress energy tensor takes the more general form
  \cite{AP} and the
  condition for the ${\cal N}=2$ supersymmetry enables us to
have simpler form for the stress energy tensor.}.

\subsection{The ${\cal N}=2$ supersymmetric linear
  $W_{\infty}[\la]$ algebra}

In \cite{Bergetalnpb,Bergetalplb}, 
the  ${\cal N}=2$ supersymmetric linear
$W_{\infty}[\la]$ algebra \footnote{In this paper, we are considering
  the linear algebra where the corresponding OPE
  does not have any quadratic or higher order terms in the currents
  of the right hand side although the currents are quadratic in the
  operators. See also the relevant work in \cite{HP}
  where the nonlinear structures occur in the context of $AdS_3$
  higher spin theory.}
is realized by the following $\beta \, \ga$ and $b \, c$
ghost systems which satisfy the OPEs
\bea
\ga(\bar{z})\, \beta(\bar{w}) = \frac{1}{(\bar{z}-\bar{w})} + \cdots,
\qquad
c(\bar{z}) \, b(\bar{w}) = \frac{1}{(\bar{z}-\bar{w})} + \cdots.
\label{ghost}
\eea
The conformal weights for
$\beta$, $\ga$, $b$ and $c$
are given by $\la$, $1-\la$, $\frac{1}{2}+\la$
and $\frac{1}{2}-\la$ respectively \footnote{
\label{lahigher}
  In
terms of  the parameter $\la_{hs}$ of the higher spin algebra
$hs[\la_{hs}]$, there exists a relation
$\la=\frac{1}{2} \, \la_{hs}$. }.
Note that the normalizations in the right hand sides of
(\ref{ghost}) are given by $+1$.

Then the higher spin currents are given by
\cite{Bergetalnpb,Bergetalplb} \footnote{We thank M. Vasiliev for pointing
  out that these are quasiprimary operators under the stress energy tensor
  $V_{\la}^{(2)+}$ ten years ago.}
\bea
V_{\la}^{(s)+} & = & \sum_{i=0}^{s-1}\, a^i(s, \la)\, \bar{\pa}^{s-1-i}\,
(( \bar{\pa}^i \, \beta ) \, \ga) +
 \sum_{i=0}^{s-1}\, a^i(s, \la+\frac{1}{2})\, \bar{\pa}^{s-1-i}\,
(( \bar{\pa}^i \, b ) \, c ),
 \nonu \\
 V_{\la}^{(s)-} & = & -\frac{(s-1+2\la)}{(2s-1)}\,
 \sum_{i=0}^{s-1}\, a^i(s, \la)\, \bar{\pa}^{s-1-i}\,
 (( \bar{\pa}^i \, \beta ) \, \ga) \nonu \\
 & + &
 \frac{(s-2\la)}{(2s-1)}\,
 \sum_{i=0}^{s-1}\, a^i(s, \la+\frac{1}{2})\, \bar{\pa}^{s-1-i}\,
 (( \bar{\pa}^i \, b ) \, c ),
 \nonu \\
Q_{\la}^{(s)+} & = & \sum_{i=0}^{s-1}\, \al^i(s, \la)\, \bar{\pa}^{s-1-i}\,
(( \bar{\pa}^i \, \beta ) \, c) -
 \sum_{i=0}^{s-2}\, \beta^i(s, \la)\, \bar{\pa}^{s-2-i}\,
 (( \bar{\pa}^i \, b ) \, \ga ),
 \nonu \\
Q_{\la}^{(s)-} & = & \sum_{i=0}^{s-1}\, \al^i(s, \la)\, \bar{\pa}^{s-1-i}\,
(( \bar{\pa}^i \, \beta ) \, c) +
 \sum_{i=0}^{s-2}\, \beta^i(s, \la)\, \bar{\pa}^{s-2-i}\,
 (( \bar{\pa}^i \, b ) \, \ga ).
 \label{VVQQ}
 \eea
 The $\la$-dependent coefficients appearing in (\ref{VVQQ}) are given by
\bea 
 a^i(s, \la) \equiv \left(\begin{array}{c}
s-1 \\  i \\
 \end{array}\right) \, \frac{(-2\la-s+2)_{s-1-i}}{(s+i)_{s-1-i}},
 \qquad 0 \leq i \leq (s-1),
 \nonu \\
 \al^i(s, \la) \equiv \left(\begin{array}{c}
s-1 \\  i \\
 \end{array}\right) \, \frac{(-2\la-s+2)_{s-1-i}}{(s+i-1)_{s-1-i}},
 \qquad 0 \leq i \leq (s-1),
 \nonu \\
  \beta^i(s, \la) \equiv \left(\begin{array}{c}
s-2 \\  i \\
  \end{array}\right) \, \frac{(-2\la-s+2)_{s-2-i}}{(s+i)_{s-2-i}},
  \qquad 0 \leq i \leq (s-2).
  \label{coeff}
 \eea
 The $\la$-dependent coefficients (\ref{coeff}) are not independent.
 Some properties of these coefficients are given by Appendix $A$
 of \cite{Bergetalnpb}. The binomial coefficients
 for parentheses are used and the rising Pochhammer
 symbol $(a)_n \equiv a(a+1) \cdots (a+n-1)$ is also used here.
 We can check that
 the ${\cal N}=2$ superconformal generators
 are given by \footnote{In next section we will present their explicit
   forms in terms of the composite operators in the ghost systems.
   When we take $N=1$ over there, then we obtain the exact results
 \cite{Bergetalnpb,Bergetalplb}.}
 \bea
 J & = & V^{(1)-}_{\la}, \nonu \\
 G^{+} & = & -\frac{\sqrt{2}}{2} \,
 (Q^{(2)+}_{\la}- Q^{(2)-}_{\la}), \nonu \\
 G^- & = & \frac{\sqrt{2}}{4}\,
 (Q^{(2)+}_{\la}+ Q^{(2)-}_{\la}), \nonu \\
 T & = & V^{(2)+}_{\la}.
   \label{SCA1}
   \eea
   The lowest $s$ value for the bosonic currents $V_{\la}^{(s) \pm}$
   is given by $s=1$. One of them  plays the role of
   the weight-$1$ current of the ${\cal N}=2$ superconformal
   algebra in (\ref{SCA1}).
 The lowest $s$ value for the fermionic currents $Q_{\la}^{(s) \pm}$
 is given by $s=1$ also.
 In \cite{Bergetalnpb,Bergetalplb}, the ${\cal N}=2$ scalar
 multiplet is denoted by
 $(Q_{\la}^{(1) +}=Q_{\la}^{(1) -}, V_{\la}^{(1)+})$.
 We can easily see that the weights for
 the composite
 operators $\beta \, \ga$, $ b \, c$, $\beta \, c$ and $b \, \ga$
 are given by $1$, $1$, $\frac{1}{2}$ and $\frac{3}{2}$
 respectively and all the $\la$ dependence is gone.
 This means that
 their weights for the  bosonic currents $V_{\la}^{(s) \pm}$
 are given by $s$ while
 the weights for the  fermionic currents $Q_{\la}^{(s) \pm}$
 are given by $(s-\frac{1}{2})$
 \footnote{In terms of the ghost systems,
   we have $V_{\la}^{(1)+}=\beta \, \ga + b \, c$
   and $Q_{\la}^{(1)+}=Q_{\la}^{(1)-}=\beta c$. In order to calculate
   some OPEs between the ghost systems and the currents in (\ref{VVQQ}),
   some partial results on the highest order poles
   between them in \cite{Bergetalnpb,MZ} are helpful.
   For example, for the calculation of the various central charges in the
 given OPEs, we have to consider the highest order poles only.}.

\section{ The ${\cal N}=2$ supersymmetric linear
  $W^{N,N}_{\infty}[\la]$ algebra}

\subsection{The matrix generalization of
  ${\cal N}=2$ supersymmetric linear
  $W_{\infty}[\la]$ algebra}

In order to describe the multiple number of
the chiral multiplets (or the Fermi multiplets),
we need to introduce the multiple number of
$\beta \, \ga$ and $b \, c$ systems \cite{CHR}
satisfying the following defining OPEs
\bea
\ga^{i,\bar{a}}(\bar{z})\, \beta^{\bar{j},b}(\bar{w}) =
\frac{1}{(\bar{z}-\bar{w})}\, \de^{i \bar{j}}\, \de^{\bar{a} b} + \cdots,
\qquad
c^{i, \bar{a}}(\bar{z}) \, b^{\bar{j},b}(\bar{w}) =
\frac{1}{(\bar{z}-\bar{w})}\, \de^{i \bar{j}}\, \de^{\bar{a} b} + \cdots.
\label{fundOPE}
\eea
The fundamental indices $a, b, \cdots $ of $SU(N)$ in (\ref{fundOPE})
runs over $a, b, \cdots = 1, 2, \cdots, N$ and 
the antifundamental indices $\bar{a}, \bar{b}, \cdots $ of $SU(N)$
runs over $\bar{a}, \bar{b}, \cdots = 1, 2, \cdots, N$.
Similarly, we can associate the indices $i, j, \cdots$
and the indices $\bar{i}, \bar{j}, \cdots$ with the corresponding
fundamental and antifundamentals of $SU(L)$ \footnote{
We consider $SU(L)$-singlet currents in this paper.}.

By multiplying the generators of $SU(N)$ into the previous
relations (\ref{VVQQ}), we obtain the following matrix
generalization of the work in \cite{Bergetalnpb,Bergetalplb}
\bea
V_{\la}^{(s)+} & = & \sum_{i=0}^{s-1}\, a^i(s, \la)\, \bar{\pa}^{s-1-i}\,
(( \bar{\pa}^i \, \beta^{\bar{l} b} ) \, \de_{l \bar{l}} \, \de_{\bar{a} b} \,
\ga^{l \bar{a}}) +
 \sum_{i=0}^{s-1}\, a^i(s, \la+\frac{1}{2})\, \bar{\pa}^{s-1-i}\,
 (( \bar{\pa}^i \, b^{\bar{l} b} ) \,  \de_{l \bar{l}} \,
\de_{\bar{a} b}\, 
 c^{l \bar{a}} ),
 \nonu \\
 V_{\la,\hat{A}}^{(s)+} & = &
 \sum_{i=0}^{s-1}\, a^i(s, \la)\, \bar{\pa}^{s-1-i}\,
 (( \bar{\pa}^i \, \beta^{\bar{l} b} ) \, \de_{l \bar{l}} \,
 t^{\hat{A}}_{b \bar{a}} \,
\ga^{l \bar{a}}) +
 \sum_{i=0}^{s-1}\, a^i(s, \la+\frac{1}{2})\, \bar{\pa}^{s-1-i}\,
 (( \bar{\pa}^i \, b^{\bar{l} b} ) \,  \de_{l \bar{l}} \,
t^{\hat{A}}_{b \bar{a} } \, 
 c^{l \bar{a}} ),
 \nonu \\
 V_{\la}^{(s)-} & = & -\frac{(s-1+2\la)}{(2s-1)}\,
 \sum_{i=0}^{s-1}\, a^i(s, \la)\, \bar{\pa}^{s-1-i}\,
 (( \bar{\pa}^i \, \beta^{\bar{l} b} ) \, \de_{l \bar{l}} \, \de_{\bar{a} b} \,
 \ga^{l \bar{a}}) \nonu \\
 & + &
 \frac{(s-2\la)}{(2s-1)}\,
 \sum_{i=0}^{s-1}\, a^i(s, \la+\frac{1}{2})\, \bar{\pa}^{s-1-i}\,
 (( \bar{\pa}^i \, b^{\bar{l} b} ) \,  \de_{l \bar{l}} \,
\de_{\bar{a} b}\, c^{l \bar{a}} ),
\nonu \\
 V_{\la,\hat{A}}^{(s)-} & = & -\frac{(s-1+2\la)}{(2s-1)}\,
 \sum_{i=0}^{s-1}\, a^i(s, \la)\, \bar{\pa}^{s-1-i}\,
 (( \bar{\pa}^i \, \beta^{\bar{l} b} ) \, \de_{l \bar{l}} \,
 t^{\hat{A}}_{b \bar{a} } \,
 \ga^{l \bar{a}}) \nonu \\
 & + &
 \frac{(s-2\la)}{(2s-1)}\,
 \sum_{i=0}^{s-1}\, a^i(s, \la+\frac{1}{2})\, \bar{\pa}^{s-1-i}\,
 (( \bar{\pa}^i \, b^{\bar{l} b} ) \,  \de_{l \bar{l}} \,
t^{\hat{A}}_{b \bar{a}}\, c^{l \bar{a}} ),
 \nonu \\
Q_{\la}^{(s)+} & = & \sum_{i=0}^{s-1}\, \al^i(s, \la)\, \bar{\pa}^{s-1-i}\,
(( \bar{\pa}^i \, \beta^{\bar{l} b} ) \, \de_{l \bar{l}} \,
\de_{\bar{a} b} \, c^{l \bar{a}}) -
 \sum_{i=0}^{s-2}\, \beta^i(s, \la)\, \bar{\pa}^{s-2-i}\,
 (( \bar{\pa}^i \, b^{\bar{l} b} ) \, \de_{l \bar{l}} \,
\de_{\bar{a} b}\, \ga^{l \bar{a}} ),
\nonu \\
Q_{\la, \hat{A}}^{(s)+} & = &
\sum_{i=0}^{s-1}\, \al^i(s, \la)\, \bar{\pa}^{s-1-i}\,
(( \bar{\pa}^i \, \beta^{\bar{l} b} ) \, \de_{l \bar{l}} \,
t^{\hat{A}}_{b \bar{a} } \, c^{l \bar{a}}) -
 \sum_{i=0}^{s-2}\, \beta^i(s, \la)\, \bar{\pa}^{s-2-i}\,
 (( \bar{\pa}^i \, b^{\bar{l} b} ) \, \de_{l \bar{l}} \,
t^{\hat{A}}_{b \bar{a} }\, \ga^{l \bar{a}} ),
 \nonu \\
Q_{\la}^{(s)-} & = & \sum_{i=0}^{s-1}\, \al^i(s, \la)\, \bar{\pa}^{s-1-i}\,
(( \bar{\pa}^i \, \beta^{\bar{l} b} ) \, \de_{l \bar{l}} \,
\de_{\bar{a} b}\, c^{l \bar{a}}) +
 \sum_{i=0}^{s-2}\, \beta^i(s, \la)\, \bar{\pa}^{s-2-i}\,
 (( \bar{\pa}^i \, b^{\bar{l} b} ) \, \de_{l \bar{l}} \,
\de_{\bar{a} b}\, \ga^{l \bar{a}} ),
\nonu \\
Q_{\la,\hat{A}}^{(s)-} & = &
\sum_{i=0}^{s-1}\, \al^i(s, \la)\, \bar{\pa}^{s-1-i}\,
(( \bar{\pa}^i \, \beta^{\bar{l} b} ) \, \de_{l \bar{l}} \,
t^{\hat{A}}_{b \bar{a}}\, c^{l \bar{a}}) +
 \sum_{i=0}^{s-2}\, \beta^i(s, \la)\, \bar{\pa}^{s-2-i}\,
 (( \bar{\pa}^i \, b^{\bar{l} b} ) \, \de_{l \bar{l}} \,
t^{\hat{A}}_{b \bar{a} }\, \ga^{l \bar{a}} ).
\label{VVQQla}
\eea
The adjoint index $\hat{A}$ runs over $\hat{A}=1, 2, \cdots,
(N^2-1)$ \footnote{Note that in \cite{CHR}, there appear the extra
factors $\pm (-1)^s$ or $\pm (-1)^{s-\frac{1}{2}}$ in various places
in the coefficients of (\ref{VVQQla}).}.

The central charge of the stress energy tensor is given by
\footnote{We consider the $L=1$ case. If we consider the general
$L$, then this $L$ factor appears in the central charge.}
\bea
c = 3 \, N \, (1-4\la).
\label{cen1}
\eea
By comparing (\ref{cen}) with (\ref{cen1}),
the deformation parameter in \cite{Bergetalnpb,Bergetalplb}
plays the role of the rank of the random coupling of the
SYK model and it is given by \footnote{It has been
  conjectured in \cite{Peng1805} from the ${\cal N}=(0,2)$
  SYK models that
  the parameter $\la_{hs}$ (See also the footnote \ref{lahigher})
  is related to
  the $q_{syk}$ and is given by $\la_{hs} =
  \frac{1}{\frac{M}{N}\, q_{syk}}$. }
\bea
\la = \frac{1}{2(q_{syk}+1)}.
\label{laq}
\eea

We can write down the generators of the ${\cal N}=2$
superconformal algebra for matrix
generalization from (\ref{SCA1}) as follows:
\bea
J & = & (-1+2\la) \, c^a \, b^a - 2\la \, \beta^a \, \ga^a,  
\nonu \\
G^+ & = & \sqrt{2} \, \ga^a \, b^a,
\nonu \\
G^- & = & -\sqrt{2} \, \la\, \beta^a \, \bar{\pa} \, \bar{\la}^a-
\frac{(-1+2\la)}{\sqrt{2}}\, \bar{\pa} \,
 \beta^a  \, \bar{\la}^a,
 \nonu \\
 T & = & (1-\la) \, \bar{\pa} \, \beta^a \ga^a-\la \, \beta^a \,
 \bar{\pa} \, \ga^a +\frac{1}{2} \, (1+2\la)\,\bar{\pa}\,
 c^a \, b^a + \frac{1}{2}\, (-1+2\la) \, c^a \, \bar{\pa}\, b^a.
 \label{SCA2}
 \eea
 For $N=1$, we observe that
 the above relations (\ref{SCA2}) are reduced to
 the ones in \cite{Bergetalnpb}.
 By realizing the following
 relations with (\ref{laq}) \footnote{In \cite{AP}, we used the
   different terminology for the bosons.
   Note that when we change the
   ordering in the first OPE of (\ref{fundOPE}), there is a minus sign
   in the right hand side while there is no minus sign in the second OPE
   of (\ref{fundOPE}) after this change.
   We should also make sure that they have the correct
 weights in terms of deformation parameter.},
 \bea
 \ga^a \leftrightarrow \bar{\pa}\, \bar{\phi}^a,
 \qquad \beta^a \leftrightarrow \phi^a, \qquad
 c^a \leftrightarrow \frac{i}{\sqrt{2}} \, \bar{\la}^a,
 \qquad
 b^a \rightarrow \frac{i}{\sqrt{2}}\, \la^a,
 \label{identi}
 \eea
 we observe that  the relations (\ref{SCA2})
 can be identified with the ones in (\ref{SCA}) together with a factor
 $\sqrt{2} i$ in $G^+$ and a factor $-\frac{i}{\sqrt{2}}$ in
 $G^-$ (The OPE $G^{+}(\bar{z}) \, G^{-}(\bar{w})$ does not change
 with these factors).
 The conformal weights for both sides in (\ref{identi})
 are consistent with
 each other.
 We expect that there is a one-to one correspondence between
 the ${\cal N}=2$ SYK model and the
 $\beta \, \ga$ and $b \, c$ ghost systems in the
 ${\cal N}=2$ supersymmetric linear $W^{N,N}_{\infty}[\la]$ algebra.
 
\subsection{
  The ${\cal N}=2$ supersymmetric linear $W^{N,N}_{\infty}[\la=0]$ algebra}

From the exact correspondence between
the chiral multiplets and the Fermi multiplets
of the ${\cal N}=2$ $(2,2)$ SYK model and
the $\beta \, \ga$ and $b \, c$ ghost systems in (\ref{identi}),
we expect that there exist the precise relations for the
higher spin currents between them.
By linear combinations among the higher spin currents in
(\ref{VVQQla})
we can write down the higher spin currents of \cite{Ahn2202}
in terms of (\ref{VVQQla}) at $\la =0$
and it turns out that for $SU(N)$-singlet currents we have
\footnote{The parameter
  $q$ here is the same as the $\la$ in \cite{Ahn2202}.
  We can associate the $\psi^{\bar{i},a}$ and
  $\bar{\psi}^{j, \bar{b}}$ of \cite{Ahn2202}
with $b^{\bar{i},a}$ and $c^{j,\bar{b}}$.}
\bea
W_{F,h} & = &
\frac{n_{W_{F,h}}}{q^{h-2}}\,
\frac{(-1)^h}{\sum_{i=0}^{h-1}\, a^i( h, \la+\frac{1}{2}=\frac{1}{2})}\,
\Bigg[\frac{(h-1+2\la)}{(2h-1)}\, V_{\la}^{(h)+} + V_{\la}^{(h)-}
\Bigg]_{\la=0},
\nonu \\
W_{B,h} & = &
 \frac{n_{W_{B,h}}}{q^{h-2}}\,
 \frac{(-1)^h}{\sum_{i=0}^{h-1}\, a^i( h, \la=0)}
\,
\Bigg[\frac{(h-2\la)}{(2h-1)}\, V_{\la}^{(h)+} - V_{\la}^{(h)-}
  \Bigg]_{\la=0},
\nonu \\
Q_{h+\frac{1}{2}} & = & \frac{1}{2} \,\frac{n_{W_{Q,h+\frac{1}{2}}}}{q^{h-1}}
\, \frac{(-1)^{h+1}  \, h }{
   \sum_{i=0}^{h-1} \, \beta^i( h+1, \la=0)}\, \Bigg[
  Q_{\la}^{(h+1)-} - Q_{\la}^{(h+1)+}\Bigg]_{\la=0},
\nonu \\
\bar{Q}_{h+\frac{1}{2}} & = & \frac{1}{2} \,
\frac{n_{W_{Q,h+\frac{1}{2}}}}{q^{h-1}} \,
 \frac{(-1)^{h+1}  }{
   \sum_{i=0}^{h} \, \al^i( h+1, \la=0)} \,
\Bigg[ Q_{\la}^{(h+1)-} +
  Q_{\la}^{(h+1)+}\Bigg]_{\la=0}.
\label{WWQQlazero}
\eea
For $h=1$ with $\la=0$,
the coefficient of the first term of $W_{F,h=1}$
in (\ref{WWQQlazero}) vanishes and
the $W_{F,h=1}$ is proportional to $V_{\la=0}^{(1)-}=-c^a \, b^a$.
See also (\ref{SCA1}) and (\ref{SCA2}).
On the other hand, the coefficient of the first term of $W_{B,h=1}$
in (\ref{WWQQlazero}) does not vanish and
the $W_{B,h=1}$ with $\la=0$
is proportional to $-\ga^a \, \beta^a$ (which holds for nonzero
$\la$).
Then the current $W_{B,h=1}$ arises only in the
$\beta \, \ga$ and $b \, c$ ghost systems.
For $h=0$, the $Q_{\frac{1}{2}}$ vanishes and the
$\bar{Q}_{\frac{1}{2}}$ is proportional to $-\beta^a \, c^a$
which does not occur in the construction of \cite{Ahn2202}.
See also the footnote \ref{extracurrents}.
Therefore, we expect that there appears
the presence of the current $W_{B,h=1}$ and the current
$\bar{Q}_{\frac{1}{2}}$ in the ${\cal N}=2$ supersymmetric
linear $W_{\infty}^{N,N}[\la =0]$ algebra \footnote{The $SU(N)$-adjoint
currents are given by Appendix $A$.}.

Furthermore, we can compare with each coefficient
appearing in the free field realization in \cite{Ahn2202}
and the one in (\ref{VVQQla}) at $\la =0$.
In order to do this, we should act the antiholomorphic
partial derivatives on the composite operators fully.
Then the binomial coefficients appear.
It turns out that there appear the following identities
\bea
\frac{n_{W_{F,h}}}{q^{h-2}} (-1)^k   \left(\begin{array}{c}
h-1 \\  k \\
\end{array}\right)^2 & = & \frac{n_{W_{F,h}}}{q^{h-2}}
\frac{1 }{\sum_{i=0}^{h-1} a^i( h, \frac{1}{2})}
\sum_{i=0}^{h-1}\, a^i( h, \frac{1}{2})
 \left(\begin{array}{c}
h-1-i \\  k \\
\end{array}\right),
 \nonu \\
 \frac{n_{W_{B,h}}}{q^{h-2}}
 \frac{(-1)^k}{(h-1)}   \left(\begin{array}{c}
h-1 \\  k \\
\end{array}\right)
 \left(\begin{array}{c}
h-1 \\  k+1 \\
\end{array}\right)
 & = &
 \frac{n_{W_{B,h}}}{q^{h-2}}
 \frac{1 }{\sum_{i=0}^{h-1} a^i( h, 0)}
\sum_{i=0}^{h-1}\, a^i( h, 0)
 \left(\begin{array}{c}
h-1-i \\  k \\
\end{array}\right),
 \nonu  \\
 \frac{n_{W_{Q,h}}(-1)^k}{q^{h-\frac{3}{2}}} 
 \left(\begin{array}{c}
h-\frac{3}{2} \\  k \\
\end{array}\right)
 \left(\begin{array}{c}
h-\frac{1}{2} \\  k \\
\end{array}\right)
 & = &
\frac{n_{W_{Q,h-\frac{1}{2}}}}{q^{h-2}}
 \frac{(-1)^h  (h-1) }{
   \sum_{i=0}^{h-2} \beta^i( h, 0)}
\sum_{i=0}^{h-2} \beta^i( h, 0)
 \left(\begin{array}{c}
h-2-i \\  k \\
\end{array}\right),
 \nonu \\
\frac{n_{W_{Q,h}} (-1)^{h-\frac{3}{2} +k} }{q^{h-\frac{3}{2}}}
  \left(\begin{array}{c}
h-\frac{3}{2} \\  k \\
\end{array}\right)
 \left(\begin{array}{c}
h-\frac{1}{2} \\  k \\
\end{array}\right)
 & = &
\frac{n_{W_{Q,h-\frac{1}{2}}}}{q^{h-2}}
 \frac{(-1)^h  }{
   \sum_{i=0}^{h-1} \al^i( h, 0)}
\sum_{i=0}^{h-1} \al^i( h, 0)
\left(\begin{array}{c}
h-1-i \\  k \\
\end{array}\right).
\nonu \\
\label{iden}
\eea
It is rather nontrivial to check these relations for
generic $h$ and $k$, but we can try to do this for several values for
these quantities. Note that in the right hand sides of (\ref{iden}),
the additional binomial coefficients occur
by expanding the antiholomorphic
partial derivatives fully as described before \footnote{
Note that the various summations in the denominators of the
right hand sides of (\ref{iden}) can be written in terms of
the fractional forms of the various gamma functions at nonzero $\la$
\cite{MZ}.}.

Therefore, the ${\cal N}=2$  SYK model
has  ${\cal N}=2$ supersymmetric linear
$W^{N,N}_{\infty}[\la=0]$ algebra where
the higher spin currents are given by (\ref{VVQQla})
and by using the relations (\ref{WWQQlazero}), the explicit
(anti)commutator relations can be read off from the previous results
in \cite{Ahn2202}(See also \cite{Odake}).
The relation between the parameters
is given by (\ref{laq}). Of course, as explained before,
in the (anti)commutators relations, we observe that there appear
the currents $W_{B,h=1}$ and 
$\bar{Q}_{\frac{1}{2}}$. Moreover, their OPEs with other higher
spin currents $W_{F,h\geq 1}$, $W_{B,h\geq 1}$, $Q_{h+\frac{1}{2}\geq
\frac{1}{2}}$
and $\bar{Q}_{h+\frac{1}{2}\geq \frac{1}{2}}$ will appear in general.
In next section, we will present the (anti)commutator relations
for nonzero $\la$. Therefore, once we put the $\la$ to be zero in these
equations, we obtain the final results.

\subsubsection{
  The realization of
  ${\cal N}=2$ supersymmetric linear $W^{N,N}_{\infty}[\la=0]$ algebra
in the ${\cal N}=2$ SYK model }

That is, in the limit of  
\bea
q_{syk} \rightarrow \infty,
\label{qsyklimit}
\eea
the ${\cal N}=2$ SYK models reveal the
${\cal N}=2$ supersymmetric linear $W^{N,N}_{\infty}[\la=0]$ algebra.
The generators are given by
\bea
&& i) \qquad W_{F,h\geq 1}, \qquad W_{B,h\geq 2}, \qquad
Q_{h+\frac{3}{2}\geq
  \frac{3}{2}}, \qquad \bar{Q}_{h+\frac{1}{2}\geq \frac{3}{2}},
\nonu \\
&& ii) \qquad W_{B,h=1}, \qquad \bar{Q}_{h+\frac{1}{2}= \frac{1}{2}}.
\label{gen}
\eea
The algebra between the currents in the first line of
(\ref{gen})
is closed and the explicit form is given by the ones
in \cite{Ahn2202}. In the right hand sides of these
(anti)commutator relations we can see only the operators
in the first line of (\ref{gen}).
Due to the presence of the operators in the second line of (\ref{gen}),
we should calculate the OPEs between these weight-$1,\frac{1}{2}$ currents
and the remaining ones in the first line of (\ref{gen})
as well as their own OPEs
in order to
describe the full algebra if we do not decouple these currents
\footnote{We will see that  we have their explicit forms
  as 
  $W_{B,1}=-\frac{1}{4}\, \ga^a \, \beta^a$ and
  $\bar{Q}_{\frac{1}{2}}=-\frac{1}{\sqrt{2}}\,
  \beta^a \, c^a$. Even they do not depend on the $\la$ parameter
  from the footnote \ref{extracurrents}.
  By construction of \cite{Ahn2202}, there is no
  $\phi^{\bar{i},a}$ dependence and its derivative
  $\bar{\pa}\, \phi^{\bar{i},a}$
appears only.}.
As we will see next section, 
once the $\la$ becomes nonzero value (a deviation from (\ref{qsyklimit})),
then this does not hold any more
because the right hand sides of these (anti)commutator relations
possess the operators in the second line of (\ref{gen}). 

\subsection{The
  ${\cal N}=2$ supersymmetric linear
  $W^{N,N}_{\infty}[\la]$ algebra }

\subsubsection{The higher spin currents for nonzero $\la$}

Let us consider the nonzero $\la$ case.
We take the previous expressions (\ref{WWQQlazero})
by considering the $\la$ dependence explicitly. Then we have
the following $SU(N)$-singlet currents
\footnote{
\label{equalcoeff}
  The $SU(N)$-adjoint
  currents are given by Appendix $A$. Each $\la$ independent
  coefficient of bosonic current is the same and
 each $\la$ independent
 coefficient of fermionic current is the same:
$\frac{n_{W_{F,h}}}{q^{h-2}}\,
 \frac{(-1)^h}{\sum_{i=0}^{h-1}\, a^i( h, \la+\frac{1}{2}=\frac{1}{2})}\,=
  \frac{n_{W_{B,h}}}{q^{h-2}}\,
 \frac{(-1)^h}{\sum_{i=0}^{h-1}\, a^i( h, \la=0)}
\,$ and $\frac{1}{2} \,\frac{n_{W_{Q,h+\frac{1}{2}}}}{q^{h-1}}
\, \frac{(-1)^{h+1}  \, h }{
   \sum_{i=0}^{h-1} \, \beta^i( h+1, \la=0)}\,= \frac{1}{2} \,
\frac{n_{W_{Q,h+\frac{1}{2}}}}{q^{h-1}} \,
 \frac{(-1)^{h+1}  }{
   \sum_{i=0}^{h} \, \al^i( h+1, \la=0)} \,$.}
\bea
W^{\la}_{F,h} & = &
\frac{n_{W_{F,h}}}{q^{h-2}}\,
\frac{(-1)^h}{\sum_{i=0}^{h-1}\, a^i( h, \la+\frac{1}{2}=\frac{1}{2})}\,
\Bigg[\frac{(h-1+2\la)}{(2h-1)}\, V_{\la}^{(h)+} + V_{\la}^{(h)-}
\Bigg],
\nonu \\
W^{\la}_{B,h} & = &
 \frac{n_{W_{B,h}}}{q^{h-2}}\,
 \frac{(-1)^h}{\sum_{i=0}^{h-1}\, a^i( h, \la=0)}
\,
\Bigg[\frac{(h-2\la)}{(2h-1)}\, V_{\la}^{(h)+} - V_{\la}^{(h)-}
  \Bigg],
\nonu \\
Q^{\la}_{h+\frac{1}{2}} & = & \frac{1}{2} \,\frac{n_{W_{Q,h+\frac{1}{2}}}}{q^{h-1}}
\, \frac{(-1)^{h+1}  \, h }{
   \sum_{i=0}^{h-1} \, \beta^i( h+1, \la=0)}\, \Bigg[
  Q_{\la}^{(h+1)-} - Q_{\la}^{(h+1)+}\Bigg],
\nonu \\
\bar{Q}^{\la}_{h+\frac{1}{2}} & = & \frac{1}{2} \,
\frac{n_{W_{Q,h+\frac{1}{2}}}}{q^{h-1}} \,
 \frac{(-1)^{h+1}  }{
   \sum_{i=0}^{h} \, \al^i( h+1, \la=0)} \,
\Bigg[ Q_{\la}^{(h+1)-} +
  Q_{\la}^{(h+1)+}\Bigg].
\label{WWQQnonzerola}
\eea
In particular, $V_{\la}^{(1)-}$ has $\ga^a \,\beta^a $ term also
for nonzero $\la$. See also the weight-$1$ current in (\ref{SCA2}).
We would like to obtain the algebra generated by these currents in
(\ref{WWQQnonzerola}).

We present the currents for low weights as follows:
\bea
W^{\la}_{F,1} & = & -\frac{1}{4}\,
\Big( V_{\la}^{(1)-} + 2\la \, V^{(1)+}_{\la}\Big),
\qquad
W^{\la}_{F,2}  = 
\Big( V_{\la}^{(2)-} + \frac{1}{3} \, (1+2\la) \, V^{(2)+}_{\la}\Big),
\nonu \\
W^{\la}_{F,3}  & = & -4 \, 
\Big( V_{\la}^{(3)-} + \frac{1}{5} \, (2+2\la) \, V^{(3)+}_{\la}\Big),
\qquad
W^{\la}_{F,4}  =16 \, 
\Big( V_{\la}^{(4)-} + \frac{1}{7} \, (3+2\la) \, V^{(4)+}_{\la}\Big),
\nonu \\
W^{\la}_{B,1} & = & -\frac{1}{4}\,
\Big( -V_{\la}^{(1)-} + (1-2\la) \, V^{(1)+}_{\la}\Big),
\qquad
W^{\la}_{B,2}  = 
\Big( -V_{\la}^{(2)-} + \frac{1}{3} \, (2-2\la) \, V^{(2)+}_{\la}\Big),
\nonu \\
W^{\la}_{B,3}  & = & -4 \, 
\Big(- V_{\la}^{(3)-} + \frac{1}{5} \, (3-2\la) \, V^{(3)+}_{\la}\Big),
\qquad
W^{\la}_{B,4}  =16 \, 
\Big( -V_{\la}^{(4)-} + \frac{1}{7} \, (4-2\la) \, V^{(4)+}_{\la}\Big),
\nonu \\
Q^{\la}_{ \frac{3}{2}} & = & \frac{1}{\sqrt{2}} \,
\Big( Q_{\la}^{ (2) -} -  Q_{\la}^{ (2) +}\Big),
\qquad
Q^{\la}_{ \frac{5}{2}}  =  - 2 \sqrt{2} \,
\Big( Q_{\la}^{ (3) -} -  Q_{\la}^{ (3) +}\Big),
\nonu \\
Q^{\la}_{ \frac{7}{2}}  & = &  8 \sqrt{2} \,
\Big( Q_{\la}^{ (4) -} -  Q_{\la}^{ (4) +}\Big),
\qquad Q^{\la}_{ \frac{9}{2}}   =   -32 \sqrt{2} \,
\Big( Q_{\la}^{ (5) -} -  Q_{\la}^{ (5) +}\Big),
\nonu \\
\bar{Q}^{\la}_{ \frac{1}{2}} & = & -\frac{1}{2\sqrt{2}} \,
\Big( \bar{Q}_{\la}^{ (1) -} +  \bar{Q}_{\la}^{ (1) +}\Big),
\qquad
\bar{Q}^{\la}_{ \frac{3}{2}}  =  \frac{1}{\sqrt{2}} \,
\Big( \bar{Q}_{\la}^{ (2) -} +  \bar{Q}_{\la}^{ (2) +}\Big),
\nonu \\
\bar{Q}^{\la}_{ \frac{5}{2}} & = & -2\sqrt{2} \,
\Big( \bar{Q}_{\la}^{ (3) -} +  \bar{Q}_{\la}^{ (3) +}\Big),
\qquad
\bar{Q}^{\la}_{ \frac{7}{2}}  =  8 \sqrt{2} \,
\Big( \bar{Q}_{\la}^{ (4) -} +  \bar{Q}_{\la}^{ (4) +}\Big),
\nonu \\
\bar{Q}^{\la}_{ \frac{9}{2}} & = & -32 \sqrt{2} \,
\Big( \bar{Q}_{\la}^{ (5) -} +  \bar{Q}_{\la}^{ (5) +}\Big),
\qquad \cdots \qquad.
\label{lowspincurrents}
\eea
According to the normalization in (\ref{WWQQnonzerola}),
as we increase the spin, we simply multiply by $-4$ (except
$\bar{Q}_{\frac{1}{2}}^{\la}$).
We will calculate the various OPEs by using these explicit
expressions (\ref{lowspincurrents}).

The generators in (\ref{SCA2}) in terms of the currents
can be written as
\bea
J & = &  V_{\la}^{(1)-}=-4\Big( (1-2\la)\, W_{F,1}-2 \la \, W_{B,1}\Big),
\nonu \\
G^{+} & = & Q^{\la}_{ \frac{3}{2}}  =  \frac{1}{\sqrt{2}} \,
\Big( Q_{\la}^{ (2) -} -  Q_{\la}^{ (2) +}\Big), 
\nonu \\
G^{-} & = & \frac{1}{2}\, \bar{Q}^{\la}_{ \frac{3}{2}}  =
\frac{1}{2\sqrt{2}} \,
\Big( \bar{Q}_{\la}^{ (2) -} +  \bar{Q}_{\la}^{ (2) +}\Big),
\nonu \\
T & = &   V_{\la}^{(2)+}=\Big( W^{\la}_{F,2} + W^{\la}_{B,2} \Big).
\label{SCA3}
\eea
At $\la=0$, the weight-$1$ current $J$ in (\ref{SCA3}) of the
${\cal N}=2$ superconformal algebra does not depend on the bosonic
$\beta \, \ga$ operators \footnote{
  \label{extracurrents} Note that we have
  $W_{B,1}^{\la}=-\frac{1}{4}\, \ga^a \, \beta^a$ and
  $\bar{Q}_{\frac{1}{2}}^{\la}=-\frac{1}{\sqrt{2}}\,
  \beta^a \, c^a$.}.

\subsubsection{The structure constants for nonzero $\la$}

Let us introduce the generalized hypergeometric function
\bea
\phi_{r}^{h_1 ,h_2}(\Lambda,a)  \equiv \ _4F_3\left[
\begin{array}{c|}
\frac{1}{2} + \Lambda \ ,  \frac{1}{2} - \Lambda  \ , \frac{1+a-r}{ 2}\ , \frac{a-r}{2}\\
\frac{3}{2}-h_1 \ , \frac{3}{2} -h_2\ , \frac{1}{2}+ h_1+h_2-r
\end{array}  \ 1\right].
\label{phi}
\eea
In general, the sum of upper four elements plus $1$
($=\frac{5}{2}+a-r$)
is not equal to the sum of lower three elements
($=\frac{7}{2}-r$) for generic $a \neq 1$
\footnote{
  For $\la =0$, we introduce
  the generalized hypergeometric function
  \bea
\hat{\phi}^{h_1,h_2}_{h}(x,y)
\!&
\equiv \!&
{}_4 F_3
\Bigg[
\begin{array}{c}
-\frac{1}{2}-x-2y, \frac{3}{2}-x+2y, -\frac{h+1}{2}+x,
-\frac{h}{2} +x \\
-h_1+\frac{3}{2},-h_2+\frac{3}{2},h_1+h_2-h-\frac{3}{2}
\end{array} ; 1
\Bigg]\ .
\nonu
\eea
By using the notation of (\ref{phi}), we have 
\bea
\hat{\phi}_{r-1-a}^{h_1 ,h_2}(0,\frac{1}{2}(-1-\La))  \equiv \ _4F_3\left[
\begin{array}{c|}
\frac{1}{2} + \Lambda \ ,  \frac{1}{2} - \Lambda  \ , \frac{1+a-r}{ 2}\ , \frac{a-r}{2}\\
\frac{3}{2}-h_1 \ , \frac{3}{2} -h_2\ , (\frac{1}{2}+ h_1+h_2-r)+a-1
\end{array}  \ 1\right].
\nonu 
\eea
The last of lower elements contains the additional $(a-1)$ which is
nonzero for $a\neq 1$. We check that for $a=1$,
the expression of (\ref{phi}) reduces to the one in \cite{PRS1990-1}
where their $s$, $r$, $i$ and $j$ in $(3.14)$ correspond to our
$-\frac{1}{2}\, (1-\Lambda)$, $\frac{1}{2}\,(r-2)$, $(h_1-2)$ and $(h_2-2)$
respectively.}.
Furthermore, we introduce the mode dependent function
\bea
N^{h_1,h_2}_{h}(m,n)
\!&
\equiv \!&
\sum_{l=0 }^{h+1}(-1)^l
\left(\begin{array}{c}
h+1 \\  l \\
\end{array}\right)
[h_1-1+m]_{h+1-l}[h_1-1-m]_l
\nonu \\
\!& \times \!& [h_2-1+n]_l [h_2-1-n]_{h+1-l}.
\label{Ndef}
\eea
The falling Pochhammer symbol
$[a]_n \equiv a(a-1) \cdots (a-n+1)$ in (\ref{Ndef})
is used.

We have found three different kinds of structure
constants in the context of the matrix generalization of
$AdS_3$ Vasiliev higher spin theory as follows \cite{AK2009}:
\bea
\mathrm{BB}^{h_1,h_2}_{r,\,\pm}(m,n; \mu )
&\equiv&
 -\frac{1  }{ (r-1)!} N_{r-2}^{h_1, h_2}(m,n) \Bigg[
\phi_{r}^{h_1 ,h_2}(\mu,1)  \pm \phi_{r}^{h_1 ,h_2}(1-\mu,1)  
\Bigg],
\nonu\\
\mathrm{BF}^{h_1,h_2+\frac{1}{2}}_{r,\,\pm}(m,\rho; \mu )
&\equiv&
 -\frac{1  }{ (r-1)!} N_{r-2}^{h_1, h_2+\frac{1}{2}}(m,\rho) \Bigg[
 \phi_{r+1}^{h_1 ,h_2+1}(\mu,\frac{3\pm1}{2})
 \nonu \\
 & \pm & \phi_{r+1}^{h_1 ,h_2+1}(1-\mu,\frac{3\pm1}{2})  
\Bigg],
\nonu\\
\mathrm{FF}^{h_1+\frac{1}{2},h_2+\frac{1}{2}}_{r,\,\pm}(\rho,\omega; \mu )
&\equiv&
-\frac{1  }{ (r-1)!}N_{r-2}^{h_1+\frac{1}{2}, h_2+\frac{1}{2}}(\rho,\omega)
\Bigg[
 \phi_{r+1}^{h_1+1 ,h_2+1}(\mu,\frac{3\pm1}{2})  \nonu \\
 & \pm & \phi_{r+1}^{h_1+1 ,h_2+1}(1-\mu,\frac{3\pm1}{2})  
 \Bigg],
\label{3struct}
\eea
where the relations (\ref{phi}) and (\ref{Ndef})
are needed \footnote{
\label{symm}
  We have the following symmetry
  between the structure constants \cite{AK2009}
  under the transformation
  $\mu \leftrightarrow 1-\mu$
\bea
\mathrm{BB}^{h_1,h_2}_{r,\,\pm}(m,n; \mu ) & = &
\pm
\mathrm{BB}^{h_1,h_2}_{r,\,\pm}(m,n; 1-\mu ),
\nonu \\
\mathrm{BF}^{h_1,h_2+\frac{1}{2}}_{r,\,\pm}(m,\rho; \mu ) &=&
\pm
\mathrm{BF}^{h_1,h_2+\frac{1}{2}}_{r,\,\pm}(m,\rho; 1-\mu ),
\nonu \\
\mathrm{FF}^{h_1+\frac{1}{2},h_2+\frac{1}{2}}_{r,\,\pm}(\rho,\omega; \mu )
& = &
\pm 
\mathrm{FF}^{h_1+\frac{1}{2},h_2+\frac{1}{2}}_{r,\,\pm}(\rho,\omega; 1-\mu ).
\nonu
\eea
This implies that
the half of the structure constants 
vanishes at $\mu=
\frac{1}{2}$ ($\la=\frac{1}{4}$ or $q_{syk}=1$).
}.

From the lesson of \cite{AK2009} where the mode dependent
structure constants for vanishing $\la$ can be written in terms of
the linear combinations of (\ref{3struct}),
we do expect that for nonzero $\la$, they satisfy as follows:
\bea
p_{F,h}^{h_1,h_2}(m,n,\la) & = & -\frac{1}{4} \Bigg[
  \mathrm{BB}^{h_1,h_2}_{h+2,\,+} +
   \mathrm{BB}^{h_1,h_2}_{h+2,\,-} \Bigg]_{\mu = 2 \la},
\nonu \\
p_{B,h}^{h_1,h_2}(m,n,\la) & = & -\frac{1}{4} \Bigg[
  \mathrm{BB}^{h_1,h_2}_{h+2,\,+} -
   \mathrm{BB}^{h_1,h_2}_{h+2,\,-} \Bigg]_{\mu = 2 \la},
\nonu \\
q_{F,2h}^{h_1,h_2+\frac{1}{2}}(m,n,\la) & = &  \Bigg[
 -\frac{1}{8} \mathrm{BF}^{h_1,h_2+\frac{1}{2}}_{2h+2,\,+} +
 \frac{(2h_1-2h-3)}{16(h+1)}\,
 \mathrm{BF}^{h_1,h_2+\frac{1}{2}}_{2h+2,\,-} \Bigg]_{\mu = 2 \la},
\nonu \\
q_{F,2h+1}^{h_1,h_2+\frac{1}{2}}(m,n,\la) & = &  \Bigg[
 \frac{1}{8} \mathrm{BF}^{h_1,h_2+\frac{1}{2}}_{2h+3,\,+} -
 \frac{(h_1-h-2)}{4(2h+3)}\,
 \mathrm{BF}^{h_1,h_2+\frac{1}{2}}_{2h+3,\,-} \Bigg]_{\mu = 2 \la},
\nonu \\
q_{B,2h}^{h_1,h_2+\frac{1}{2}}(m,n,\la) & = &  \Bigg[
 -\frac{1}{8} \mathrm{BF}^{h_1,h_2+\frac{1}{2}}_{2h+2,\,+} -
 \frac{(2h_1-2h-3)}{16(h+1)}\,
 \mathrm{BF}^{h_1,h_2+\frac{1}{2}}_{2h+2,\,-} \Bigg]_{\mu = 2 \la},
\nonu \\
q_{B,2h+1}^{h_1,h_2+\frac{1}{2}}(m,n,\la) & = &  \Bigg[
 -\frac{1}{8} \mathrm{BF}^{h_1,h_2+\frac{1}{2}}_{2h+3,\,+} -
 \frac{(h_1-h-2)}{4(2h+3)}\,
 \mathrm{BF}^{h_1,h_2+\frac{1}{2}}_{2h+3,\,-} \Bigg]_{\mu = 2 \la},
\nonu \\
o_{F,2h}^{h_1+\frac{1}{2},h_2+\frac{1}{2}}(m,n,\la) & = &  \Bigg[
 -\mathrm{FF}^{h_1+\frac{1}{2},h_2+\frac{1}{2}}_{2h+1,\,+} -
 \frac{2(h_1+h_2-h)}{(2h+1)}\,
 \mathrm{FF}^{h_1+\frac{1}{2},h_2+\frac{1}{2}}_{2h+1,\,-} \Bigg]_{\mu = 2 \la},
\nonu \\
o_{F,2h+1}^{h_1+\frac{1}{2},h_2+\frac{1}{2}}(m,n,\la) & = &  \Bigg[
 \mathrm{FF}^{h_1+\frac{1}{2},h_2+\frac{1}{2}}_{2h+2,\,+} +
 \frac{2(h_1+h_2-h)-1}{2(h+1)}\,
 \mathrm{FF}^{h_1+\frac{1}{2},h_2+\frac{1}{2}}_{2h+2,\,-} \Bigg]_{\mu = 2 \la},
\nonu \\
o_{B,2h}^{h_1+\frac{1}{2},h_2+\frac{1}{2}}(m,n,\la) & = &  \Bigg[
 -\mathrm{FF}^{h_1+\frac{1}{2},h_2+\frac{1}{2}}_{2h+1,\,+} +
 \frac{2(h_1+h_2-h)}{(2h+1)}\,
 \mathrm{FF}^{h_1+\frac{1}{2},h_2+\frac{1}{2}}_{2h+1,\,-} \Bigg]_{\mu = 2 \la},
\nonu \\
o_{B,2h+1}^{h_1+\frac{1}{2},h_2+\frac{1}{2}}(m,n,\la) & = &  \Bigg[
- \mathrm{FF}^{h_1+\frac{1}{2},h_2+\frac{1}{2}}_{2h+2,\,+} +
 \frac{2(h_1+h_2-h)-1}{(2h+2)}\,
 \mathrm{FF}^{h_1+\frac{1}{2},h_2+\frac{1}{2}}_{2h+2,\,-} \Bigg]_{\mu = 2 \la}.
\label{structla}
\eea
We have checked that the above relations for several $h_1$, $h_2$  and
$h$ are satisfied in the specific OPE examples.
That is, the structure constants are indeed the right hand sides
of (\ref{structla}).

\subsubsection{ The example of the explicit OPE
$ W_{F,4}^{\la}(\bar{z}) \, W_{F,4}^{\la}(\bar{w})$ for nonzero $\la$}

For example, for $h_1=4$ and $h_2=4$, we can calculate the OPE
$W_{F,4}^{\la}(\bar{z}) \, W_{F,4}^{\la}(\bar{w})$
by using (\ref{WWQQnonzerola}), (\ref{VVQQla}) and (\ref{fundOPE})
and reexpressing each pole in terms of
$W_{F,h}^{\la}(\bar{w})$ with $h=2,4,6$ and their derivatives
as follows \footnote{In the OPEs we are considering in this paper,
  the number $N$ is fixed by $N=3$ which is the smallest number
  having the nontrivial $d$ symbol of $SU(N)$. The number $L$ is fixed
by $L=1$ which is introduced around the equation (\ref{fundOPE}).}:
\bea
&& W_{F,4}^{\la}(\bar{z}) \, W_{F,4}^{\la}(\bar{w})  = 
\frac{1}{(\bar{z}-\bar{w})^8}\, \Bigg[
  -\frac{768}{5}  (112 \lambda ^6-280 \lambda ^4+147 \lambda ^2-9
  )  \Bigg]\nonu \\
&& +
  \frac{1}{(\bar{z}-\bar{w})^6}\, \Bigg[
    \frac{2048}{5} (\lambda -1) (\lambda +1) (2 \lambda -3)
              (2 \lambda +3) \Bigg] \,  W_{F,2}^{\la}(\bar{w})
  \nonu \\
  & &+ \frac{1}{(\bar{z}-\bar{w})^5}\, \frac{1}{2}\, \Bigg[
    \frac{2048}{5} (\lambda -1) (\lambda +1) (2 \lambda -3)
              (2 \lambda +3) \Bigg] \,  \bar{\pa}\, W_{F,2}^{\la}(\bar{w})
  \nonu \\
    & &+ \frac{1}{(\bar{z}-\bar{w})^4}\, \Bigg[ \frac{3}{20}\, 
    \frac{2048}{5} (\lambda -1) (\lambda +1) (2 \lambda -3)
    (2 \lambda +3)  \,  \bar{\pa}^2\, W_{F,2}^{\la}
- \frac{96}{5}  (4 \lambda ^2-19) \,
  W_{F,4}^{\la}  \Bigg](\bar{w})
  \nonu \\
    & &+ \frac{1}{(\bar{z}-\bar{w})^3}\, \Bigg[ \frac{1}{30}\, 
    \frac{2048}{5} (\lambda -1) (\lambda +1) (2 \lambda -3)
    (2 \lambda +3)  \,  \bar{\pa}^3\, W_{F,2}^{\la}
- \frac{1}{2}\, \frac{96}{5}  (4 \lambda ^2-19) \,
 \bar{\pa}\,  W_{F,4}^{\la}  \Bigg](\bar{w})
  \nonu \\
     & &+ \frac{1}{(\bar{z}-\bar{w})^2}\, \Bigg[ \frac{1}{168}\, 
    \frac{2048}{5} (\lambda -1) (\lambda +1) (2 \lambda -3)
    (2 \lambda +3)  \,  \bar{\pa}^4\, W_{F,2}^{\la}
 - \frac{5}{36}\, \frac{96}{5}  (4 \lambda ^2-19) \,
 \bar{\pa}^2\,  W_{F,4}^{\la}  \nonu \\
 && + 6 \,  W_{F,6}^{\la} \Bigg](\bar{w})
 + \frac{1}{(\bar{z}-\bar{w})}\, \Bigg[ \frac{1}{1120}\, 
    \frac{2048}{5} (\lambda -1) (\lambda +1) (2 \lambda -3)
    (2 \lambda +3)  \,  \bar{\pa}^5\, W_{F,2}^{\la}
  \nonu \\
 & &- \frac{1}{36}\, \frac{96}{5}  (4 \lambda ^2-19) \,
  \bar{\pa}^3\,  W_{F,4}^{\la}  + \frac{1}{2} \, 6 \,   \bar{\pa}\,
  W_{F,6}^{\la} \Bigg](\bar{w}) + \cdots
  \nonu \\
  & &=  \frac{1}{(\bar{z}-\bar{w})^8}\,
  \Bigg[
  -\frac{768}{5}  (112 \lambda ^6-280 \lambda ^4+147 \lambda ^2-9
  )  \Bigg]
  -
  p_{F,4}^{4,4}(\bar{\pa}_{\bar{z}},\bar{\pa}_{\bar{w}},\la)
  \Bigg[ \frac{ W_{F,2}^{\la}(\bar{w})}{(\bar{z}-\bar{w})} \Bigg]
  \nonu \\
  && -
  p_{F,2}^{4,4}(\bar{\pa}_{\bar{z}},\bar{\pa}_{\bar{w}},\la)
  \Bigg[ \frac{ W_{F,4}^{\la}(\bar{w})}{(\bar{z}-\bar{w})} \Bigg]
 -  p_{F,0}^{4,4}(\bar{\pa}_{\bar{z}},\bar{\pa}_{\bar{w}},\la)
\Bigg[ \frac{ W_{F,6}^{\la}(\bar{w})}{(\bar{z}-\bar{w})} \Bigg]
+ \cdots.
  \label{4-4ope}
\eea
In the second relation of (\ref{4-4ope}), we
reexpress the structure constants in terms of
the differential operators
$ p_{F,6-h}^{4,4}(\bar{\pa}_{\bar{z}},\bar{\pa}_{\bar{w}},\la)$
with $h=2,4,6$.
By using the first equation of (\ref{structla}) for fixed
$h_1=h_2=4$,
we obtain
\bea
&& p_{F,4}^{4,4}(m,n,\la)  = 
\frac{64}{525} (\lambda -1) (\lambda +1) (2 \lambda -3)
(2 \lambda +3) 
\label{p444poly}
 \\
&& \times  (m-n) (3 m^4-2 m^3 n+4 m^2 n^2-39 m^2-2 m n^3+
20 m n+3 n^4-39 n^2+108).
\nonu
\eea
From this (\ref{p444poly}),
we can read off the corresponding differential operator
by taking the terms having a degree $(h+1)=5$ as
follows:
\bea
p_{F,4}^{4,4}(\bar{\pa}_{\bar{z}},\bar{\pa}_{\bar{w}},\la) & = &
\frac{64}{525}\times 9 \times \frac{1}{9} \,
(\lambda -1) (\lambda +1) (2 \lambda -3)
(2 \lambda +3) \,
\nonu \\
& \times & 
\Big(3 \bar{\pa}_{\bar{z}}^5-5 \bar{\pa}_{\bar{z}}^4 \bar{\pa}_{\bar{w}}+
6 \bar{\pa}_{\bar{z}}^3 \bar{\pa}_{\bar{w}}^2-
6 \bar{\pa}_{\bar{z}}^2 \bar{\pa}_{\bar{w}}^3
+5 \bar{\pa}_{\bar{z}} \bar{\pa}_{\bar{w}}^4-3 \bar{\pa}_{\bar{w}}^5 \Big).
\label{pfdiff}
\eea
Then we can calculate the quantity
$-p_{F,4}^{4,4}(\bar{\pa}_{\bar{z}},\bar{\pa}_{\bar{w}},\la)
\, \Big[\frac{W^{\la}_{F,2}(\bar{w})}{(\bar{z}-\bar{w})}
  \Big]$ by acting (\ref{pfdiff}) on the operator
$\Big[\frac{W^{\la}_{F,2}(\bar{w})}{(\bar{z}-\bar{w})}\Big]$
and this will lead to the corresponding terms in (\ref{4-4ope}).
Similarly, we can calculate
$ p_{F,2}^{4,4}(m,n,\la)=-\frac{8}{15}\, (-19)\, (-\frac{1}{19})\,
(4 \la^2-19) (m-n) (m^2-m n+n^2-7)$.
Then we can determine
$p_{F,2}^{4,4}(\bar{\pa}_{\bar{z}},\bar{\pa}_{\bar{w}},\la)=
-\frac{8}{15}\,  (-19)\, (-\frac{1}{19})\,
(4 \la^2-19) \, \Big( \bar{\pa}_{\bar{z}}^3-2
\bar{\pa}_{\bar{z}}^2 \bar{\pa}_{\bar{w}}+
2 \bar{\pa}_{\bar{z}} \bar{\pa}_{\bar{w}}^2- \bar{\pa}_{\bar{w}}^3 \Big) 
$ and this leads to the current $W^{\la}_{F,4}(\bar{w})$
and its derivatives in (\ref{4-4ope}) by performing
$-p_{F,2}^{4,4}(\bar{\pa}_{\bar{z}},\bar{\pa}_{\bar{w}},\la)
\, \Big[\frac{W^{\la}_{F,4}(\bar{w})}{(\bar{z}-\bar{w})}
  \Big]$. Finally, after calculating
$p_{F,0}^{4,4}(\bar{\pa}_{\bar{z}},\bar{\pa}_{\bar{w}},\la)=
3(m-n)$, the result of
$-p_{F,0}^{4,4}(\bar{\pa}_{\bar{z}},\bar{\pa}_{\bar{w}},\la)
\, \Big[\frac{W^{\la}_{F,6}(\bar{w})}{(\bar{z}-\bar{w})}
  \Big]$ provides the corresponding terms in (\ref{4-4ope})
\footnote{The corresponding commutator relation can be obtained
  by using the formula in \cite{CFT,Blumenhagenetal}.
  For example, we can obtain that
  $p_{ijk}(m,n)$ in $(2.54)$ of \cite{CFT} is given by
  $\frac{1}{3360} \, (m-n) \, 
  (3 m^4-2 m^3 n+4 m^2 n^2-39 m^2-2 m n^3+20 m n+3 n^4-39 n^2+108)$.
  After multiplying the coefficient
$\frac{2048}{5} (\lambda -1) (\lambda +1) (2 \lambda -3)
  (2 \lambda +3)$, we obtain
the above $p_{F,4}^{4,4}(m,n,\la)$ in (\ref{p444poly}).}.

In Appendix $B$
we present other various OPEs between the $SU(N)$-singlet currents
for fixed $h_1$ and $h_2$.
In this way, we make sure that  the structure constants
in (\ref{structla}) are correct ones.

\subsubsection{ The complete (anti)commutator relations
  between the $SU(N)$-singlet currents for
nonzero $\la$}

From the analysis of previous section together with
the similar descriptions in Appendix $B$, we conclude that
the final complete (anti)commutator relations between the
$SU(N)$-singlet currents  for
nonzero $\la$ with the insertion of
$q$ dependence appropriately
can be summarized by \footnote{In this paper,
  we do not present the explicit form for the
  central charges  $c_{W_{\mathrm{F},h_1}}(m,\la)$, $c_{W_{\mathrm{B},h_1}}(m,\la)$
  and $ c_{Q\bar{Q}_{h_1+\frac{1}{2}}}(r,\la)$. We calculated them for
  fixed $h_1$ and $h_2$ in previous section, and Appendix $B$.
  The calculations can be performed by taking the procedure
  in \cite{PRS1990}. Or we can use the partial results in \cite{MZ}
  to obtain them explicitly.}
\bea
\big[(W^{\la}_{\mathrm{F},h_1})_m,(W^{\la}_{\mathrm{F},h_2})_n\big] 
\!&=& \!
\sum^{h_1+h_2-3}_{h= 0, \mbox{\footnotesize even}} \, q^h\,
p_{\mathrm{F}}^{h_1,h_2, h}(m,n,\la)
\, (   W^{\la}_{\mathrm{F},h_1+h _2-2-h} )_{m+n}\nonu \\
\!& + \!&
N\, c_{W_{\mathrm{F},h_1}}(m,\la) \,
\delta^{h_1 h_2}\,q^{2(h_1-2)}\,\delta_{m+n},
\nonu \\
\big[(W^{\la}_{\mathrm{B},h_1})_m,(W^{\la}_{\mathrm{B},h_2})_n\big] 
\!&=& \!
\sum^{h_1+h_2-4}_{h= 0, \mbox{\footnotesize even}} \, q^h\,
p_{\mathrm{B}}^{h_1,h_2, h}(m,n,\la)
\, (   W^{\la}_{\mathrm{B},h_1+h_2-2-h} )_{m+n} \nonu \\
\!& + \!& \Bigg[ q^{h}\,
p_{\mathrm{B}}^{h_1,h_2, h}(m,n,\la)
\, (   W^{\la}_{\mathrm{B},h_1+h_2-2-h} )_{m+n} \Bigg]_{h=h_1+h_2-3}
\nonu \\
\!& + \!&
N\, c_{W_{\mathrm{B},h_1}}(m,\la) \,
\delta^{h_1 h_2}\,q^{2(h_1-2)}\,\delta_{m+n},
\nonu \\
\big[(W^{\la}_{\mathrm{F},h_1})_m,(Q^{\la}_{h_2+\frac{1}{2}})_r\big] 
\!&= &\!
\sum^{h_1+h_2-3}_{h=-1}\, q^h \,
q_{\mathrm{F}}^{h_1,h_2+\frac{1}{2}, h}(m,r,\la) 
\, (Q^{\la}_{h_1+h_2-\frac{3}{2}-h})_{m+r}\ ,
\nonu \\
\big[(W^{\la}_{\mathrm{B},h_1})_m,(Q^{\la}_{h_2+\frac{1}{2}})_r\big] 
\!&= &\!
\sum^{h_1+h_2-3}_{h=-1}\, q^h \,
q_{\mathrm{B}}^{h_1,h_2+\frac{1}{2}, h}(m,r,\la) 
\, (Q^{\la}_{h_1+h_2-\frac{3}{2}-h})_{m+r}\ ,
\nonu \\
\big[(W^{\la}_{\mathrm{F},h_1})_m,(\bar{Q}^{\la}_{h_2+\frac{1}{2}})_r\big] 
\!&= &\!
\sum^{h_1+h_2-3}_{h=-1}\, q^h \, (-1)^h\,
q_{\mathrm{F}}^{h_1,h_2+\frac{1}{2}, h}(m,r,\la) 
\, (\bar{Q}^{\la}_{h_1+h_2-\frac{3}{2}-h})_{m+r}
\nonu \\
\!&+ &\!  \Bigg[ q^{h} \, (-1)^{h}\,
q_{\mathrm{F}}^{h_1,h_2+\frac{1}{2}, h}(m,r,\la) 
\, (\bar{Q}^{\la}_{h_1+h_2-\frac{3}{2}-h})_{m+r}\Bigg]_{h=h_1+h_2-2} \ ,
\nonu \\
\big[(W^{\la}_{\mathrm{B},h_1})_m,(\bar{Q}^{\la}_{h_2+\frac{1}{2}})_r\big] 
\!&= &\!
\sum^{h_1+h_2-3}_{h=-1}\, q^h \,
(-1)^h \, q_{\mathrm{B}}^{h_1,h_2+\frac{1}{2}, h}(m,r,\la) 
\, (\bar{Q}^{\la}_{h_1+h_2-\frac{3}{2}-h})_{m+r}\ 
\nonu \\
\!&+ &\!
\Bigg[ q^h \,
(-1)^h \, q_{\mathrm{B}}^{h_1,h_2+\frac{1}{2}, h}(m,r,\la) 
\, (\bar{Q}^{\la}_{h_1+h_2-\frac{3}{2}-h})_{m+r} \Bigg]_{h=h_1+h_2-2} \ ,
\nonu \\
\{(Q^{\la}_{h_1+\frac{1}{2}})_r,(\bar{Q}^{\la}_{h_2+\frac{1}{2}})_s\} 
\!&=&\!
\sum^{h_1+h_2-1}_{h=0} \,
q^h 
\, o_{\mathrm{F}}^{h_1+\frac{1}{2},h_2+\frac{1}{2},h}(r,s,\la) \,
(W^{\la}_{\mathrm{F},h_1+h_2-h})_{r+s}
\nonu \\
\!& + \!& \sum^{h_1+h_2-2}_{h=0} \,
q^h \,
o_{\mathrm{B}}^{h_1+\frac{1}{2},h_2+\frac{1}{2},h}(r,s,\la) \,
(W^{\la}_{\mathrm{B},h_1+h_2-h})_{r+s} 
\nonu \\
\!& + \!& \Bigg[
q^h \,
o_{\mathrm{B}}^{h_1+\frac{1}{2},h_2+\frac{1}{2},h}(r,s,\la) \,
(W^{\la}_{\mathrm{B},h_1+h_2-h})_{r+s} \Bigg]_{h=h_1+h_2-1} 
\nonu \\
\!&+\!&  N\, c_{Q\bar{Q}_{h_1+\frac{1}{2}}}(r,\la)
\,  \delta^{h_1 h_2} \, q^{2(h_1+\frac{1}{2}-1)}
\delta_{r+s}.
\label{final1}
\eea
In the right hand sides of (\ref{final1}), we emphasize that
the additional weights $\frac{1}{2}, 1$ currents
appear by taking the square brackets \footnote{We can still use
  the (anti)commutator relations
  for $\la$=0 by allowing the corresponding
  upper limits in the summation over the dummy variable
  $h$ properly at the four places. Each single term can combine with
  each summation term because the $\la $ and
  mode dependent structure
  constants in each single term can be written in terms of the
  same structure constants in each summation term.
}.
Of course we assume that the possible lowest weights for
$W^{\la}_{\mathrm{F},h}$, $W^{\la}_{\mathrm{B},h}$,
$Q^{\la}_{h+\frac{1}{2}}$ and $\bar{Q}^{\la}_{h+\frac{1}{2}}$
are given by $h=1$, $h=2$, $h+\frac{1}{2}=\frac{3}{2}$
and  $h+\frac{1}{2}=\frac{3}{2}$ respectively.
In other words, among the field contents in \cite{PRS1990},
the above  weights $\frac{1}{2}, 1$ currents occur
in the right hand sides of the (anti)commutator relations at nonzero
$\la$. 
In order to fully describe the complete structure of
the
  ${\cal N}=2$ supersymmetric linear
$W^{N,N}_{\infty}[\la]$ algebra, we need
to calculate the OPEs between the additional weights $\frac{1}{2},1$
currents 
and the remaining currents.
Also their own OPEs should be calculated.

In Appendix $C$,
we present some OPEs
containing the additional currents $\bar{Q}^{\la}_{\frac{1}{2}}$
or $W^{\la}_{\mathrm{B},1}$ for fixed $h_1$ and $h_2$.
It turns out that the OPEs containing the weight-$\frac{1}{2}$
current  $\bar{Q}^{\la}_{\frac{1}{2}}$
have the previous known structure constants
while  the OPEs containing the weight-$1$
current $W^{\la}_{\mathrm{B},1}$, at first sight,
do not have their structure constants
which can be written in terms of the known expressions
appearing in (\ref{final1}), although there are explicit
$\la$-dependent terms in their OPEs.

In particular, we can check the following relations \footnote{
\label{laonehalf}  Similar relations can be checked as follows:
$
p_{\mathrm{F}}^{h_1,h_2, h=h_1+h_2-3}(m,n,\la=\frac{1}{2})  = 
q_{\mathrm{F}}^{h_1,h_2+\frac{1}{2}, h=h_1+h_2-2}(m,r,\la=\frac{1}{2}) = 
q_{\mathrm{B}}^{h_1,h_2+\frac{1}{2}, h=h_1+h_2-2}(m,r,\la=\frac{1}{2})
= 
o_{\mathrm{F}}^{h_1+\frac{1}{2},h_2+\frac{1}{2}, h=h_1+h_2-1}(r,s,\la=\frac{1}{2}) = 
0
$.
We reproduce the subalgebra
of 
the  ${\cal N}=2$ supersymmetric linear
$W^{N,N}_{\infty}[\la=\frac{1}{2}]$ algebra which is isomorphic
to the  ${\cal N}=2$ supersymmetric linear
$W^{N,N}_{\infty}[\la=0]$ algebra as in the introduction.
In this case, the bosonic subalgebra
is given by $W_{\infty}^N[\la=\frac{1}{2}]$ generated by
$W_{F,h}^{\la=\frac{1}{2}}$ and $W_{1+\infty}^N[\la=\frac{1}{2}]$
 generated by
 $W_{B,h}^{\la=\frac{1}{2}}$.
Then by the contraction limit for the parameter $q$,
 we obtain the $w_{1+\infty}$ algebra from the latter.
 If the decoupling of $W_{B,h=1}^{\la=\frac{1}{2}}$
 in the latter occurs as in the subsection \ref{DEC},
 then this bosonic subalgebra becomes
 $W_{\infty}^N[\la=\frac{1}{2}]$.
}
\bea
p_{\mathrm{B}}^{h_1,h_2, h=h_1+h_2-3}(m,n,\la=0) & = &
q_{\mathrm{F}}^{h_1,h_2+\frac{1}{2}, h=h_1+h_2-2}(m,r,\la=0)\nonu
\\& = &
q_{\mathrm{B}}^{h_1,h_2+\frac{1}{2}, h=h_1+h_2-2}(m,r,\la=0)
\nonu \\
&= &
o_{\mathrm{B}}^{h_1+\frac{1}{2},h_2+\frac{1}{2}, h=h_1+h_2-1}(r,s,\la=0)
\nonu \\
& = &
0.
\label{vanishing}
\eea
Then according to (\ref{vanishing}),
the square brackets in 
the
above (anti)commutator relations (\ref{final1})
vanish at $\la=0$ and we reproduce the subalgebra
of 
the  ${\cal N}=2$ supersymmetric linear
$W^{N,N}_{\infty}[\la=0]$ algebra \cite{Ahn2202}. The bosonic subalgebra
is given by $W_{1+\infty}^N[\la=0]$ generated by
$W_{F,h}^{\la=0}$ and $W_{\infty}^N[\la=0]$
 generated by
 $W_{B,h}^{\la=0}$.
 Then by the appropriate limit for the parameter $q$,
 we will obtain the $w_{1+\infty}$ algebra from the former.
 If the decoupling of $W_{F,h=1}^{\la=0}$
 in the former occurs like as the
 footnote \ref{decouplinginthefer},
 then this bosonic subalgebra becomes
 $W_{\infty}^N[\la=0]$.

In Appendix $D$,
the 
remaining (anti)commutator relations
of ${\cal N}=2$ supersymmetric linear $W_{\infty}^{N,N}[\la]$ algebra
are summarized, by considering the $SU(N)$ adjoint index
$\hat{A}$ properly.

\subsubsection{ The decoupling of $\bar{Q}^{\la}_{\frac{1}{2}}$
  and $W^{\la}_{\mathrm{B},1}$
\label{DEC}}

At nonzero $\la$, the (anti)commutator relations imply that
the weight-$1,\frac{1}{2}$ currents occur in the right hand sides.
Now we can try to decouple them.
Let us consider the fifth equation of (\ref{final1}) by taking
$h_1=1$ and $h_2=1$. Then we can calculate
the OPE $W_{F,1}^{\la}(\bar{z}) \, \bar{Q}^{\la}_{\frac{3}{2}}(\bar{w})$.
By requiring that the new weight-$\frac{3}{2}$ current
should remove the unwanted current $\bar{Q}^{\la}_{\frac{1}{2}}$
\footnote{We need to admit that the new currents are not quasiprimary
operators.},
we have
\bea
\bar{Q}^{\la}_{new,\frac{3}{2}}=\bar{Q}^{\la}_{\frac{3}{2}}- 4 \la\, \bar{\pa}\,
\bar{Q}^{\la}_{\frac{1}{2}}.
\label{new3half}
\eea
Now we go to the sixth equation of (\ref{final1})
and substitute (\ref{new3half}) into that equation
in order to obtain the new weight-$2$ current.
It turns out that
by taking
\bea
W_{new,B,2}^{\la} = W_{B,2}^{\la} - 4 \, \la\,\bar{\pa} \,
 W_{B,1}^{\la},
\label{new2}
\eea
 we can remove the unwanted 
current $\bar{Q}^{\la}_{\frac{1}{2}}$.
We can check that the $W_{B,1}^{\la}$ dependence disappears
when we calculate the OPE $
Q^{\la}_{\frac{3}{2}}(\bar{z})\,
\bar{Q}^{\la}_{new,\frac{3}{2}}(\bar{w})$.
Similarly, we can calculate
\bea
\bar{Q}^{\la}_{new,\frac{5}{2}}=\bar{Q}^{\la}_{\frac{5}{2}}- \frac{8}{3}\,
\la\, (1 +2\la) \, \bar{\pa}^2\,
\bar{Q}^{\la}_{\frac{1}{2}},
\label{new5half}
\eea
by considering the OPE
$W_{F,1}^{\la}(\bar{z}) \, \bar{Q}^{\la}_{\frac{5}{2}}(\bar{w})$
and removing the  unwanted 
current $\bar{Q}^{\la}_{\frac{1}{2}}$.
Moreover we can obtain
\bea
W_{new,B,3}^{\la} =
W_{B,3}^{\la} - \frac{8}{3} \, \la \, (1+2\la)\,
\,\bar{\pa}^2 \,
 W_{B,1}^{\la},
\label{new3}
\eea
by considering the OPE
$W_{B,3}^{\la}(\bar{z}) \, \bar{Q}^{\la}_{\frac{5}{2}}(\bar{w})$
and removing the  unwanted 
current $\bar{Q}^{\la}_{\frac{1}{2}}$.
In this way we determine the new currents,
(\ref{new3half}), (\ref{new2}), (\ref{new5half}) and (\ref{new3}).
We expect that we can continue to perform this procedure
and remove the above weight-$1, \frac{1}{2}$ currents
\footnote{As described before, by using the partial results in
  \cite{MZ}, we can obtain the new currents for any weight $h$
  by calculating the highest order pole in any OPE in order to remove
the weigh-$\frac{1}{2}$ current.}.

Therefore, in principle, eventually we obtain
the complete (anti)commutator relations with modified
$\la$-dependent  known structure constants, as a subalgebra, where the
unwanted  weight-$1, \frac{1}{2}$ currents disappear completely
\footnote{
\label{decouplinginthefer}
  We can decouple the $W_{F,1}^{\la}$ and
  $\bar{Q}^{\la}_{\frac{1}{2}}$ by introducing the new currents
  $\bar{Q}^{\la}_{new,\frac{3}{2}}=\bar{Q}^{\la}_{\frac{3}{2}}- 2 (2\la-1)\,
  \bar{\pa}\,
  \bar{Q}^{\la}_{\frac{1}{2}}$ and
  $W_{new,F,2}^{\la} = W_{F,2}^{\la} - 2 ( 2\la-1 )\,\bar{\pa} \,
  W_{F,1}^{\la}$ in the context of
the  ${\cal N}=2$ supersymmetric linear
$W^{N,N}_{\infty}[\la=\frac{1}{2}]$ algebra together with the footnote
\ref{laonehalf}.
The higher weight currents can be constructed similarly.}.

\subsubsection{
  The realization of
  ${\cal N}=2$ supersymmetric
linear
  $W^{N,N}_{\infty}[\la=\frac{1}{4}]$ algebra
in the ${\cal N}=2$ SYK model }

As noted in the footnote \ref{symm},
at $\la=\frac{1}{4}$ \footnote{At this point,
 ${\cal N}=2$ supersymmetric
linear
$W^{N,N}_{\infty}[\la=\frac{1}{4}]$ algebra is self isomorphic
because the solution of $\la =\frac{1}{2}-\la$ provides
the $\la =\frac{1}{4}$, as in
the introduction.}, all the
second terms in the right hand sides  of
the structure constants (\ref{structla}) vanish.
In the ${\cal N}=2$ SYK models, this is equivalent to take
the following limit
\bea
q_{syk} \rightarrow 1.
\label{qequalone}
\eea
The interaction is quadratic.
As observed in \cite{Bergetalnpb}, there exists a subalgebra
generated by \footnote{We thank M. Vasiliev for discussion on this matter
further.}
\bea
&& V^{(s),+}_{\la}, s=2,4,6, \cdots, \qquad
V^{(s),-}_{\la}, s=1,3,5, \cdots, 
\nonu \\
&& Q^{(s),+}_{\la}, s=1,3,5, \cdots, \qquad
Q^{(s),-}_{\la}, s=1,3,5, \cdots, 
\label{case1}
\eea
or by
\bea
&& V^{(s),+}_{\la}, s=2,4,6, \cdots, \qquad
V^{(s),-}_{\la}, s=1,3,5, \cdots, 
\nonu \\
&& Q^{(s),+}_{\la}, s=2,4,6, \cdots, \qquad
Q^{(s),-}_{\la}, s=2,4,6, \cdots.
\label{case2}
\eea
There is no supersymmetry in the first case (\ref{case1})
\cite{Bergetalnpb} \footnote{
It would be interesting to study this case in the context of
the SYK models because there is no supersymmetry.}.
We calculate some OPEs for fixed $h_1$ and $h_2$
in order to see this behavior explicitly
in Appendix $E$.
In the basis of \cite{Bergetalnpb}, we obtain the following
relations from (\ref{WWQQnonzerola})
\bea
 V^{(h),+}_{\la} & = & \frac{1}{\Bigg[\frac{n_{W_{F,h}}}{q^{h-2}}\,
     \frac{(-1)^h}{\sum_{i=0}^{h-1}\, a^i( h, \la+\frac{1}{2}=\frac{1}{2})}
     \Bigg]}\,
 W_{F,h}^{\la} + \frac{1}{\Bigg[ \frac{n_{W_{B,h}}}{q^{h-2}}\,
 \frac{(-1)^h}{\sum_{i=0}^{h-1}\, a^i( h, \la=0)} \Bigg]}\, W_{B,h}^{\la},
 \label{fourrelations} \\
 V^{(h),-}_{\la} & = & \frac{\frac{(h-2\la)}{(2h-1)}}{\Bigg[\frac{n_{W_{F,h}}}{q^{h-2}}\,
     \frac{(-1)^h}{\sum_{i=0}^{h-1}\, a^i( h, \la+\frac{1}{2}=\frac{1}{2})}
     \Bigg]}\,
 W_{F,h}^{\la} - \frac{\frac{(h-1+2\la)}{(2h-1)}}{\Bigg[ \frac{n_{W_{B,h}}}{q^{h-2}}\,
 \frac{(-1)^h}{\sum_{i=0}^{h-1}\, a^i( h, \la=0)} \Bigg]}\, W_{B,h}^{\la},
 \nonu \\
 Q^{(h+1),+}_{\la} & = & \frac{1}{2} \Bigg(-\frac{1}{
 \Bigg[\frac{1}{2} \,\frac{n_{W_{Q,h+\frac{1}{2}}}}{q^{h-1}}
\, \frac{(-1)^{h+1}  \, h  }{
   \sum_{i=0}^{h-1} \, \beta^i( h+1, \la=0)}\Big]}\, Q^{\la}_{h+\frac{1}{2}}
 + \frac{1}{\Big[ \frac{1}{2} \,
\frac{n_{W_{Q,h+\frac{1}{2}}}}{q^{h-1}} \,
 \frac{(-1)^{h+1}  }{
   \sum_{i=0}^{h} \, \al^i( h+1, \la=0)}\Big]}\,
 \bar{Q}^{\la}_{h+\frac{1}{2}} \Bigg),
 \nonu \\
 \bar{ Q}^{(h+1),-}_{\la} & = & \frac{1}{2} \Bigg(\frac{1}{
 \Bigg[\frac{1}{2} \,\frac{n_{W_{Q,h+\frac{1}{2}}}}{q^{h-1}}
\, \frac{(-1)^{h+1}  \, h  }{
   \sum_{i=0}^{h-1} \, \beta^i( h+1, \la=0)}\Big]}\, Q^{\la}_{h+\frac{1}{2}}
 + \frac{1}{\Big[ \frac{1}{2} \,
\frac{n_{W_{Q,h+\frac{1}{2}}}}{q^{h-1}} \,
 \frac{(-1)^{h+1}  }{
   \sum_{i=0}^{h} \, \al^i( h+1, \la=0)}\Big]}\,
 \bar{Q}^{\la}_{h+\frac{1}{2}} \Bigg).
 \nonu
 \eea

 Then we can calculate the commutator relation
 $\big[  (V^{(h_1),+}_{\la})_m,  (V^{(h_2),+}_{\la})_n\big]$.
 By substituting the first equation of (\ref{fourrelations}) into
 this commutator, we obtain
 the following coefficient function of $(V^{(h_1+h_2-2-h),-}_{\la})_{m+n}$:
 \bea
&& \Bigg[\frac{1}{\frac{n_{W_{F,h_1}}}{q^{h_1-2}}\,
     \frac{(-1)^{h_1}}{\sum_{i=0}^{h_1-1}\, a^i( h_1, \frac{1}{2})}}
     \Bigg]\Bigg[\frac{1}{\frac{n_{W_{F,h_2}}}{q^{h_2-2}}\,
     \frac{(-1)^{h_2}}{\sum_{i=0}^{h_2-1}\, a^i( h_2, \frac{1}{2})}}
   \Bigg]\nonu \\
 && \times 
\Bigg[\frac{n_{W_{F,h_1+h_2-2-h}}}{q^{h_1+h_2-2-h-2}}\,
    \frac{(-1)^{h_1+h_2-2-h}}{\sum_{i=0}^{h_1+h_2-2-h-1}\, a^i(
      h_1+h_2-2-h, \frac{1}{2})}
  \Bigg] \,  q^h\,
p_{\mathrm{F}}^{h_1,h_2, h}(m,n,\la)
\nonu \\
&& - \Bigg[ \frac{1}{ \frac{n_{W_{B,h_1}}}{q^{h_1-2}}\,
    \frac{(-1)^{h_1}}{\sum_{i=0}^{h_1-1}\, a^i( h_1, 0)}} \Bigg]
\Bigg[ \frac{1}{ \frac{n_{W_{B,h_2}}}{q^{h_2-2}}\,
 \frac{(-1)^{h_2}}{\sum_{i=0}^{h_2-1}\, a^i( h_2, 0)}} \Bigg]
\nonu \\
&& \times \Bigg[ \frac{n_{W_{B,h_1+h_2-2-h}}}{q^{h_1+h_2-2-h-2}}\,
  \frac{(-1)^{h_1+h_2-2-h}}{\sum_{i=0}^{h_1+h_2-2-h-1}\,
    a^i( h_1+h_2-2-h, 0)} \Bigg]\, q^h\,
p_{\mathrm{B}}^{h_1,h_2, h}(m,n,\la).
\label{relationone}
 \eea
 Here $p_{\mathrm{F}}^{h_1,h_2, h}(m,n,\la)$ and
 $p_{\mathrm{B}}^{h_1,h_2, h}(m,n,\la)$ are given by the first terms
 in (\ref{structla}).
 Now we can check the above coefficient (\ref{relationone})
 vanishes at $\la =\frac{1}{4}$ (corresponding to (\ref{qequalone}) in the
 ${\cal N}=2$ SYK models)
 implying that we can decouple the
 currents $V_{\la}^{(h),-}$ with even $h$ and therefore, we do not have
 these currents in (\ref{case1}) or (\ref{case2}). 
For even $h_1$ and $h_2$, the combination $(h_2+h_2-2)$ is even.
 
 Similarly,
 the decoupling of the
 currents $V_{\la}^{(h),+}$ with odd $h$
 can be analyzed as follows.
The commutator relation
 $\big[  (V^{(h_1),+}_{\la})_m,  (V^{(h_2),-}_{\la})_n\big]$ can be obtained 
by substituting the first and second equations of
(\ref{fourrelations}) into
 this commutator, we obtain
 the following coefficient function of $(V^{(h_1+h_2-2-h),+}_{\la})_{m+n}$
 as follows:
 \bea
 && \Bigg[\frac{(h_2-2\la)}{(2h_2-1)}\Bigg]\,
\Bigg[\frac{(h_1+h_2-2-h-1+2\la)}{2(h_1+h_2-2-h)-1}\Bigg]\,
 \Bigg[\frac{1}{\frac{n_{W_{F,h_1}}}{q^{h_1-2}}\,
     \frac{(-1)^{h_1}}{\sum_{i=0}^{h_1-1}\, a^i( h_1, \frac{1}{2})}}
     \Bigg]\Bigg[\frac{1}{\frac{n_{W_{F,h_2}}}{q^{h_2-2}}\,
     \frac{(-1)^{h_2}}{\sum_{i=0}^{h_2-1}\, a^i( h_2, \frac{1}{2})}}
   \Bigg]\nonu \\
 && \times 
\Bigg[\frac{n_{W_{F,h_1+h_2-2-h}}}{q^{h_1+h_2-2-h-2}}\,
    \frac{(-1)^{h_1+h_2-2-h}}{\sum_{i=0}^{h_1+h_2-2-h-1}\, a^i(
      h_1+h_2-2-h, \frac{1}{2})}
  \Bigg] \,  q^h\,
p_{\mathrm{F}}^{h_1,h_2, h}(m,n,\la)
\nonu \\
&& -  \Bigg[\frac{(h_2-1+2\la)}{(2h_2-1)}\Bigg]\,
\Bigg[\frac{(h_1+h_2-2-h-2\la)}{2(h_1+h_2-2-h)-1}\Bigg]\,
    \Bigg[ \frac{1}{ \frac{n_{W_{B,h_1}}}{q^{h_1-2}}\,        
    \frac{(-1)^{h_1}}{\sum_{i=0}^{h_1-1}\, a^i( h_1, 0)}} \Bigg]
\Bigg[ \frac{1}{ \frac{n_{W_{B,h_2}}}{q^{h_2-2}}\,
 \frac{(-1)^{h_2}}{\sum_{i=0}^{h_2-1}\, a^i( h_2, 0)}} \Bigg]
\nonu \\
&& \times \Bigg[ \frac{n_{W_{B,h_1+h_2-2-h}}}{q^{h_1+h_2-2-h-2}}\,
  \frac{(-1)^{h_1+h_2-2-h}}{\sum_{i=0}^{h_1+h_2-2-h-1}\,
    a^i( h_1+h_2-2-h, 0)} \Bigg]\, q^h\,
p_{\mathrm{B}}^{h_1,h_2, h}(m,n,\la).
 \label{relationtwo}
 \eea
 The above coefficient (\ref{relationtwo})
 vanishes at $\la =\frac{1}{4}$
 and  we can decouple the
 currents $V_{\la}^{(h),+}$ with odd $h$ and therefore, we do not have
 these currents in (\ref{case1}) or (\ref{case2}).
 For even $h_1$ and odd $h_2$, the combination
 $(h_2+h_2-2)$ is odd.
 In Appendix $E$, we will see more details on this matter.
 

\subsubsection{The relation with celestial holography}

We have found the matrix generalization of
the ${\cal N}=2$ supersymmetric $W_{\infty}$ algebra
\cite{PRS1990} by adding the additional parameter $\la$.
Then we can follow the procedure of \cite{PRSS} by using
the topological twisting \cite{Witten1988,EY}. The bosonic
$SU(N)$-singlet current of weight $h$ can be given by
$W_{B,h}^{\la}$, $W_{F,h}^{\la}$, $\bar{\pa}\, W_{B,h-1}^{\la}$ and
$\bar{\pa}\, W_{F,h-1}^{\la}$.
The corresponding $SU(N)$-adjoint current
can be constructed by multiplying the $SU(N)$ generators
into the above four kinds of operators. For the fermionic
currents, we take $Q^{\la}_{h+\frac{1}{2}}$ and
$Q^{\la,\hat{A}}_{h+\frac{1}{2}}$. Then the seven OPEs
between these currents (or the corresponding (anti)commutator
relations) can be determined explicitly. The structure constants
found in \cite{PRSS,Ahn2202} can be generalized to
$\la$ dependent ones where the explicit expressions are given by
(\ref{structla}).
When we apply the two-dimensional algebra to the
${\cal N}=1$ supersymmetric Einstein-Yang-Mills theory, it is crucial
to realize that the mode dependent function (\ref{Ndef})
is obtained by performing the nontrivial contour integrals
\cite{MRSV}.
Then the OPEs between the graviton, the gravitino, the gluon and
the gluino
can be  obtained and the corresponding structure constants
are given in (\ref{structla}) with $\la$ dependence.

\section{ Conclusions and outlook}
We derived that the parameter of ${\cal N}=2$ SYK models 
can be realized by the one in the ${\cal N}=2$
supersymmetric linear $W^{N,N}_{\infty}[\la]$ algebra through the
equation (\ref{laq}).
The complete results for
the ${\cal N}=2$
supersymmetric linear $W^{N,N}_{\infty}[\la]$ algebra
 are summarized by (\ref{final1}) and Appendix (\ref{Final}). 

It is an open problem to compare the present results with
the ones in \cite{Bergetalnpb,Bergetalplb} and to observe how they
coincide with each other analytically. So far, we have considered
the ${\cal N}=2$ SYK models and it is interesting to study
the ${\cal N}=(0,2)$ SYK models and check whether there exists
a higher spin realization or not. There is a partial work
in \cite{AP} on the limit of $q_{syk} \rightarrow \frac{N}{M}$
in this
direction.
It is also interesting problem to generalize the work of \cite{FSTZ}
to the case having the above ${\cal N}=2$
supersymmetric linear $W^{N,N}_{\infty}[\la]$ algebra.

\vspace{.7cm}

\centerline{\bf Acknowledgments}

We
would like to
thank
Y. Hikida on \cite{CHR},
M.H. Kim on the generalized hypergeometric function
and M. Vasiliev on \cite{Bergetalnpb,Bergetalplb}
for discussions.
This work was supported by
the National Research Foundation of Korea(NRF) grant
funded by the Korea government(MSIT)(No. 2020R1F1A1066893).

\newpage

\appendix

\renewcommand{\theequation}{\Alph{section}\mbox{.}\arabic{equation}}

\section{
The $SU(N)$-adjoint higher spin currents }

As done in (\ref{WWQQlazero}), we can check that
the  $SU(N)$-adjoint higher spin currents
at vanishing $\la$
can be obtained as follows:
\bea
W_{F,h}^{\hat{A}} & = &
\frac{n_{W_{F,h}}}{q^{h-2}}\,
\frac{(-1)^h}{\sum_{i=0}^{h-1}\, a^i( h, \la+\frac{1}{2}=\frac{1}{2})}\,
\Bigg[\frac{(h-1+2\la)}{(2h-1)}\, V_{\la,\hat{A}}^{(h)+} + V_{\la,\hat{A}}^{(h)-}
\Bigg]_{\la=0},
\nonu \\
W_{B,h}^{\hat{A}} & = &
 \frac{n_{W_{B,h}}}{q^{h-2}}\,
 \frac{(-1)^h}{\sum_{i=0}^{h-1}\, a^i( h, \la=0)}
\,
\Bigg[\frac{(h-2\la)}{(2h-1)}\, V_{\la,\hat{A}}^{(h)+} - V_{\la,\hat{A}}^{(h)-}
  \Bigg]_{\la=0},
\nonu \\
Q^{\hat{A}}_{h+\frac{1}{2}} & = & \frac{1}{2} \,\frac{n_{W_{Q,h+\frac{1}{2}}}}{q^{h-1}}
\, \frac{(-1)^{h+1}  \, h }{
   \sum_{i=0}^{h-1} \, \beta^i( h+1, \la=0)}\, \Bigg[
  Q_{\la,\hat{A}}^{(h+1)-} - Q_{\la,\hat{A}}^{(h+1)+}\Bigg]_{\la=0},
\nonu \\
\bar{Q}^{\hat{A}}_{h+\frac{1}{2}} & = & \frac{1}{2} \,
\frac{n_{W_{Q,h+\frac{1}{2}}}}{q^{h-1}} \,
 \frac{(-1)^{h+1}  }{
   \sum_{i=0}^{h} \, \al^i( h+1, \la=0)} \,
\Bigg[ Q_{\la,\hat{A}}^{(h+1)-} +
  Q_{\la,\hat{A}}^{(h+1)+}\Bigg]_{\la=0}.
\label{appenda1}
\eea
Similarly, for nonzero $\la$, we take the following higher
spin currents together with (\ref{VVQQla})
\bea
W_{F,h}^{\la,\hat{A}} & = &
\frac{n_{W_{F,h}}}{q^{h-2}}\,
\frac{(-1)^h}{\sum_{i=0}^{h-1}\, a^i( h, \la+\frac{1}{2}=\frac{1}{2})}\,
\Bigg[\frac{(h-1+2\la)}{(2h-1)}\, V_{\la,\hat{A}}^{(h)+} + V_{\la,\hat{A}}^{(h)-}
\Bigg],
\nonu \\
W_{B,h}^{\la,\hat{A}} & = &
 \frac{n_{W_{B,h}}}{q^{h-2}}\,
 \frac{(-1)^h}{\sum_{i=0}^{h-1}\, a^i( h, \la=0)}
\,
\Bigg[\frac{(h-2\la)}{(2h-1)}\, V_{\la,\hat{A}}^{(h)+} - V_{\la,\hat{A}}^{(h)-}
  \Bigg],
\nonu \\
Q^{\la,\hat{A}}_{h+\frac{1}{2}} & = &
\frac{1}{2} \,\frac{n_{W_{Q,h+\frac{1}{2}}}}{q^{h-1}}
\, \frac{(-1)^{h+1}  \, h }{
   \sum_{i=0}^{h-1} \, \beta^i( h+1, \la=0)}\, \Bigg[
  Q_{\la,\hat{A}}^{(h+1)-} - Q_{\la,\hat{A}}^{(h+1)+}\Bigg],
\nonu \\
\bar{Q}^{\la,\hat{A}}_{h+\frac{1}{2}} & = & \frac{1}{2} \,
\frac{n_{W_{Q,h+\frac{1}{2}}}}{q^{h-1}} \,
 \frac{(-1)^{h+1}  }{
   \sum_{i=0}^{h} \, \al^i( h+1, \la=0)} \,
\Bigg[ Q_{\la,\hat{A}}^{(h+1)-} +
  Q_{\la,\hat{A}}^{(h+1)+}\Bigg].
\label{appenda2}
\eea
In the normalization of Appendix (\ref{appenda2}), we take
the same normalization of Appendix (\ref{appenda1}). That is,
the overall factor does not depend on the $\la$ explicitly.

\section{ The partial OPEs in the 
${\cal N}=2$ supersymmetric linear
  $W^{N,N}_{\infty}[\la]$ algebra  
}

We present six examples of (\ref{final1}) for fixed $h_1$
and $h_2$.

\subsection{The OPE  $W_{B,4}^{\la}(\bar{z}) \, W_{B,4}^{\la}(\bar{w})$}

From (\ref{WWQQnonzerola}), (\ref{VVQQla}) and (\ref{fundOPE}),
we can calculate the OPE between the weight-$4$ currents
as follows:
\bea
&& W_{B,4}^{\la}(\bar{z}) \, W_{B,4}^{\la}(\bar{w})  = 
\frac{1}{(\bar{z}-\bar{w})^8}\, \Bigg[
\frac{3072}{5} (28 \lambda ^6-84 \lambda ^5+35 \lambda ^4+70 \lambda ^3-42 \lambda ^2-7 \lambda +3)
\Bigg]\nonu \\
&& +
  \frac{1}{(\bar{z}-\bar{w})^6}\, \Bigg[
 \frac{2048}{5} (\lambda -2) (\lambda +1) (2 \lambda -3) (2 \lambda +1)
    \Bigg] \,  W_{B,2}^{\la}(\bar{w})
  \nonu \\
  & &+ \frac{1}{(\bar{z}-\bar{w})^5}\, \frac{1}{2}\, \Bigg[
     \frac{2048}{5} (\lambda -2) (\lambda +1) (2 \lambda -3) (2 \lambda +1)
     \Bigg] \,  \bar{\pa}\, W_{B,2}^{\la}(\bar{w})
  \nonu \\
    & &+ \frac{1}{(\bar{z}-\bar{w})^4}\, \Bigg[ \frac{3}{20}
    \frac{2048}{5} (\lambda -2) (\lambda +1) (2 \lambda -3)
    (2 \lambda +1)     \bar{\pa}^2\, W_{B,2}^{\la}
    \nonu \\
    && -  \frac{192}{5}  (2 \lambda ^2-2 \lambda -9)
  W_{B,4}^{\la}  \Bigg](\bar{w})
  \nonu \\
    & &+ \frac{1}{(\bar{z}-\bar{w})^3} \Bigg[ \frac{1}{30} 
    \frac{2048}{5} (\lambda -2) (\lambda +1) (2 \lambda -3)
    (2 \lambda +1)      \bar{\pa}^3 W_{B,2}^{\la}
    \nonu \\
    && - \frac{1}{2}\frac{192}{5}  (2 \lambda ^2-2 \lambda -9) 
 \bar{\pa}  W_{B,4}^{\la}  \Bigg](\bar{w})
  \nonu \\
     & &+ \frac{1}{(\bar{z}-\bar{w})^2}\, \Bigg[ \frac{1}{168}\, 
 \frac{2048}{5} (\lambda -2) (\lambda +1) (2 \lambda -3) (2 \lambda +1)    \,  \bar{\pa}^4\, W_{B,2}^{\la}
 \nonu \\
 && - \frac{5}{36}\,\frac{192}{5}  (2 \lambda ^2-2 \lambda -9) \,
 \bar{\pa}^2\,  W_{B,4}^{\la} + 6 \,  W_{B,6}^{\la} \Bigg](\bar{w})
  \nonu \\
  &&
 + \frac{1}{(\bar{z}-\bar{w})}\, \Bigg[ \frac{1}{1120}\, 
   \frac{2048}{5} (\lambda -2) (\lambda +1) (2 \lambda -3) (2 \lambda +1)  \,  \bar{\pa}^5\, W_{B,2}^{\la}
  \nonu \\
 & &- \frac{1}{36}\,\frac{192}{5}  (2 \lambda ^2-2 \lambda -9) \,
  \bar{\pa}^3\,  W_{B,4}^{\la}  + \frac{1}{2} \, 6 \,   \bar{\pa}\,
  W_{B,6}^{\la} \Bigg](\bar{w}) + \cdots
  \nonu \\
  & &=  \frac{1}{(\bar{z}-\bar{w})^8}\,
  \Bigg[
\frac{3072}{5} (28 \lambda ^6-84 \lambda ^5+35 \lambda ^4+70 \lambda ^3-42 \lambda ^2-7 \lambda +3)
\Bigg]\nonu \\
  && -
  p_{B,4}^{4,4}(\bar{\pa}_{\bar{z}},\bar{\pa}_{\bar{w}},\la)
  \Bigg[ \frac{ W_{B,2}^{\la}(\bar{w})}{(\bar{z}-\bar{w})} \Bigg]
-
  p_{B,2}^{4,4}(\bar{\pa}_{\bar{z}},\bar{\pa}_{\bar{w}},\la)
  \Bigg[ \frac{ W_{B,4}^{\la}(\bar{w})}{(\bar{z}-\bar{w})} \Bigg]
 -  p_{B,0}^{4,4}(\bar{\pa}_{\bar{z}},\bar{\pa}_{\bar{w}},\la)
\Bigg[ \frac{ W_{B,6}^{\la}(\bar{w})}{(\bar{z}-\bar{w})} \Bigg]
\nonu \\
&& + \cdots.
\label{first}
\eea
It is straightforward to calculate this result
because we are 
considering the linear algebra and collect each pole in terms
of the various descendant terms and new quasiprimary operator
inside the Thielemans package \cite{Thielemans}. Then all the structure constants
can be determined and depend on the $\la$ explicitly as above.
In Appendix (\ref{first}), we also present the structure constants
in terms of (\ref{structla}) after inserting the derivatives.
As we expect, up to minus sign,  we observe that the
above OPE behaves as 
the first and third terms of the second equation in (\ref{final1})
in the sense that there are three terms with correct
mode dependent structure constants.
Due to the $(-1)^{h-1}$ factor when we change from the second
equation of (\ref{final1}) to the above OPE and there are
$h=0,2,4$ cases which are even, we have all the minus signs
in the above OPE.

Then how we can see the existence of the second term of the
second equation of (\ref{final1})?
We can read off the weight from 
the condition $h=h_1+h_2-3$ with $h_1+h_2-2-h=1$
and we take $h_1=4$ and $h_2=3$. Then we observe the weight
for the $W_{B,h_1+h_2-2-h}^{\la}$ is equal to $1$.
Then we can calculate the following OPE
\bea
&& W_{B,4}^{\la}(\bar{z}) \, W_{B,3}^{\la}(\bar{w})  = 
\frac{1}{(\bar{z}-\bar{w})^7}\, \Bigg[
384 (\lambda -1) \lambda  (2 \lambda -3) (2 \lambda -1) (2 \lambda +1)
\Bigg]
\nonu \\
&& +
\frac{1}{(\bar{z}-\bar{w})^6}\, \Bigg[
  512 (\lambda -1) \lambda  (2 \lambda -3) (2 \lambda +1)
 \, W_{B,1}^{\la} \Bigg](\bar{w})
\nonu \\
&& +
\frac{1}{(\bar{z}-\bar{w})^5}\, \Bigg[
  512 (\lambda -1) \lambda  (2 \lambda -3) (2 \lambda +1)
 \, \bar{\pa}\, W_{B,1}^{\la} \Bigg](\bar{w})
\nonu \\
&& +
\frac{1}{(\bar{z}-\bar{w})^4}\, \Bigg[
 \frac{1}{2}\, 512 (\lambda -1) \lambda  (2 \lambda -3) (2 \lambda +1)
 \, \bar{\pa}^2 \, W_{B,1}^{\la}
-\frac{384}{5}  (\lambda -2) (\lambda +1) \, W_{B,3}^{\la}
 \Bigg](\bar{w})
\nonu \\
&& +
\frac{1}{(\bar{z}-\bar{w})^3}\, \Bigg[
 \frac{1}{6}\, 512 (\lambda -1) \lambda  (2 \lambda -3) (2 \lambda +1)
 \, \bar{\pa}^3 \, W_{B,1}^{\la}
 -\frac{2}{3}\, \frac{384}{5}  (\lambda -2) (\lambda +1) \,
 \bar{\pa}\, W_{B,3}^{\la}
 \Bigg](\bar{w})
\nonu \\
&& +
\frac{1}{(\bar{z}-\bar{w})^2}\, \Bigg[
 \frac{1}{24}\, 512 (\lambda -1) \lambda  (2 \lambda -3) (2 \lambda +1)
 \, \bar{\pa}^4 \, W_{B,1}^{\la}
 -\frac{5}{21}\, \frac{384}{5}  (\lambda -2) (\lambda +1) \,
 \bar{\pa}^2 \, W_{B,3}^{\la}
 \nonu \\
 && + 5 \, W_{B,5}^{\la} \Bigg](\bar{w})
\nonu \\
&& +
\frac{1}{(\bar{z}-\bar{w})}\, \Bigg[
 \frac{1}{120}\, 512 (\lambda -1) \lambda  (2 \lambda -3) (2 \lambda +1)
 \, \bar{\pa}^5 \, W_{B,1}^{\la}
 -\frac{5}{84}\, \frac{384}{5}  (\lambda -2) (\lambda +1) \,
 \bar{\pa}^3 \, W_{B,3}^{\la}
 \nonu \\
 && + \frac{3}{5} \, 5 \, \bar{\pa}\, W_{B,5}^{\la} \Bigg](\bar{w}) +\cdots
\nonu \\
&& =
\frac{1}{(\bar{z}-\bar{w})^7}\, \Bigg[
384 (\lambda -1) \lambda  (2 \lambda -3) (2 \lambda -1) (2 \lambda +1)
\Bigg]
\nonu \\
  && -
  p_{B,4}^{4,3}(\bar{\pa}_{\bar{z}},\bar{\pa}_{\bar{w}},\la)
  \Bigg[ \frac{ W_{B,1}^{\la}(\bar{w})}{(\bar{z}-\bar{w})} \Bigg]
-
  p_{B,2}^{4,3}(\bar{\pa}_{\bar{z}},\bar{\pa}_{\bar{w}},\la)
  \Bigg[ \frac{ W_{B,3}^{\la}(\bar{w})}{(\bar{z}-\bar{w})} \Bigg]
 -  p_{B,0}^{4,3}(\bar{\pa}_{\bar{z}},\bar{\pa}_{\bar{w}},\la)
\Bigg[ \frac{ W_{B,5}^{\la}(\bar{w})}{(\bar{z}-\bar{w})} \Bigg]
\nonu \\
&& + \cdots.
\label{first-1}
\eea
Therefore, in this example, we observe the three terms of
the second equation of (\ref{final1}).
Note that all the structure constants associated with
the weight-$1$ current $ W_{B,1}^{\la}(\bar{w})$ in Appendix
(\ref{first-1})
contain the $\la$ factor explicitly.
This implies that when we take the vanishing limit
for the $\la$, the weight-$1$ current and its descendant terms
disappear.

\subsection{The OPE  $W_{F,4}^{\la}(\bar{z}) \, Q_{\frac{7}{2}}^{\la}(\bar{w})$}

Let us consider the third equation of (\ref{final1})
with $h_1=4$ and $h_2=3$. Then we obtain the following
result as done before
\bea
&& W_{F,4}^{\la}(\bar{z}) \, Q_{\frac{7}{2}}^{\la}(\bar{w})  = 
  \frac{1}{(\bar{z}-\bar{w})^6}\, \Bigg[
\frac{256}{5} (\lambda -1) (\lambda +1) (2 \lambda -3) (2 \lambda +1) (2 \lambda +3)
    \Bigg] \,  Q_{\frac{3}{2}}^{\la}(\bar{w})
  \nonu \\
  && +
 \frac{1}{(\bar{z}-\bar{w})^5}\, \Bigg[
   \frac{2}{3}
   \frac{256}{5} (\lambda -1) (\lambda +1) (2 \lambda -3) (2 \lambda +1) (2 \lambda +3) \,  \bar{\pa}\, Q_{\frac{3}{2}}^{\la}
   \nonu \\
   && -\frac{256}{25} (\lambda -4) (\lambda +1) (2 \lambda -3)
   (2 \lambda +3)\,  Q_{\frac{5}{2}}^{\la} \Bigg](\bar{w})
 \nonu \\
 && +
 \frac{1}{(\bar{z}-\bar{w})^4}\, \Bigg[
   \frac{1}{4}
   \frac{256}{5} (\lambda -1) (\lambda +1) (2 \lambda -3) (2 \lambda +1) (2 \lambda +3) \,  \bar{\pa}^2\, Q_{\frac{3}{2}}^{\la}
   \nonu \\
   && -\frac{3}{5} \,
   \frac{256}{25} (\lambda -4) (\lambda +1) (2 \lambda -3)
   (2 \lambda +3)\,  \bar{\pa}\, Q_{\frac{5}{2}}^{\la}-
   \frac{16}{25} (2 \lambda +3) (4 \lambda ^2+18 \lambda -61)
   \, Q_{\frac{7}{2}}^{\la} 
   \Bigg](\bar{w})
 \nonu \\
  && +
 \frac{1}{(\bar{z}-\bar{w})^3}\, \Bigg[
   \frac{1}{15}
   \frac{256}{5} (\lambda -1) (\lambda +1) (2 \lambda -3) (2 \lambda +1) (2 \lambda +3) \,  \bar{\pa}^3\, Q_{\frac{3}{2}}^{\la}
   \nonu \\
   && -\frac{1}{5} \,
   \frac{256}{25} (\lambda -4) (\lambda +1) (2 \lambda -3)
   (2 \lambda +3)\,  \bar{\pa}^2\, Q_{\frac{5}{2}}^{\la}-
   \frac{4}{7}\,
   \frac{16}{25} (2 \lambda +3) (4 \lambda ^2+18 \lambda -61)
   \, \bar{\pa}\, Q_{\frac{7}{2}}^{\la} 
   \nonu \\
   && +
\frac{4}{5} (2 \lambda ^2-5 \lambda -27)\,  Q_{\frac{9}{2}}^{\la} 
   \Bigg](\bar{w})
 \nonu \\
  && +
 \frac{1}{(\bar{z}-\bar{w})^2}\, \Bigg[
   \frac{1}{72}
   \frac{256}{5} (\lambda -1) (\lambda +1) (2 \lambda -3) (2 \lambda +1) (2 \lambda +3) \,  \bar{\pa}^4\, Q_{\frac{3}{2}}^{\la}
   \nonu \\
   && -\frac{1}{21} \,
   \frac{256}{25} (\lambda -4) (\lambda +1) (2 \lambda -3)
   (2 \lambda +3)\,  \bar{\pa}^3\, Q_{\frac{5}{2}}^{\la}-
   \frac{5}{28}\,
   \frac{16}{25} (2 \lambda +3) (4 \lambda ^2+18 \lambda -61)
   \, \bar{\pa}^2\, Q_{\frac{7}{2}}^{\la} 
   \nonu \\
   && +\frac{5}{9}\,
\frac{4}{5} (2 \lambda ^2-5 \lambda -27)\, \bar{\pa}\, Q_{\frac{9}{2}}^{\la} 
+
\frac{1}{10} (2 \lambda +27) \, Q_{\frac{11}{2}}^{\la}  
\Bigg](\bar{w})
 \nonu \\
 && +
 \frac{1}{(\bar{z}-\bar{w})}\, \Bigg[
   \frac{1}{420}
   \frac{256}{5} (\lambda -1) (\lambda +1) (2 \lambda -3) (2 \lambda +1) (2 \lambda +3) \,  \bar{\pa}^5\, Q_{\frac{3}{2}}^{\la}
   \nonu \\
   && -\frac{1}{112} \,
   \frac{256}{25} (\lambda -4) (\lambda +1) (2 \lambda -3)
   (2 \lambda +3)\,  \bar{\pa}^4\, Q_{\frac{5}{2}}^{\la}
   \nonu \\
   && -
   \frac{5}{126}\,
   \frac{16}{25} (2 \lambda +3) (4 \lambda ^2+18 \lambda -61)
   \, \bar{\pa}^3\, Q_{\frac{7}{2}}^{\la} 
   \nonu \\
   && +\frac{1}{6}\,
\frac{4}{5} (2 \lambda ^2-5 \lambda -27)\, \bar{\pa}^2\, Q_{\frac{9}{2}}^{\la} 
+\frac{6}{11}\,
\frac{1}{10} (2 \lambda +27) \, \bar{\pa}\, Q_{\frac{11}{2}}^{\la}  
-\frac{1}{4} \, Q_{\frac{13}{2}}^{\la} \Bigg](\bar{w})+\cdots
 \nonu  \\
  & &=  -
  q_{F,4}^{4,\frac{7}{2}}(\bar{\pa}_{\bar{z}},\bar{\pa}_{\bar{w}},\la)
  \Bigg[ \frac{ Q_{\frac{3}{2}}^{\la}(\bar{w})}{(\bar{z}-\bar{w})} \Bigg]
+
  q_{F,3}^{4,\frac{7}{2}}(\bar{\pa}_{\bar{z}},\bar{\pa}_{\bar{w}},\la)
  \Bigg[ \frac{ Q_{\frac{5}{2}}^{\la}(\bar{w})}{(\bar{z}-\bar{w})} \Bigg]
   -  q_{F,2}^{4,\frac{7}{2}}(\bar{\pa}_{\bar{z}},\bar{\pa}_{\bar{w}},\la)
  \Bigg[ \frac{ Q_{\frac{7}{2}}^{\la}(\bar{w})}{(\bar{z}-\bar{w})} \Bigg]
  \nonu \\
  && +  q_{F,1}^{4,\frac{7}{2}}(\bar{\pa}_{\bar{z}},\bar{\pa}_{\bar{w}},\la)
  \Bigg[ \frac{ Q_{\frac{9}{2}}^{\la}(\bar{w})}{(\bar{z}-\bar{w})} \Bigg]
 -
  q_{F,0}^{4,\frac{7}{2}}(\bar{\pa}_{\bar{z}},\bar{\pa}_{\bar{w}},\la)
  \Bigg[ \frac{ Q_{\frac{11}{2}}^{\la}(\bar{w})}{(\bar{z}-\bar{w})} \Bigg]
  +q_{F,-1}^{4,\frac{7}{2}}(\bar{\pa}_{\bar{z}},\bar{\pa}_{\bar{w}},\la)
  \Bigg[ \frac{ Q_{\frac{13}{2}}^{\la}(\bar{w})}{(\bar{z}-\bar{w})} \Bigg]
  \nonu \\
  && + \cdots.
  \label{second}
  \eea
  As we expect, there are six nontrivial terms with
  $h=-1$, $0$, $1$, $2$, $3$ and $4$.
  Due to the previous overall factor $(-1)^{h-1}$,
  we have minus signs for the even $h$. For odd $h$
  we have plus signs in the above OPE in
  Appendix (\ref{second}).
  
\subsection{The OPE  $W_{B,4}^{\la}(\bar{z}) \, Q_{\frac{7}{2}}^{\la}(\bar{w})$}

Let us move on the fourth equation of (\ref{final1})
and we can calculate the corresponding OPE for $h_1=4$
and $h_2=3$ in our notation and this leads to the following OPE
\bea
&& W_{B,4}^{\la}(\bar{z}) \, Q_{\frac{7}{2}}^{\la}(\bar{w})  = 
\frac{1}{(\bar{z}-\bar{w})^6}\, \Bigg[
-  \frac{512}{5}  (\lambda -2) (\lambda -1) (\lambda +1) (2 \lambda -3) (2 \lambda +1)
    \Bigg] \,  Q_{\frac{3}{2}}^{\la}(\bar{w})
  \nonu \\
  && +
 \frac{1}{(\bar{z}-\bar{w})^5}\, \Bigg[
  - \frac{2}{3}
  \frac{512}{5}  (\lambda -2) (\lambda -1) (\lambda +1) (2 \lambda -3) (2 \lambda +1)  \,  \bar{\pa}\, Q_{\frac{3}{2}}^{\la}
   \nonu \\
   && +\frac{256}{25} (\lambda -2) (\lambda +1) (2 \lambda -3) (2 \lambda +7) \,  Q_{\frac{5}{2}}^{\la} \Bigg](\bar{w})
 \nonu \\
 && +
 \frac{1}{(\bar{z}-\bar{w})^4}\, \Bigg[
-   \frac{1}{4}
  \frac{512}{5}  (\lambda -2) (\lambda -1) (\lambda +1) (2 \lambda -3) (2 \lambda +1)  \,  \bar{\pa}^2\, Q_{\frac{3}{2}}^{\la}
   \nonu \\
   && +\frac{3}{5} \,\frac{256}{25} (\lambda -2) (\lambda +1) (2 \lambda -3) (2 \lambda +7)
   \,  \bar{\pa}\, Q_{\frac{5}{2}}^{\la}
+\frac{32}{25} (\lambda -2) \left(4 \lambda ^2-22 \lambda -51\right)
   \, Q_{\frac{7}{2}}^{\la} 
   \Bigg](\bar{w})
 \nonu \\
  && +
 \frac{1}{(\bar{z}-\bar{w})^3}\, \Bigg[
 -  \frac{1}{15}
  \frac{512}{5}  (\lambda -2) (\lambda -1) (\lambda +1) (2 \lambda -3) (2 \lambda +1)   \,  \bar{\pa}^3\, Q_{\frac{3}{2}}^{\la}
   \nonu \\
   && +\frac{1}{5} \,
\frac{256}{25} (\lambda -2) (\lambda +1) (2 \lambda -3) (2 \lambda +7)  \,
\bar{\pa}^2\, Q_{\frac{5}{2}}^{\la}
+\frac{4}{7} \, \frac{32}{25} (\lambda -2) (4 \lambda ^2-22 \lambda -51)
   \, \bar{\pa}\, Q_{\frac{7}{2}}^{\la} 
   \nonu \\
   && - \frac{4}{5}  (2 \lambda ^2+3 \lambda -29)
\,  Q_{\frac{9}{2}}^{\la} 
   \Bigg](\bar{w})
 \nonu \\
  && +
 \frac{1}{(\bar{z}-\bar{w})^2}\, \Bigg[
 -  \frac{1}{72}  \frac{512}{5}  (\lambda -2) (\lambda -1) (\lambda +1) (2 \lambda -3) (2 \lambda +1) 
  \,  \bar{\pa}^4\, Q_{\frac{3}{2}}^{\la}
   \nonu \\
   && +\frac{1}{21} \,
   \frac{256}{25} (\lambda -2) (\lambda +1) (2 \lambda -3) (2 \lambda +7)
   \, \bar{\pa}^3 \, Q_{\frac{5}{2}}^{\la}+
   \frac{5}{28}\,
  \frac{32}{25} (\lambda -2) (4 \lambda ^2-22 \lambda -51)
   \, \bar{\pa}^2\, Q_{\frac{7}{2}}^{\la} 
   \nonu \\
   && -\frac{5}{9}\,
 \frac{4}{5}  (2 \lambda ^2+3 \lambda -29)\, \bar{\pa}\, Q_{\frac{9}{2}}^{\la} 
+
\frac{1}{5} (- \lambda +14) \, Q_{\frac{11}{2}}^{\la}  
\Bigg](\bar{w})
 \nonu \\
 && +
 \frac{1}{(\bar{z}-\bar{w})}\, \Bigg[
-   \frac{1}{420}
 \frac{512}{5}  (\lambda -2) (\lambda -1) (\lambda +1) (2 \lambda -3) (2 \lambda +1)    \,  \bar{\pa}^5\, Q_{\frac{3}{2}}^{\la}
   \nonu \\
   && +\frac{1}{112} \,
\frac{256}{25} (\lambda -2) (\lambda +1) (2 \lambda -3) (2 \lambda +7)  \,  \bar{\pa}^4\, Q_{\frac{5}{2}}^{\la}+
   \frac{5}{126}\,
 \frac{32}{25} (\lambda -2) (4 \lambda ^2-22 \lambda -51)
   \, \bar{\pa}^3\, Q_{\frac{7}{2}}^{\la} 
   \nonu \\
   && -\frac{1}{6}\,
 \frac{4}{5}  (2 \lambda ^2+3 \lambda -29)
   \, \bar{\pa}^2 \, Q_{\frac{9}{2}}^{\la} 
+\frac{6}{11}\,
\frac{1}{5} (- \lambda +14)
\, \bar{\pa}\, Q_{\frac{11}{2}}^{\la}  
+\frac{1}{4} \, Q_{\frac{13}{2}}^{\la} \Bigg](\bar{w})+\cdots
 \nonu  \\
  & &=  -
  q_{B,4}^{4,\frac{7}{2}}(\bar{\pa}_{\bar{z}},\bar{\pa}_{\bar{w}},\la)
  \Bigg[ \frac{ Q_{\frac{3}{2}}^{\la}(\bar{w})}{(\bar{z}-\bar{w})} \Bigg]
+
  q_{B,3}^{4,\frac{7}{2}}(\bar{\pa}_{\bar{z}},\bar{\pa}_{\bar{w}},\la)
  \Bigg[ \frac{ Q_{\frac{5}{2}}^{\la}(\bar{w})}{(\bar{z}-\bar{w})} \Bigg]
   -  q_{B,2}^{4,\frac{7}{2}}(\bar{\pa}_{\bar{z}},\bar{\pa}_{\bar{w}},\la)
  \Bigg[ \frac{ Q_{\frac{7}{2}}^{\la}(\bar{w})}{(\bar{z}-\bar{w})} \Bigg]
  \nonu \\
  && +  q_{B,1}^{4,\frac{7}{2}}(\bar{\pa}_{\bar{z}},\bar{\pa}_{\bar{w}},\la)
  \Bigg[ \frac{ Q_{\frac{9}{2}}^{\la}(\bar{w})}{(\bar{z}-\bar{w})} \Bigg]
 -
  q_{B,0}^{4,\frac{7}{2}}(\bar{\pa}_{\bar{z}},\bar{\pa}_{\bar{w}},\la)
  \Bigg[ \frac{ Q_{\frac{11}{2}}^{\la}(\bar{w})}{(\bar{z}-\bar{w})} \Bigg]
  +q_{B,-1}^{4,\frac{7}{2}}(\bar{\pa}_{\bar{z}},\bar{\pa}_{\bar{w}},\la)
  \Bigg[ \frac{ Q_{\frac{13}{2}}^{\la}(\bar{w})}{(\bar{z}-\bar{w})} \Bigg] + \cdots.
  \nonu \\
  \label{third}
  \eea
  Again, the alternating signs appear due to the previous analysis
  in Appendix (\ref{third}). For even $h$, there is a minus sign.
  
\subsection{The OPE  $W_{F,4}^{\la}(\bar{z}) \,
  \bar{Q}_{\frac{7}{2}}^{\la}(\bar{w})$}

Let us continue to calculate the OPE
associated with the fifth equation of (\ref{final1})
and it turns out that 
\bea
&& W_{F,4}^{\la}(\bar{z}) \, \bar{Q}_{\frac{7}{2}}^{\la}(\bar{w})  =
\frac{1}{(\bar{z}-\bar{w})^7}\,\Bigg[
-\frac{8192}{5} \, (\lambda -1) \lambda  (\lambda +1) (2 \lambda -3) (2 \lambda -1) (2 \lambda +1)\Bigg]\,  \bar{Q}_{\frac{1}{2}}^{\la}(\bar{w})
\nonu \\ 
&&+  \frac{1}{(\bar{z}-\bar{w})^6}\, \Bigg[
  -\frac{8192}{5} \, (\lambda -1) \lambda  (\lambda +1) (2 \lambda -3) (2 \lambda -1) (2 \lambda +1)\,  \bar{\pa}\, \bar{Q}_{\frac{1}{2}}^{\la}
  \nonu \\
  && +\frac{512}{5} (\lambda -1) (\lambda +1) (2 \lambda -3) (2 \lambda +1) (2 \lambda +3)\,
 \bar{Q}_{\frac{3}{2}}^{\la}
    \Bigg](\bar{w})
  \nonu \\
  && +
  \frac{1}{(\bar{z}-\bar{w})^5}\, \Bigg[
    -\frac{1}{2} \,
    \frac{8192}{5} \, (\lambda -1) \lambda  (\lambda +1) (2 \lambda -3) (2 \lambda -1) (2 \lambda +1)\,  \bar{\pa}^2 \, \bar{Q}_{\frac{1}{2}}^{\la}
    \nonu \\
 &&+  \frac{2}{3}
 \frac{512}{5} (\lambda -1) (\lambda +1) (2 \lambda -3) (2 \lambda +1) (2 \lambda +3)  \,  \bar{\pa}\, \bar{Q}_{\frac{3}{2}}^{\la}
   \nonu \\
   && +\frac{256}{25} (\lambda -4) (\lambda +1) (2 \lambda -3)
   (2 \lambda +3)\,  \bar{Q}_{\frac{5}{2}}^{\la} \Bigg](\bar{w})
 \nonu \\
 && +
 \frac{1}{(\bar{z}-\bar{w})^4}\, \Bigg[
 -\frac{1}{6} \,
 \frac{8192}{5} \, (\lambda -1) \lambda  (\lambda +1) (2 \lambda -3) (2 \lambda -1) (2 \lambda +1)\,  \bar{\pa}^3 \, \bar{Q}_{\frac{1}{2}}^{\la}
 \nonu \\
 &&  \frac{1}{4}
\frac{512}{5} (\lambda -1) (\lambda +1) (2 \lambda -3) (2 \lambda +1) (2 \lambda +3)
 \,  \bar{\pa}^2\, \bar{Q}_{\frac{3}{2}}^{\la}
   \nonu \\
   && +\frac{3}{5} \,
   \frac{256}{25} (\lambda -4) (\lambda +1) (2 \lambda -3)
   (2 \lambda +3)\,  \bar{\pa}\, \bar{Q}_{\frac{5}{2}}^{\la}-
   \frac{16}{25} (2 \lambda +3) (4 \lambda ^2+18 \lambda -61)
   \, \bar{Q}_{\frac{7}{2}}^{\la} 
   \Bigg](\bar{w})
 \nonu \\
  && +
 \frac{1}{(\bar{z}-\bar{w})^3}\, \Bigg[
   -  \frac{1}{24}\,
   \frac{8192}{5} \, (\lambda -1) \lambda  (\lambda +1) (2 \lambda -3) (2 \lambda -1) (2 \lambda +1)\,  \bar{\pa}^4 \, \bar{Q}_{\frac{1}{2}}^{\la}
   \nonu \\
   && +
   \frac{1}{15}
\frac{512}{5} (\lambda -1) (\lambda +1) (2 \lambda -3) (2 \lambda +1) (2 \lambda +3)
   \,  \bar{\pa}^3\, \bar{Q}_{\frac{3}{2}}^{\la}
   \nonu \\
   && +\frac{1}{5} \,
  \frac{256}{25} (\lambda -4) (\lambda +1) (2 \lambda -3)
   (2 \lambda +3)
   \,  \bar{\pa}^2\, \bar{Q}_{\frac{5}{2}}^{\la}-
   \frac{4}{7}\,
  \frac{16}{25} (2 \lambda +3) (4 \lambda ^2+18 \lambda -61)  
   \, \bar{\pa}\, \bar{Q}_{\frac{7}{2}}^{\la} 
   \nonu \\
   && -
\frac{4}{5} (2 \lambda ^2-5 \lambda -27)\,  \bar{Q}_{\frac{9}{2}}^{\la} 
   \Bigg](\bar{w})
 \nonu \\
  && +
 \frac{1}{(\bar{z}-\bar{w})^2}\, \Bigg[
   - \frac{1}{120} \,
 \frac{8192}{5} \, (\lambda -1) \lambda  (\lambda +1) (2 \lambda -3) (2 \lambda -1) (2 \lambda +1)\,  \bar{\pa}^5 \, \bar{Q}_{\frac{1}{2}}^{\la}
   \nonu \\
   && +
   \frac{1}{72}
\frac{512}{5} (\lambda -1) (\lambda +1) (2 \lambda -3) (2 \lambda +1) (2 \lambda +3)
   \,  \bar{\pa}^4\, \bar{Q}_{\frac{3}{2}}^{\la}
   \nonu \\
   && +\frac{1}{21} \,
 \frac{256}{25} (\lambda -4) (\lambda +1) (2 \lambda -3)
   (2 \lambda +3)
   \,  \bar{\pa}^3\, \bar{Q}_{\frac{5}{2}}^{\la}-
   \frac{5}{28}\,
   \frac{16}{25} (2 \lambda +3) (4 \lambda ^2+18 \lambda -61)
   \, \bar{\pa}^2\, \bar{Q}_{\frac{7}{2}}^{\la} 
   \nonu \\
   && -\frac{5}{9}\,
\frac{4}{5} (2 \lambda ^2-5 \lambda -27)\, \bar{\pa}\, \bar{Q}_{\frac{9}{2}}^{\la} 
+
\frac{1}{10} (2 \lambda +27) \, \bar{Q}_{\frac{11}{2}}^{\la}  
\Bigg](\bar{w})
 \nonu \\
 && +
 \frac{1}{(\bar{z}-\bar{w})}\, \Bigg[
   -\frac{1}{720}\,
\frac{8192}{5} \, (\lambda -1) \lambda  (\lambda +1) (2 \lambda -3) (2 \lambda -1) (2 \lambda +1)\,  \bar{\pa}^6 \, \bar{Q}_{\frac{1}{2}}^{\la}
   \nonu \\
   &&+
   \frac{1}{420}
\frac{512}{5} (\lambda -1) (\lambda +1) (2 \lambda -3) (2 \lambda +1) (2 \lambda +3)
   \,  \bar{\pa}^5\, \bar{Q}_{\frac{3}{2}}^{\la}
   \nonu \\
   && +\frac{1}{112} \,
 \frac{256}{25} (\lambda -4) (\lambda +1) (2 \lambda -3)
   (2 \lambda +3)
   \,  \bar{\pa}^4\, \bar{Q}_{\frac{5}{2}}^{\la}-
   \frac{5}{126}\,
   \frac{16}{25} (2 \lambda +3) (4 \lambda ^2+18 \lambda -61)
   \, \bar{\pa}^3\, \bar{Q}_{\frac{7}{2}}^{\la} 
   \nonu \\
   && -\frac{1}{6}\,
\frac{4}{5} (2 \lambda ^2-5 \lambda -27)\, \bar{\pa}^2\, \bar{Q}_{\frac{9}{2}}^{\la} 
+\frac{6}{11}\,
\frac{1}{10} (2 \lambda +27) \, \bar{\pa}\, \bar{Q}_{\frac{11}{2}}^{\la}  
+\frac{1}{4} \, \bar{Q}_{\frac{13}{2}}^{\la} \Bigg](\bar{w})+\cdots
 \nonu  \\
 & &=
 -
  q_{F,5}^{4,\frac{7}{2}}(\bar{\pa}_{\bar{z}},\bar{\pa}_{\bar{w}},\la)
  \Bigg[ \frac{ \bar{Q}_{\frac{1}{2}}^{\la}(\bar{w})}{(\bar{z}-\bar{w})} \Bigg]
 -
  q_{F,4}^{4,\frac{7}{2}}(\bar{\pa}_{\bar{z}},\bar{\pa}_{\bar{w}},\la)
  \Bigg[ \frac{ \bar{Q}_{\frac{3}{2}}^{\la}(\bar{w})}{(\bar{z}-\bar{w})} \Bigg]
-
  q_{F,3}^{4,\frac{7}{2}}(\bar{\pa}_{\bar{z}},\bar{\pa}_{\bar{w}},\la)
  \Bigg[ \frac{ \bar{Q}_{\frac{5}{2}}^{\la}(\bar{w})}{(\bar{z}-\bar{w})} \Bigg]
  \nonu \\
  && -  q_{F,2}^{4,\frac{7}{2}}(\bar{\pa}_{\bar{z}},\bar{\pa}_{\bar{w}},\la)
  \Bigg[ \frac{ \bar{Q}_{\frac{7}{2}}^{\la}(\bar{w})}{(\bar{z}-\bar{w})} \Bigg]
 -  q_{F,1}^{4,\frac{7}{2}}(\bar{\pa}_{\bar{z}},\bar{\pa}_{\bar{w}},\la)
  \Bigg[ \frac{ \bar{Q}_{\frac{9}{2}}^{\la}(\bar{w})}{(\bar{z}-\bar{w})} \Bigg]
 -
  q_{F,0}^{4,\frac{7}{2}}(\bar{\pa}_{\bar{z}},\bar{\pa}_{\bar{w}},\la)
  \Bigg[ \frac{ \bar{Q}_{\frac{11}{2}}^{\la}(\bar{w})}{(\bar{z}-\bar{w})}
    \Bigg] \nonu \\
  &&
  -q_{F,-1}^{4,\frac{7}{2}}(\bar{\pa}_{\bar{z}},\bar{\pa}_{\bar{w}},\la)
  \Bigg[ \frac{ \bar{Q}_{\frac{13}{2}}^{\la}(\bar{w})}{(\bar{z}-\bar{w})} \Bigg] + \cdots.
  \label{fourth}
  \eea
  Note that in the fifth equation of (\ref{final1}), there
  exists $(-1)^h$ factor. So when we write down the OPE as above,
  this factor combines with the previous factor $(-1)^{h-1}$.
  This implies that there are no $h$ dependence in the $(-1)$
  factor. We are left with the final $(-1)$ factor
  which appears in Appendix (\ref{fourth}).
  As described before, the structure constants
  associated with the weight-$\frac{1}{2}$ current
  contain the $\la$ factor and this leads to the fact that
  this weight-$\frac{1}{2}$ current disappear when we take
  the vanishing $\la$ limit \footnote{There is a factor $(2\la-1)$
    which vanishes at $\la=\frac{1}{2}$. See also the footnote
    \ref{laonehalf}.}.
  
\subsection{The OPE  $W_{B,4}^{\la}(\bar{z}) \, \bar{Q}_{\frac{7}{2}}^{\la}(\bar{w})$}

For the sixth equation of (\ref{final1}), we can calculate
the following OPE
\bea
&& W_{B,4}^{\la}(\bar{z}) \, \bar{Q}_{\frac{7}{2}}^{\la}(\bar{w})  =
\frac{1}{(\bar{z}-\bar{w})^7}\,\Bigg[
\frac{8192}{5} \, (\lambda -1) \lambda  (\lambda +1) (2 \lambda -3) (2 \lambda -1) (2 \lambda +1)\Bigg]\,  \bar{Q}_{\frac{1}{2}}^{\la}(\bar{w})
\nonu \\ 
&&+  \frac{1}{(\bar{z}-\bar{w})^6}\, \Bigg[
  \frac{8192}{5} \, (\lambda -1) \lambda  (\lambda +1) (2 \lambda -3) (2 \lambda -1) (2 \lambda +1)\,  \bar{\pa}\, \bar{Q}_{\frac{1}{2}}^{\la}
  \nonu \\
  &&-\frac{1024}{5} \,
  (\lambda -2) (\lambda -1) (\lambda +1) (2 \lambda -3) (2 \lambda +1)\,
 \bar{Q}_{\frac{3}{2}}^{\la}
    \Bigg](\bar{w})
  \nonu \\
  && +
  \frac{1}{(\bar{z}-\bar{w})^5}\, \Bigg[
    \frac{1}{2} \,
    \frac{8192}{5} \, (\lambda -1) \lambda  (\lambda +1) (2 \lambda -3) (2 \lambda -1) (2 \lambda +1)\,  \bar{\pa}^2 \, \bar{Q}_{\frac{1}{2}}^{\la}
    \nonu \\
 &&-  \frac{2}{3}
 \frac{1024}{5} \,
  (\lambda -2) (\lambda -1) (\lambda +1) (2 \lambda -3) (2 \lambda +1) \,  \bar{\pa}\, \bar{Q}_{\frac{3}{2}}^{\la}
   \nonu \\
   &&-\frac{256}{25}\, (\lambda -2) (\lambda +1) (2 \lambda -3) (2 \lambda +7)\,  \bar{Q}_{\frac{5}{2}}^{\la} \Bigg](\bar{w})
 \nonu \\
 && +
 \frac{1}{(\bar{z}-\bar{w})^4}\, \Bigg[
 \frac{1}{6} \,
 \frac{8192}{5} \, (\lambda -1) \lambda  (\lambda +1) (2 \lambda -3) (2 \lambda -1) (2 \lambda +1)\,  \bar{\pa}^3 \, \bar{Q}_{\frac{1}{2}}^{\la}
 \nonu \\
 &&  -\frac{1}{4}
\frac{1024}{5} \,
  (\lambda -2) (\lambda -1) (\lambda +1) (2 \lambda -3) (2 \lambda +1)
 \,  \bar{\pa}^2\, \bar{Q}_{\frac{3}{2}}^{\la}
   \nonu \\
   && -\frac{3}{5} \,
   \frac{256}{25}\, (\lambda -2) (\lambda +1) (2 \lambda -3) (2 \lambda +7) \,  \bar{\pa}\, \bar{Q}_{\frac{5}{2}}^{\la}
+\frac{32}{25} (\lambda -2) \left(4 \lambda ^2-22 \lambda -51\right)
   \, \bar{Q}_{\frac{7}{2}}^{\la} 
   \Bigg](\bar{w})
 \nonu \\
  && +
 \frac{1}{(\bar{z}-\bar{w})^3}\, \Bigg[
     \frac{1}{24}\,
   \frac{8192}{5} \, (\lambda -1) \lambda  (\lambda +1) (2 \lambda -3) (2 \lambda -1) (2 \lambda +1)\,  \bar{\pa}^4 \, \bar{Q}_{\frac{1}{2}}^{\la}
   \nonu \\
   && -
   \frac{1}{15}
\frac{1024}{5} \,
  (\lambda -2) (\lambda -1) (\lambda +1) (2 \lambda -3) (2 \lambda +1)
   \,  \bar{\pa}^3\, \bar{Q}_{\frac{3}{2}}^{\la}
   \nonu \\
   && -\frac{1}{5} \,
  \frac{256}{25}\, (\lambda -2) (\lambda +1) (2 \lambda -3) (2 \lambda +7)
   \,  \bar{\pa}^2\, \bar{Q}_{\frac{5}{2}}^{\la}+
   \frac{4}{7}\,
 \frac{32}{25} (\lambda -2) (4 \lambda ^2-22 \lambda -51)
   \, \bar{\pa}\, \bar{Q}_{\frac{7}{2}}^{\la} 
   \nonu \\
   && +
\frac{4}{5} (2 \lambda ^2+3 \lambda -29)
   \,  \bar{Q}_{\frac{9}{2}}^{\la} 
   \Bigg](\bar{w})
 \nonu \\
  && +
 \frac{1}{(\bar{z}-\bar{w})^2}\, \Bigg[
    \frac{1}{120} \,
 \frac{8192}{5} \, (\lambda -1) \lambda  (\lambda +1) (2 \lambda -3) (2 \lambda -1) (2 \lambda +1)\,  \bar{\pa}^5 \, \bar{Q}_{\frac{1}{2}}^{\la}
   \nonu \\
   && -
   \frac{1}{72}
\frac{1024}{5} \,
  (\lambda -2) (\lambda -1) (\lambda +1) (2 \lambda -3) (2 \lambda +1)
   \,  \bar{\pa}^4\, \bar{Q}_{\frac{3}{2}}^{\la}
   \nonu \\
   && -\frac{1}{21} \,
  \frac{256}{25}\, (\lambda -2) (\lambda +1) (2 \lambda -3) (2 \lambda +7) 
   \,  \bar{\pa}^3\, \bar{Q}_{\frac{5}{2}}^{\la}+
   \frac{5}{28}\,
\frac{32}{25} (\lambda -2) (4 \lambda ^2-22 \lambda -51)
   \, \bar{\pa}^2\, \bar{Q}_{\frac{7}{2}}^{\la} 
   \nonu \\
   && +\frac{5}{9}\,
\frac{4}{5} (2 \lambda ^2+3 \lambda -29)
   \, \bar{\pa}\, \bar{Q}_{\frac{9}{2}}^{\la} 
+
\frac{1}{5} (- \lambda +14) \, \bar{Q}_{\frac{11}{2}}^{\la}  
\Bigg](\bar{w})
 \nonu \\
 && +
 \frac{1}{(\bar{z}-\bar{w})}\, \Bigg[
   \frac{1}{720}\,
\frac{8192}{5} \, (\lambda -1) \lambda  (\lambda +1) (2 \lambda -3) (2 \lambda -1) (2 \lambda +1)\,  \bar{\pa}^6 \, \bar{Q}_{\frac{1}{2}}^{\la}
   \nonu \\
   &&-
   \frac{1}{420}
\frac{1024}{5} \,
  (\lambda -2) (\lambda -1) (\lambda +1) (2 \lambda -3) (2 \lambda +1)
   \,  \bar{\pa}^5\, \bar{Q}_{\frac{3}{2}}^{\la}
   \nonu \\
   && -\frac{1}{112} \,
 \frac{256}{25}\, (\lambda -2) (\lambda +1) (2 \lambda -3) (2 \lambda +7) 
   \,  \bar{\pa}^4\, \bar{Q}_{\frac{5}{2}}^{\la}+
   \frac{5}{126}\,
\frac{32}{25} (\lambda -2) (4 \lambda ^2-22 \lambda -51)
   \, \bar{\pa}^3\, \bar{Q}_{\frac{7}{2}}^{\la} 
   \nonu \\
   && +\frac{1}{6}\,
\frac{4}{5} (2 \lambda ^2+3 \lambda -29)
   \, \bar{\pa}^2\, \bar{Q}_{\frac{9}{2}}^{\la} 
+\frac{6}{11}\,
\frac{1}{5} (- \lambda +14) 
\, \bar{\pa}\, \bar{Q}_{\frac{11}{2}}^{\la}  
-\frac{1}{4} \, \bar{Q}_{\frac{13}{2}}^{\la} \Bigg](\bar{w})+\cdots
 \nonu  \\
 & &=
 -
  q_{B,5}^{4,\frac{7}{2}}(\bar{\pa}_{\bar{z}},\bar{\pa}_{\bar{w}},\la)
  \Bigg[ \frac{ \bar{Q}_{\frac{1}{2}}^{\la}(\bar{w})}{(\bar{z}-\bar{w})} \Bigg]
 -
  q_{B,4}^{4,\frac{7}{2}}(\bar{\pa}_{\bar{z}},\bar{\pa}_{\bar{w}},\la)
  \Bigg[ \frac{ \bar{Q}_{\frac{3}{2}}^{\la}(\bar{w})}{(\bar{z}-\bar{w})} \Bigg]
-
  q_{B,3}^{4,\frac{7}{2}}(\bar{\pa}_{\bar{z}},\bar{\pa}_{\bar{w}},\la)
  \Bigg[ \frac{ \bar{Q}_{\frac{5}{2}}^{\la}(\bar{w})}{(\bar{z}-\bar{w})} \Bigg]
  \nonu \\
  && -  q_{B,2}^{4,\frac{7}{2}}(\bar{\pa}_{\bar{z}},\bar{\pa}_{\bar{w}},\la)
  \Bigg[ \frac{ \bar{Q}_{\frac{7}{2}}^{\la}(\bar{w})}{(\bar{z}-\bar{w})} \Bigg]
 -  q_{B,1}^{4,\frac{7}{2}}(\bar{\pa}_{\bar{z}},\bar{\pa}_{\bar{w}},\la)
  \Bigg[ \frac{ \bar{Q}_{\frac{9}{2}}^{\la}(\bar{w})}{(\bar{z}-\bar{w})} \Bigg]
 -
  q_{B,0}^{4,\frac{7}{2}}(\bar{\pa}_{\bar{z}},\bar{\pa}_{\bar{w}},\la)
  \Bigg[ \frac{ \bar{Q}_{\frac{11}{2}}^{\la}(\bar{w})}{(\bar{z}-\bar{w})}
    \Bigg] \nonu \\
  &&
  -q_{B,-1}^{4,\frac{7}{2}}(\bar{\pa}_{\bar{z}},\bar{\pa}_{\bar{w}},\la)
  \Bigg[ \frac{ \bar{Q}_{\frac{13}{2}}^{\la}(\bar{w})}{(\bar{z}-\bar{w})} \Bigg] + \cdots.
  \label{fifth}
  \eea
  In Appendix (\ref{fifth}),
  there are overall minus signs as mentioned before
  because the sixth equation of (\ref{final1}) has the
  $(-1)^h$ factor.   
  Again, the presence of the weight-$\frac{1}{2}$ current
  appears at the nonzero $\la$
  \footnote{We observe that there is a factor $(2\la-1)$
    which vanishes at $\la=\frac{1}{2}$.
  See the footnote
    \ref{laonehalf}.}. 
  
\subsection{The OPE  $Q_{\frac{7}{2}}^{\la}(\bar{z}) \,
  \bar{Q}_{\frac{7}{2}}^{\la}(\bar{w})$}

Now let us look at the final equation of (\ref{final1})
and we consider the following OPE for $h_1=h_2=3$ in
our notation 
\bea
&& Q_{\frac{7}{2}}^{\la}(\bar{z}) \,
\bar{Q}_{\frac{7}{2}}^{\la}(\bar{w})=
\frac{1}{(\bar{z}-\bar{w})^7} \Bigg[
 - \frac{3072}{5}  (4 \lambda -1)
  (12 \lambda ^4-12 \lambda ^3-13 \lambda ^2+8 \lambda +3)
  \Bigg]
\nonu \\
&& +
\frac{1}{(\bar{z}-\bar{w})^6}\, \Bigg[
  \frac{4096}{5} (\lambda -1) (\lambda +1) (2 \lambda -3)
  (2 \lambda +1) (2\la-1)\,
 W_{F,1}^{\la} \nonu \\
 && +
 \frac{8192}{5} (\lambda -1) \la
 (\lambda +1) (2 \lambda -3) (2 \lambda +1)\,
 W_{B,1}^{\la}
  \Bigg](\bar{w})
\nonu \\
&& +
\frac{1}{(\bar{z}-\bar{w})^5}\, \Bigg[
 \frac{1}{2}\, \frac{4096}{5} (\lambda -1) (\lambda +1) (2 \lambda -3)
  (2 \lambda +1) (2\la-1)\,
 \bar{\pa} \, W_{F,1}^{\la} \nonu \\
 && + \frac{1}{2} \,
 \frac{8192}{5} (\lambda -1) \la
 (\lambda +1) (2 \lambda -3) (2 \lambda +1)\,
 \bar{\pa} \,  W_{B,1}^{\la} \nonu \\
 && +
\frac{1024}{25} (\lambda -1) (\lambda +1) (2 \lambda -3) (2 \lambda +17)
 \, W_{F,2}^{\la}
 \nonu \\
 && + \frac{1024}{25} (\lambda -9) (\lambda +1) (2 \lambda -3) (2 \lambda +1)
 \,  W_{B,2}^{\la}
  \Bigg](\bar{w})
\nonu \\
&& +
\frac{1}{(\bar{z}-\bar{w})^4}\, \Bigg[
 \frac{1}{6}\, \frac{4096}{5} (\lambda -1) (\lambda +1) (2 \lambda -3)
  (2 \lambda +1) (2\la-1)\,
 \bar{\pa}^2 \, W_{F,1}^{\la} \nonu \\
 && + \frac{1}{6} \,
 \frac{8192}{5} (\lambda -1) \la
 (\lambda +1) (2 \lambda -3) (2 \lambda +1)\,
 \bar{\pa}^2 \,  W_{B,1}^{\la} \nonu \\
 && + \frac{1}{2}\,
\frac{1024}{25} (\lambda -1) (\lambda +1) (2 \lambda -3) (2 \lambda +17)
 \,
 \bar{\pa} \, W_{F,2}^{\la}
 \nonu \\
 && +\frac{1}{2}\,
 \frac{1024}{25} (\lambda -9) (\lambda +1) (2 \lambda -3) (2 \lambda +1)
 \,  \bar{\pa} \, W_{B,2}^{\la}
 \nonu \\
 &&-\frac{256}{25}  (2 \lambda -3)
 (4 \lambda ^2-6 \lambda -25)\, W_{F,3}^{\la}
   - \frac{512}{25}  (\lambda +1)
 (4 \lambda ^2+2 \lambda -27)
 W_{B,3}^{\la}
 \Bigg](\bar{w})
\nonu \\
&& +
\frac{1}{(\bar{z}-\bar{w})^3}\, \Bigg[
 \frac{1}{24}\, \frac{4096}{5} (\lambda -1) (\lambda +1) (2 \lambda -3)
  (2 \lambda +1) (2\la-1)\,
 \bar{\pa}^3 \, W_{F,1}^{\la} \nonu \\
 && + \frac{1}{24} \,
 \frac{8192}{5} (\lambda -1) \la
 (\lambda +1) (2 \lambda -3) (2 \lambda +1)\,
 \bar{\pa}^3 \,  W_{B,1}^{\la} \nonu \\
 && + \frac{3}{20}\,
\frac{1024}{25} (\lambda -1) (\lambda +1) (2 \lambda -3) (2 \lambda +17)
 \,
 \bar{\pa}^2 \, W_{F,2}^{\la}
 \nonu \\
 && +\frac{3}{20}\,
 \frac{1024}{25} (\lambda -9) (\lambda +1) (2 \lambda -3) (2 \lambda +1)
 \,  \bar{\pa}^2 \, W_{B,2}^{\la}
 \nonu \\
 &&-\frac{1}{2}\,
 \frac{256}{25}  (2 \lambda -3) (4 \lambda ^2-6 \lambda -
 25)\, \bar{\pa} \, W_{F,3}^{\la}
- \frac{1}{2}\, \frac{512}{25}  (\lambda +1)
 (4 \lambda ^2+2 \lambda -27)
 \, \bar{\pa} \, W_{B,3}^{\la} \nonu \\
&&
- \frac{64}{25}  (4 \lambda ^2+18 \lambda -61)\,
 W_{F,4}^{\la} 
-\frac{64}{25}  (4 \lambda ^2-22 \lambda -51)\,
  W_{B,4}^{\la}
 \Bigg](\bar{w})
\nonu \\
&& +
\frac{1}{(\bar{z}-\bar{w})^2}\, \Bigg[
 \frac{1}{120}\, \frac{4096}{5} (\lambda -1) (\lambda +1) (2 \lambda -3)
  (2 \lambda +1) (2\la-1)\,
 \bar{\pa}^4 \, W_{F,1}^{\la} \nonu \\
 && + \frac{1}{120} \,
 \frac{8192}{5} (\lambda -1) \la
 (\lambda +1) (2 \lambda -3) (2 \lambda +1)\,
 \bar{\pa}^4 \,  W_{B,1}^{\la} \nonu \\
 && + \frac{1}{30}\,
\frac{1024}{25} (\lambda -1) (\lambda +1) (2 \lambda -3) (2 \lambda +17)
 \,
 \bar{\pa}^3 \, W_{F,2}^{\la}
 \nonu \\
 && +\frac{1}{30}\,
 \frac{1024}{25} (\lambda -9) (\lambda +1) (2 \lambda -3) (2 \lambda +1)
 \,  \bar{\pa}^3 \, W_{B,2}^{\la}
 \nonu \\
 &&-\frac{1}{7}\,
 \frac{256}{25}  (2 \lambda -3) (4 \lambda ^2-6 \lambda -
 25)\, \bar{\pa}^2 \, W_{F,3}^{\la}
- \frac{1}{7}\, \frac{512}{25}  (\lambda +1)
 (4 \lambda ^2+2 \lambda -27)
 \, \bar{\pa}^2 \, W_{B,3}^{\la} \nonu \\
&&
- \frac{1}{2} \, \frac{64}{25}  (4 \lambda ^2+18 \lambda -61)\,
 \bar{\pa} \, W_{F,4}^{\la} 
-\frac{1}{2} \, \frac{64}{25}  (4 \lambda ^2-22 \lambda -51)\,
 \bar{\pa} \,  W_{B,4}^{\la}
 \nonu \\
 && + \frac{8}{5} (2 \lambda -13) \,W_{F,5}^{\la} 
+\frac{16 (\lambda +6)}{5}\, 
 W_{B,5}^{\la}
 \Bigg](\bar{w})
\nonu \\
&& +
\frac{1}{(\bar{z}-\bar{w})}\, \Bigg[
 \frac{1}{720}\, \frac{4096}{5} (\lambda -1) (\lambda +1) (2 \lambda -3)
  (2 \lambda +1) (2\la-1)\,
 \bar{\pa}^5 \, W_{F,1}^{\la} \nonu \\
 && + \frac{1}{720} \,
 \frac{8192}{5} (\lambda -1) \la
 (\lambda +1) (2 \lambda -3) (2 \lambda +1)\,
 \bar{\pa}^5 \,  W_{B,1}^{\la} \nonu \\
 && + \frac{1}{168}\,
\frac{1024}{25} (\lambda -1) (\lambda +1) (2 \lambda -3) (2 \lambda +17)
 \,
 \bar{\pa}^4 \, W_{F,2}^{\la}
 \nonu \\
 && +\frac{1}{168}\,
 \frac{1024}{25} (\lambda -9) (\lambda +1) (2 \lambda -3) (2 \lambda +1)
 \,  \bar{\pa}^4 \, W_{B,2}^{\la}
 \nonu \\
 &&-\frac{5}{168}\,
 \frac{256}{25}  (2 \lambda -3) (4 \lambda ^2-6 \lambda -
 25)\, \bar{\pa}^3 \, W_{F,3}^{\la}
- \frac{5}{168}\, \frac{512}{25}  (\lambda +1)
 (4 \lambda ^2+2 \lambda -27)
 \, \bar{\pa}^3 \, W_{B,3}^{\la} \nonu \\
&&
- \frac{5}{36} \, \frac{64}{25}  (4 \lambda ^2+18 \lambda -61)\,
 \bar{\pa}^2 \, W_{F,4}^{\la} 
-\frac{5}{36} \, \frac{64}{25}  (4 \lambda ^2-22 \lambda -51)\,
 \bar{\pa}^2 \,  W_{B,4}^{\la}
 \nonu \\
 && + \frac{1}{2}\, \frac{8}{5} (2 \lambda -13) \,
 \bar{\pa} \, W_{F,5}^{\la} 
+\frac{1}{2} \, \frac{16 (\lambda +6)}{5}\, 
 \bar{\pa} \, W_{B,5}^{\la}
 + 2\, W_{F,6}^{\la} + 2\,  W_{B,6}^{\la}\Bigg](\bar{w})
+\cdots
\nonu \\
&& =
\frac{1}{(\bar{z}-\bar{w})^7} \Bigg[
  -\frac{3072}{5}  (4 \lambda -1)
  (12 \lambda ^4-12 \lambda ^3-13 \lambda ^2+8 \lambda +3)
  \Bigg]\nonu \\
&&
-
  o_{F,5}^{\frac{7}{2},\frac{7}{2}}(\bar{\pa}_{\bar{z}},\bar{\pa}_{\bar{w}},\la)
  \Bigg[ \frac{
      W_{F,1}^{\la}(\bar{w})}{(\bar{z}-\bar{w})} \Bigg]
  -
  o_{B,5}^{\frac{7}{2},\frac{7}{2}}(\bar{\pa}_{\bar{z}},\bar{\pa}_{\bar{w}},\la)
  \Bigg[ \frac{
      W_{B,1}^{\la}(\bar{w})}{(\bar{z}-\bar{w})} \Bigg]
  +o_{F,4}^{\frac{7}{2},\frac{7}{2}}(\bar{\pa}_{\bar{z}},\bar{\pa}_{\bar{w}},\la)
  \Bigg[ \frac{
      W_{F,2}^{\la}(\bar{w})}{(\bar{z}-\bar{w})} \Bigg]
  \nonu \\
  && +
  o_{B,4}^{\frac{7}{2},\frac{7}{2}}(\bar{\pa}_{\bar{z}},\bar{\pa}_{\bar{w}},\la)
  \Bigg[ \frac{
      W_{B,2}^{\la}(\bar{w})}{(\bar{z}-\bar{w})} \Bigg]
   -o_{F,3}^{\frac{7}{2},\frac{7}{2}}(\bar{\pa}_{\bar{z}},\bar{\pa}_{\bar{w}},\la)
  \Bigg[ \frac{
      W_{F,3}^{\la}(\bar{w})}{(\bar{z}-\bar{w})} \Bigg]
  -
  o_{B,3}^{\frac{7}{2},\frac{7}{2}}(\bar{\pa}_{\bar{z}},\bar{\pa}_{\bar{w}},\la)
  \Bigg[ \frac{
      W_{B,3}^{\la}(\bar{w})}{(\bar{z}-\bar{w})} \Bigg]
  \nonu \\
&&  +
 o_{F,2}^{\frac{7}{2},\frac{7}{2}}(\bar{\pa}_{\bar{z}},\bar{\pa}_{\bar{w}},\la)
  \Bigg[ \frac{
      W_{F,4}^{\la}(\bar{w})}{(\bar{z}-\bar{w})} \Bigg]
  +
  o_{B,2}^{\frac{7}{2},\frac{7}{2}}(\bar{\pa}_{\bar{z}},\bar{\pa}_{\bar{w}},\la)
  \Bigg[ \frac{
      W_{B,4}^{\la}(\bar{w})}{(\bar{z}-\bar{w})} \Bigg]
  -
 o_{F,1}^{\frac{7}{2},\frac{7}{2}}(\bar{\pa}_{\bar{z}},\bar{\pa}_{\bar{w}},\la)
  \Bigg[ \frac{
      W_{F,5}^{\la}(\bar{w})}{(\bar{z}-\bar{w})} \Bigg]
  \nonu \\
  && -
  o_{B,1}^{\frac{7}{2},\frac{7}{2}}(\bar{\pa}_{\bar{z}},\bar{\pa}_{\bar{w}},\la)
  \Bigg[ \frac{
      W_{B,5}^{\la}(\bar{w})}{(\bar{z}-\bar{w})} \Bigg]
   +
 o_{F,0}^{\frac{7}{2},\frac{7}{2}}(\bar{\pa}_{\bar{z}},\bar{\pa}_{\bar{w}},\la)
  \Bigg[ \frac{
      W_{F,6}^{\la}(\bar{w})}{(\bar{z}-\bar{w})} \Bigg]
  +
  o_{B,0}^{\frac{7}{2},\frac{7}{2}}(\bar{\pa}_{\bar{z}},\bar{\pa}_{\bar{w}},\la)
  \Bigg[ \frac{
      W_{B,6}^{\la}(\bar{w})}{(\bar{z}-\bar{w})} \Bigg] + \cdots.
  \nonu \\
  \label{sixth}
  \eea
  There appear four different kinds of terms in Appendix
  (\ref{sixth}).
  In this case, for even $h$, there are plus signs.
  Note that there is a factor $(4\la-1)$ which vanishes at
  $\la=\frac{1}{4}$. Moreover, the $\la$ factor appears in the
  weight-$1$ current $ W_{B,1}^{\la}(\bar{w})$
  (and its descendant terms)
  \footnote{There is a factor $(2\la-1)$
  which vanishes at $\la=\frac{1}{2}$  in the
  weight-$1$ current $ W_{F,1}^{\la}(\bar{w})$
  (and its descendant terms). See also the footnote
    \ref{laonehalf}.}.

  \section{Some OPEs containing the
    $\bar{Q}^{\la}_{\frac{1}{2}}$ or the  $ W_{B,1}^{\la}$}
  
  We consider
  the OPEs
  corresponding to the (anti)commutator relations in (\ref{final1})
  where the left hand sides contain
  $\bar{Q}^{\la}_{\frac{1}{2}}$ or the  $ W_{B,1}^{\la}$.
  For the first four cases, we have
  $\bar{Q}^{\la}_{\frac{1}{2}}$ and for the remaining ones
 we have  $ W_{B,1}^{\la}$. 
  
  \subsection{The OPE $ W_{F,4}^{\la}(\bar{z}) \, \bar{Q}^{\la}_{\frac{1}{2}}(\bar{w})$}

  Let us consider the fifth equation of (\ref{final1}).
  We calculate the following OPE by taking
  the second current as $\bar{Q}^{\la}_{\frac{1}{2}}$
  \bea
 &&  W_{F,4}^{\la}(\bar{z}) \, \bar{Q}^{\la}_{\frac{1}{2}}(\bar{w})
  = -\frac{1}{(\bar{z}-\bar{w})^4}\,
  \frac{8}{5}  (\lambda -1) (2 \lambda -3) (2 \lambda -1)\,
  \bar{Q}^{\la}_{\frac{1}{2}}(\bar{w})
  \nonu \\
  &&+
  \frac{1}{(\bar{z}-\bar{w})^3}\,
  \Bigg[
-4\, \frac{8}{5}  (\lambda -1) (2 \lambda -3) (2 \lambda -1)\,
  \bar{\pa} \, \bar{Q}^{\la}_{\frac{1}{2}}
+\frac{16}{5} (\lambda -1) (2 \lambda -3)\, \bar{Q}^{\la}_{\frac{3}{2}}
  \Bigg](\bar{w})
  \nonu \\
  && +\frac{1}{(\bar{z}-\bar{w})^2}\,
  \Bigg[
-5 \, \frac{8}{5}  (\lambda -1) (2 \lambda -3) (2 \lambda -1)\,
\bar{\pa}^2 \, \bar{Q}^{\la}_{\frac{1}{2}}
+\frac{5}{3}\,
\frac{16}{5} (\lambda -1) (2 \lambda -3)\, \bar{\pa} \,
\bar{Q}^{\la}_{\frac{3}{2}}
\nonu \\
&& +(3-2 \lambda)\, \bar{Q}^{\la}_{\frac{5}{2}}
    \Bigg](\bar{w})
  \nonu \\
  && +\frac{1}{(\bar{z}-\bar{w})}\,
  \Bigg[
-\frac{10}{3} \, \frac{8}{5}  (\lambda -1) (2 \lambda -3) (2 \lambda -1)\,
\bar{\pa}^3 \, \bar{Q}^{\la}_{\frac{1}{2}}
+\frac{5}{4}\,
\frac{16}{5} (\lambda -1) (2 \lambda -3)\, \bar{\pa}^2 \,
\bar{Q}^{\la}_{\frac{3}{2}}
\nonu \\
&& +\frac{6}{5}\, (3-2 \lambda)\, \bar{Q}^{\la}_{\frac{5}{2}}
+\frac{1}{2}\, \bar{Q}^{\la}_{\frac{7}{2}}
    \Bigg](\bar{w}) + \cdots
  \nonu \\
  && = -
  q_{F,2}^{4,\frac{1}{2}}(\bar{\pa}_{\bar{z}},\bar{\pa}_{\bar{w}},\la)
  \Bigg[ \frac{ \bar{Q}_{\frac{1}{2}}^{\la}(\bar{w})}{(\bar{z}-\bar{w})} \Bigg]
 -2
  q_{F,1}^{4,\frac{1}{2}}(\bar{\pa}_{\bar{z}},\bar{\pa}_{\bar{w}},\la)
  \Bigg[ \frac{ \bar{Q}_{\frac{3}{2}}^{\la}(\bar{w})}{(\bar{z}-\bar{w})} \Bigg]
   -2
  q_{F,0}^{4,\frac{1}{2}}(\bar{\pa}_{\bar{z}},\bar{\pa}_{\bar{w}},\la)
  \Bigg[ \frac{ \bar{Q}_{\frac{5}{2}}^{\la}(\bar{w})}{(\bar{z}-\bar{w})} \Bigg]
  \nonu \\
  && -2
  q_{F,-1}^{4,\frac{1}{2}}(\bar{\pa}_{\bar{z}},\bar{\pa}_{\bar{w}},\la)
  \Bigg[ \frac{ \bar{Q}_{\frac{7}{2}}^{\la}(\bar{w})}{(\bar{z}-\bar{w})} \Bigg]
+\cdots.
  \label{First}
  \eea
  It turns out that
  we can express the OPE in Appendix (\ref{First})
  as the one in Appendix (\ref{fourth}) except the numerical values
  $-2$ rather than $-1$. If we rescale $\bar{Q}_{\frac{1}{2}}^{\la}(\bar{w})$
  by $\frac{1}{2}$, then there appears $2$ in the first term of
  Appendix (\ref{fourth}) and the above $2$'s in Appendix (\ref{First})
  disappear.
Even at $\la=0$, this OPE arises.
  
  \subsection{The OPE
  $ W_{B,4}^{\la}(\bar{z}) \, \bar{Q}^{\la}_{\frac{1}{2}}(\bar{w})$}

   Let us consider the sixth equation of (\ref{final1}).
  We calculate the following OPE by taking
  the second current as $\bar{Q}^{\la}_{\frac{1}{2}}$
  \bea
 &&  W_{B,4}^{\la}(\bar{z}) \, \bar{Q}^{\la}_{\frac{1}{2}}(\bar{w})
  = \frac{1}{(\bar{z}-\bar{w})^4}\,
\frac{16}{5} \lambda  (\lambda +1) (2 \lambda +1)\,
  \bar{Q}^{\la}_{\frac{1}{2}}(\bar{w})
  \nonu \\
  &&+
  \frac{1}{(\bar{z}-\bar{w})^3}\,
  \Bigg[
4\,\frac{16}{5} \lambda  (\lambda +1) (2 \lambda +1) \,
  \bar{\pa} \, \bar{Q}^{\la}_{\frac{1}{2}}
  -\frac{16}{5}  (\lambda +1) (2 \lambda +1)\,
  \bar{Q}^{\la}_{\frac{3}{2}}
  \Bigg](\bar{w})
  \nonu \\
  && +\frac{1}{(\bar{z}-\bar{w})^2}\,
  \Bigg[
5 \,\frac{16}{5} \lambda  (\lambda +1) (2 \lambda +1)\,
\bar{\pa}^2 \, \bar{Q}^{\la}_{\frac{1}{2}}
-\frac{5}{3}\,\frac{16}{5} \,
 (\lambda +1) (2 \lambda +1)
\, \bar{\pa} \,
\bar{Q}^{\la}_{\frac{3}{2}}
\nonu \\
&& +2(1+ \lambda)\, \bar{Q}^{\la}_{\frac{5}{2}}
    \Bigg](\bar{w})
  \nonu \\
 && +\frac{1}{(\bar{z}-\bar{w})}\,
  \Bigg[
\frac{10}{3} \,\frac{16}{5} \lambda  (\lambda +1) (2 \lambda +1)\,
\bar{\pa}^3 \, \bar{Q}^{\la}_{\frac{1}{2}}
-5 \,\frac{16}{5} \,
 (\lambda +1) (2 \lambda +1)
\, \bar{\pa}^2 \,
\bar{Q}^{\la}_{\frac{3}{2}}
\nonu \\
&& +\frac{6}{5}\, 2(1+ \lambda)\, \bar{\pa} \,
\bar{Q}^{\la}_{\frac{5}{2}}
-\frac{1}{2}\,   \bar{Q}^{\la}_{\frac{7}{2}} \Bigg](\bar{w})
  +\cdots
  \nonu  \\
   && = -
  q_{B,2}^{4,\frac{1}{2}}(\bar{\pa}_{\bar{z}},\bar{\pa}_{\bar{w}},\la)
  \Bigg[ \frac{ \bar{Q}_{\frac{1}{2}}^{\la}(\bar{w})}{(\bar{z}-\bar{w})} \Bigg]
 -2
  q_{B,1}^{4,\frac{1}{2}}(\bar{\pa}_{\bar{z}},\bar{\pa}_{\bar{w}},\la)
  \Bigg[ \frac{ \bar{Q}_{\frac{3}{2}}^{\la}(\bar{w})}{(\bar{z}-\bar{w})} \Bigg]
   -2
  q_{B,0}^{4,\frac{1}{2}}(\bar{\pa}_{\bar{z}},\bar{\pa}_{\bar{w}},\la)
  \Bigg[ \frac{ \bar{Q}_{\frac{5}{2}}^{\la}(\bar{w})}{(\bar{z}-\bar{w})} \Bigg]
  \nonu \\
  && -2
  q_{B,-1}^{4,\frac{1}{2}}(\bar{\pa}_{\bar{z}},\bar{\pa}_{\bar{w}},\la)
  \Bigg[ \frac{ \bar{Q}_{\frac{7}{2}}^{\la}(\bar{w})}{(\bar{z}-\bar{w})} \Bigg]
+\cdots.
  \label{Second}
  \eea
 The OPE in Appendix (\ref{Second}) looks similar to
 the one in Appendix (\ref{fifth}). Again, by the rescaling of
 the  $\bar{Q}_{\frac{1}{2}}^{\la}(\bar{w})$, we can do the previous analysis.
 Due to the $\la$ factor in front of  $\bar{Q}_{\frac{1}{2}}^{\la}(\bar{w})$,
 we observe that this term (and its descendant terms) vanishes
 at $\la=0$.
 
  \subsection{The OPE
$ Q_{\frac{7}{2}}^{\la}(\bar{z}) \, \bar{Q}^{\la}_{\frac{1}{2}}(\bar{w})$
  }

    Let us consider the final equation of (\ref{final1}),
  where  
  the corresponding second current is given by
  $\bar{Q}^{\la}_{\frac{1}{2}}$. It turns out that
   \bea
 &&  Q_{\frac{7}{2}}^{\la}(\bar{z}) \, \bar{Q}^{\la}_{\frac{1}{2}}(\bar{w})
  = \frac{1}{(\bar{z}-\bar{w})^4}\,
 \Bigg[ -\frac{48}{5}  (6 \lambda ^2-3 \lambda +1)\Bigg] 
 \nonu \\
 && + \frac{1}{(\bar{z}-\bar{w})^3}\,
 \Bigg[
  \frac{64}{5} (\lambda +1) (2 \lambda +1)\, W^{\la}_{F,1} 
+\frac{64}{5} (\lambda -1) (2 \lambda -3)\, W^{\la}_{B,1}
   \Bigg](\bar{w})
 \nonu \\
 && + \frac{1}{(\bar{z}-\bar{w})^2}\,
 \Bigg[
   8\,  \frac{64}{5} (\lambda +1) (2 \lambda +1)\, \bar{\pa}\,
   W^{\la}_{F,1} 
   +8 \, \frac{64}{5} (\lambda -1) (2 \lambda -3)\, \bar{\pa}\,
   W^{\la}_{B,1}
   \nonu \\
   && + \frac{8}{5}\, (1+\la)\,  W^{\la}_{F,2}
  -\frac{32}{5}  (2 \lambda -3) \,  W^{\la}_{B,2}
   \Bigg](\bar{w})
 \nonu \\
 && + \frac{1}{(\bar{z}-\bar{w})}\,
 \Bigg[
   \frac{20}{3}\,  \frac{64}{5} (\lambda +1) (2 \lambda +1)\, \bar{\pa}^2\,
   W^{\la}_{F,1} 
   +\frac{20}{3} \, \frac{64}{5} (\lambda -1) (2 \lambda -3)\,
   \bar{\pa}^2\,
   W^{\la}_{B,1}
   \nonu \\
   && + \frac{5}{4}\, \frac{8}{5}\, (1+\la)\,  \bar{\pa}\, W^{\la}_{F,2}
   -\frac{5}{4}\, \frac{32}{5}  (2 \lambda -3) \,
   \bar{\pa} \, W^{\la}_{B,2} + 4 \, W^{\la}_{F,3} + 4 \,
    W^{\la}_{B,3}
   \Bigg](\bar{w}) + \cdots
 \nonu \\
 &&=
 2
  o_{F,2}^{\frac{7}{2},\frac{1}{2}}(\bar{\pa}_{\bar{z}},\bar{\pa}_{\bar{w}},\la)
  \Bigg[ \frac{W_{F,1}^{\la}(\bar{w})}{(\bar{z}-\bar{w})} \Bigg]
 +2
  o_{B,2}^{\frac{7}{2},\frac{1}{2}}(\bar{\pa}_{\bar{z}},\bar{\pa}_{\bar{w}},\la)
  \Bigg[ \frac{W_{B,1}^{\la}(\bar{w})}{(\bar{z}-\bar{w})} \Bigg]-
  2
  o_{F,1}^{\frac{7}{2},\frac{1}{2}}(\bar{\pa}_{\bar{z}},\bar{\pa}_{\bar{w}},\la)
  \Bigg[ \frac{W_{F,2}^{\la}(\bar{w})}{(\bar{z}-\bar{w})} \Bigg]
  \nonu \\
  && -2
  o_{B,1}^{\frac{7}{2},\frac{1}{2}}(\bar{\pa}_{\bar{z}},\bar{\pa}_{\bar{w}},\la)
  \Bigg[ \frac{W_{B,2}^{\la}(\bar{w})}{(\bar{z}-\bar{w})} \Bigg]
  +  2
  o_{F,0}^{\frac{7}{2},\frac{1}{2}}(\bar{\pa}_{\bar{z}},\bar{\pa}_{\bar{w}},\la)
  \Bigg[ \frac{W_{F,3}^{\la}(\bar{w})}{(\bar{z}-\bar{w})} \Bigg]
 +2
  o_{B,0}^{\frac{7}{2},\frac{1}{2}}(\bar{\pa}_{\bar{z}},\bar{\pa}_{\bar{w}},\la)
  \Bigg[ \frac{W_{B,3}^{\la}(\bar{w})}{(\bar{z}-\bar{w})} \Bigg]
  \nonu \\
  && +\cdots.
\label{Third}
  \eea
  By multiplying $\frac{1}{2}$ both sides, then
  we can absorb the numerical value $2$ in the right hand side of
  Appendix (\ref{Third}). Then the behavior of this OPE
  looks similar to the one in Appendix (\ref{sixth}).

 \subsection{The OPE $ W_{B,1}^{\la}(\bar{z}) \, \bar{Q}^{\la}_{\frac{1}{2}}(\bar{w})$}

 When we take
 the first current as
 $ W_{B,1}$ further corresponding to
 the sixth equation of (\ref{final1})
then we obtain 
 \bea
W_{B,1}^{\la}(\bar{z}) \, \bar{Q}^{\la}_{\frac{1}{2}}(\bar{w})
 & = & \frac{1}{(\bar{z}-\bar{w})}\,
 \Bigg[-\frac{1}{4} \, \bar{Q}^{\la}_{\frac{1}{2}} \Bigg](\bar{w}) +
 \cdots \nonu \\
 & = &
 -q_{B,-1}^{1,\frac{1}{2}}(\bar{\pa}_{\bar{z}},\bar{\pa}_{\bar{w}},\la)
  \Bigg[ \frac{\bar{Q}_{\frac{1}{2}}^{\la}(\bar{w})}{(\bar{z}-\bar{w})} \Bigg]
   +\cdots.
 \label{Fourth}
 \eea
 Now we see that the similar behavior in Appendix (\ref{Fourth})
 arises, compared to the ones in
 Appendix (\ref{fifth}) or Appendix (\ref{Second}).
 
  From now on,  
  we consider
  that the first current is given by
  $ W_{B,1}^{\la}$.

 \subsection{The OPE $  W_{B,1}^{\la}(\bar{z}) \, W^{\la}_{B,1}(\bar{w})$}

 We obtain the following OPE
 corresponding to the second equation of (\ref{final1})
 \bea
&&  W_{B,1}^{\la}(\bar{z}) \, W^{\la}_{B,1}(\bar{w})
 = \frac{1}{(\bar{z}-\bar{w})^2}\,\Bigg[ -\frac{3}{16}\Bigg]+ \cdots.
 \label{rem1}
 \eea

 \subsection{The OPE $ W_{B,1}^{\la}(\bar{z}) \, W^{\la}_{B,4}(\bar{w})$}

 When the second current is given by
 $ W^{\la}_{B,4}(\bar{w})$, we obtain the following OPE
 \bea
&&  W_{B,1}^{\la}(\bar{z}) \, W^{\la}_{B,4}(\bar{w})
 = \frac{1}{(\bar{z}-\bar{w})^5}\,
 \Bigg[-\frac{24}{5}
   (2 \lambda -1) (2 \lambda ^2-2 \lambda +1) \Bigg]
 \nonu \\
 && +
\frac{1}{(\bar{z}-\bar{w})^4}\,
\Bigg[\frac{96}{5} (2 \lambda ^2-2 \lambda +1)
 W_{B,1}^{\la}
 \Bigg](\bar{w})
\nonu \\
&& +
\frac{1}{(\bar{z}-\bar{w})^3}\,
\Bigg[-\frac{96}{5} (2 \lambda ^2-2 \lambda +1)
 \, \bar{\pa} \, W_{B,1}^{\la} + 8(2\la-1) \, W_{B,2}^{\la}
 \Bigg](\bar{w})
\nonu \\
&& +
\frac{1}{(\bar{z}-\bar{w})^2}\,
\Bigg[\frac{1}{8}\,
  \frac{96}{5} (2 \lambda ^2-2 \lambda +1)
  \, \bar{\pa}^2 \, W_{B,1}^{\la} -\frac{1}{4}\,
  8(2\la-1) \, \bar{\pa}\,
  W_{B,2}^{\la}
+ 3 \,  W_{B,3}^{\la}
  \Bigg](\bar{w})
\nonu \\
&& +\cdots.
 \label{rem2}
 \eea

 \subsection{ The OPE $ W_{B,1}^{\la}(\bar{z}) \, Q^{\la}_{\frac{7}{2}}(\bar{w})$}

 For the fourth equation of (\ref{final1}), we can calculate
 the following OPE
 \bea
&&  W_{B,1}^{\la}(\bar{z}) \, Q^{\la}_{\frac{7}{2}}(\bar{w})
 = \frac{1}{(\bar{z}-\bar{w})^3}\,
 \Bigg[-\frac{4}{5}  (\lambda +1) (2 \lambda +1)\,
 Q^{\la}_{\frac{3}{2}}
 \Bigg](\bar{w})\nonu \\
 && +
  \frac{1}{(\bar{z}-\bar{w})^2}\,
 \Bigg[\frac{1}{3}\, \frac{4}{5}  (\lambda +1) (2 \lambda +1)\,
 \bar{\pa}\, Q^{\la}_{\frac{3}{2}}
 -\frac{4}{5} (\la +1)\,  Q^{\la}_{\frac{5}{2}}
 \Bigg](\bar{w})+
 \frac{1}{(\bar{z}-\bar{w})}\,
 \Bigg[
 \frac{1}{4} \,  Q^{\la}_{\frac{7}{2}}
 \Bigg](\bar{w}) \nonu \\
 && +
 \cdots.
 \label{rem3}
 \eea
 
 \subsection{The OPE $ W_{B,1}^{\la}(\bar{z}) \, \bar{Q}^{\la}_{\frac{7}{2}}(\bar{w})$}

 For the sixth equation of (\ref{final1}), we can calculate
 the following OPE
 \bea
&&  W_{B,1}^{\la}(\bar{z}) \, \bar{Q}^{\la}_{\frac{7}{2}}(\bar{w})
 = \frac{1}{(\bar{z}-\bar{w})^4}\,
 \Bigg[ -\frac{8}{5}  (\lambda -1)
   (2 \lambda -3) (2 \lambda -1)\, \bar{Q}^{\la}_{\frac{1}{2}}\Bigg](\bar{w})
 \nonu \\
&&+  \frac{1}{(\bar{z}-\bar{w})^3}\,
 \Bigg[  2\, \frac{8}{5}  (\lambda -1)
   (2 \lambda -3) (2 \lambda -1)\, \bar{\pa}\,
   \bar{Q}^{\la}_{\frac{1}{2}}
-\frac{12}{5}  (\lambda -1) (2 \lambda -3)\, \bar{Q}^{\la}_{\frac{3}{2}}
   \Bigg](\bar{w})
 \nonu \\
 &&+  \frac{1}{(\bar{z}-\bar{w})^2}\,
 \Bigg[  \frac{1}{2}\, \frac{8}{5}  (\lambda -1)
   (2 \lambda -3) (2 \lambda -1)\, \bar{\pa}^2 \,
   \bar{Q}^{\la}_{\frac{1}{2}}
   +\frac{1}{3}\,
   \frac{12}{5}  (\lambda -1) (2 \lambda -3)\,
   \bar{\pa}\, \bar{Q}^{\la}_{\frac{3}{2}}\nonu \\
   && -
\frac{3}{5}  (2 \lambda -3)\, \bar{Q}^{\la}_{\frac{5}{2}}
   \Bigg](\bar{w})
 -  \frac{1}{(\bar{z}-\bar{w})}\, \frac{1}{4}\,
 \bar{Q}^{\la}_{\frac{7}{2}}(\bar{w}) +\cdots.
 \label{rem4}
 \eea 

 In Appendices (\ref{rem1}), (\ref{rem2}), (\ref{rem3}),
 and (\ref{rem4}), 
we cannot express the structure constants in terms of (\ref{structla}).
 
  \section{The remaining (anti)commutator relations
  of ${\cal N}=2$ supersymmetric linear $W_{\infty}^{N,N}[\la]$ algebra}
  
  The remaining $19$ (anti)commutator relations are
  determined by 
  \bea
\big[(W^{\la}_{\mathrm{F},h_1})_m,(W^{\la,\hat{A}}_{\mathrm{F},h_2})_n\big] 
\!&=& \!
\sum^{h_1+h_2-3}_{h= 0, \mbox{\footnotesize even}} \, q^h\,
p_{\mathrm{F}}^{h_1,h_2, h}(m,n,\la)
\, (   W^{\la,\hat{A}}_{\mathrm{F},h_1+h_2-2-h} )_{m+n}\ ,
\nonu \\
\big[(W^{\la,\hat{A}}_{\mathrm{F},h_1})_m,(W^{\la,\hat{B}}_{\mathrm{F},h_2})_n\big] 
\!&=& \!
-\sum^{h_1+h_2-3}_{h= -1,
\mbox{\footnotesize odd}} \, q^h\,
p_{\mathrm{F}}^{h_1,h_2, h}(m,n,\la)
\, \frac{i}{2}\, f^{\hat{A} \hat{B}  \hat{C}} \, (   W^{\la,\hat{C}}_{\mathrm{F},h_1+h_2-2-h} )_{m+n}
\nonu \\
\!& + \!&
c_{W_{\mathrm{F},h_1}} (m,\la) \,
\delta^{\hat{A} \hat{B}}\, \delta^{h_1 h_2}\,q^{2(h_1-2)}\,\delta_{m+n}
\nonu \\
\!&+\!& \sum^{h_1+h_2-3}_{h= 0, \mbox{\footnotesize even}} \, q^h\,
p_{\mathrm{F}}^{h_1,h_2, h}(m,n,\la)
\, \Big( \frac{1}{2}\, d^{\hat{A} \hat{B} \hat{C}} \,
(   W^{\la,\hat{C}}_{\mathrm{F},h_1+h_2-2-h} )_{m+n} \nonu \\
\!& + \!& \frac{1}{N}\, \delta^{\hat{A} \hat{B}}\,
(   W^{\la}_{\mathrm{F},h_1+h_2-2-h} )_{m+n} \Big)\ ,
\nonu \\
\big[(W^{\la}_{\mathrm{B},h_1})_m,(W^{\la,\hat{A}}_{\mathrm{B},h_2})_n\big] 
\!&=& \!
\sum^{h_1+h_2-4}_{h= 0, \mbox{\footnotesize even}} \, q^h\,
p_{\mathrm{B}}^{h_1,h_2, h}(m,n,\la)
\, (   W^{\la,\hat{A}}_{\mathrm{B},h_1+h_2-2-h} )_{m+n} \nonu \\
\!&+& \! \Bigg[  q^h\,
p_{\mathrm{B}}^{h_1,h_2, h}(m,n,\la)
\, (   W^{\la,\hat{A}}_{\mathrm{B},h_1+h_2-2-h} )_{m+n}\Bigg]_{h=h_1+h_2-3} \ ,
\nonu \\
\big[(W^{\la,\hat{A}}_{\mathrm{B},h_1})_m,(W^{\la,\hat{B}}_{\mathrm{B},h_2})_n\big] 
\!&=& \!
-\sum^{h_1+h_2-4}_{h= -1, \mbox{\footnotesize odd}} \,
q^h\, p_{\mathrm{B}}^{h_1,h_2, h}(m,n,\la)
\, \frac{i}{2}\, f^{\hat{A} \hat{B}  \hat{C}} \,
(   W^{\la,\hat{C}}_{\mathrm{B},h_1+h_2-2-h} )_{m+n}
\nonu \\
\!& + \!&
\Bigg[-
q^h\, p_{\mathrm{B}}^{h_1,h_2, h}(m,n,\la)
\, \frac{i}{2}\, f^{\hat{A} \hat{B}  \hat{C}} \,
(   W^{\la,\hat{C}}_{\mathrm{B},h_1+h_2-2-h} )_{m+n} \Bigg]_{h_1+h_2-3}
\nonu \\
\!& + \!&
c_{W_{\mathrm{B},h_1}}(m,\la) \,
\delta^{\hat{A} \hat{B}}\,
\delta^{h_1 h_2}\,q^{2(h_1-2)}\,\delta_{m+n} 
\nonu \\
\!&+\!& \sum^{h_1+h_2-4}_{h= 0, \mbox{\footnotesize even}} \, q^h\,
p_{\mathrm{B}}^{h_1,h_2, h}(m,n,\la)
\, \Bigg( \frac{1}{2}\, d^{\hat{A} \hat{B} \hat{C}} \,
(   W^{\la,\hat{C}}_{\mathrm{B},h_1+h_2-2-h} )_{m+n} \nonu \\
\!& + \!& \frac{1}{N}\, \delta^{\hat{A} \hat{B}}\,
(   W^{\la}_{\mathrm{B},h_1+h_2-2-h} )_{m+n} \Bigg)\ 
\nonu \\
\!&+\!&  \Bigg[ q^h\,
p_{\mathrm{B}}^{h_1,h_2, h}(m,n,\la)
\, \Bigg( \frac{1}{2}\, d^{\hat{A} \hat{B} \hat{C}} \,
(   W^{\la,\hat{C}}_{\mathrm{B},h_1+h_2-2-h} )_{m+n} \nonu \\
\!& + \!& \frac{1}{N}\, \delta^{\hat{A} \hat{B}}\,
(   W^{\la}_{\mathrm{B},h_1+h_2-2-h} )_{m+n}\Bigg) \Bigg]_{h=h_1+h_2-3} \ ,
\nonu \\
\big[(W^{\la}_{\mathrm{F},h_1})_m,(Q^{\la,\hat{A}}_{h_2+\frac{1}{2}})_r\big] 
\!&= &\!
\sum^{h_1+h_2-3}_{h=-1}\, q^h \, q_{\mathrm{F}}^{h_1,h_2+\frac{1}{2}, h}(m,r,\la) 
\, (Q^{\la,\hat{A}}_{h_1+h_2-\frac{3}{2}-h})_{m+r}\ ,
\nonu \\
\big[(W^{\la,\hat{A}}_{\mathrm{F},h_1})_m,(Q^{\la}_{h_2+\frac{1}{2}})_r\big] 
\!&= &\!
\sum^{h_1+h_2-3}_{h=-1}\, q^h \, q_{\mathrm{F}}^{h_1,h_2+\frac{1}{2}, h}(m,r,\la) 
\, (Q^{\la,\hat{A}}_{h_1+h_2-\frac{3}{2}-h})_{m+r}\ ,
\nonu \\
\big[(W^{\la,\hat{A}}_{\mathrm{F},h_1})_m,(Q^{\la,\hat{B}}_{h_2+\frac{1}{2}})_r\big] 
\!&=& \!
\sum^{h_1+h_2-3}_{h= -1} \, q^h\, q_{\mathrm{F}}^{h_1,h_2+\frac{1}{2}, h}(m,r,\la)
\, \frac{i}{2}\, f^{\hat{A} \hat{B}  \hat{C}} \, (
Q^{\la,\hat{C}}_{h_1+h_2-\frac{3}{2}-h} )_{m+r}
\nonu \\
\!&+\!& \sum^{h_1+h_2-3}_{h= -1} \, q^h\,
q_{\mathrm{F}}^{h_1,h_2+\frac{1}{2}, h}(m,r,\la)
\, \Bigg( \frac{1}{2}\, d^{\hat{A} \hat{B} \hat{C}} \,
(   Q^{\la,\hat{C}}_{h_1+h_2-\frac{3}{2}-h} )_{m+r} \nonu \\
\!& + \!& \frac{1}{N}\, \delta^{\hat{A} \hat{B}}\,
(   Q^{\la}_{h_1+h_2-\frac{3}{2}-h} )_{m+r} \Bigg)\ ,
\nonu \\
\big[(W^{\la}_{\mathrm{B},h_1})_m,(Q^{\la,\hat{A}}_{h_2+\frac{1}{2}})_r\big] 
\!&= &\!
\sum^{h_1+h_2-3}_{h=-1}\, q^h \,
q_{\mathrm{B}}^{h_1,h_2+\frac{1}{2}, h}(m,r,\la) 
\, (Q^{\la,\hat{A}}_{h_1+h_2-\frac{3}{2}-h})_{m+r}\ ,
\nonu\\
\big[(W^{\la,\hat{A}}_{\mathrm{B},h_1})_m,(Q^{\la}_{h_2+\frac{1}{2}})_r\big] 
\!&= &\!
\sum^{h_1+h_2-3}_{h=-1}\, q^h \,
q_{\mathrm{B}}^{h_1,h_2+\frac{1}{2}, h}(m,r,\la) 
\, (Q^{\la,\hat{A}}_{h_1+h_2-\frac{3}{2}-h})_{m+r}\ ,
\nonu \\
\big[(W^{\la,\hat{A}}_{\mathrm{B},h_1})_m,(Q^{\la,\hat{B}}_{h_2+\frac{1}{2}})_r\big] 
\!&=& \!
-\sum^{h_1+h_2-3}_{h= -1} \, q^h\,
q_{\mathrm{B}}^{h_1,h_2+\frac{1}{2}, h}(m,r,\la)
\, \frac{i}{2}\, f^{\hat{A} \hat{B}  \hat{C}} \,
(   Q^{\la,\hat{C}}_{h_1+h_2-\frac{3}{2}-h} )_{m+r}
\nonu \\
\!&+\!& \sum^{h_1+h_2-3}_{h= -1} \,
q^h\, q_{\mathrm{B}}^{h_1,h_2+\frac{1}{2}, h}(m,r,\la)
\, \Bigg( \frac{1}{2}\, d^{\hat{A} \hat{B} \hat{C}} \,
(   Q^{\la,\hat{C}}_{h_1+h_2-\frac{3}{2}-h} )_{m+r} \nonu \\
\!& + \!& \frac{1}{N}\, \delta^{\hat{A} \hat{B}}\,
(   Q^{\la}_{h_1+h_2-\frac{3}{2}-h} )_{m+r} \Bigg)\ ,
\nonu \\
\big[(W^{\la}_{\mathrm{F},h_1})_m,(\bar{Q}^{\la,\hat{A}}_{h_2+\frac{1}{2}})_r\big] 
\!&= &\!
\sum^{h_1+h_2-3}_{h=-1}\, q^h \,  (-1)^h\,
q_{\mathrm{F}}^{h_1,h_2+\frac{1}{2}, h}(m,r,\la) 
\, (\bar{Q}^{\la,\hat{A}}_{h_1+h_2-\frac{3}{2}-h})_{m+r}\ 
\nonu\\
\!&+ &\! \Bigg[ q^h \,  (-1)^h\, q_{\mathrm{F}}^{h_1,h_2+\frac{1}{2}, h}(m,r,\la) 
\, (\bar{Q}^{\la,\hat{A}}_{h_1+h_2-\frac{3}{2}-h})_{m+r}\Bigg]_{h=h_1=h_2-2} \ ,
\nonu \\
\big[(W^{\la,\hat{A}}_{\mathrm{F},h_1})_m,(\bar{Q}^{\la}_{h_2+\frac{1}{2}})_r\big] 
\!&= &\!
\sum^{h_1+h_2-3}_{h=-1}\, q^h \,  (-1)^h\,
q_{\mathrm{F}}^{h_1,h_2+\frac{1}{2}, h}(m,r,\la) 
\, (\bar{Q}^{\la,\hat{A}}_{h_1+h_2-\frac{3}{2}-h})_{m+r}\ 
\nonu \\
\!&+ &\! \Bigg[  q^h \,  (-1)^h\, q_{\mathrm{F}}^{h_1,h_2+\frac{1}{2}, h}(m,r,\la) 
\, (\bar{Q}^{\la,\hat{A}}_{h_1+h_2-\frac{3}{2}-h})_{m+r}\Bigg]_{h=h_2+h_2-2} \ ,
\nonu \\
\big[(W^{\la,\hat{A}}_{\mathrm{F},h_1})_m,
  (\bar{Q}^{\la,\hat{B}}_{h_2+\frac{1}{2}})_r\big] 
\!&=& \!
-\sum^{h_1+h_2-3}_{h=-1} \, q^h\,  (-1)^h\,
q_{\mathrm{F}}^{h_1,h_2+\frac{1}{2}, h}(m,r,\la)
\, \frac{i}{2}\, f^{\hat{A} \hat{B}  \hat{C}} \,
( \bar{Q}^{\la,\hat{C}}_{h_1+h_2-\frac{3}{2}-h} )_{m+r}
\nonu \\
\!&+& \! \Bigg[ - q^h\,  (-1)^h\,
  q_{\mathrm{F}}^{h_1,h_2+\frac{1}{2}, h}(m,r,\la)
\, \frac{i}{2}\, f^{\hat{A} \hat{B}  \hat{C}} \,
( \bar{Q}^{\la,\hat{C}}_{h_1+h_2-\frac{3}{2}-h} )_{m+r} \Bigg]_{h=h_1+h_2-2} 
\nonu \\
\!&+\!& \sum^{h_1+h_2-3}_{h= -1} \, q^h\,  (-1)^h\,
q_{\mathrm{F}}^{h_1,h_2+\frac{1}{2}, h}(m,r,\la)
\, \Bigg( \frac{1}{2}\, d^{\hat{A} \hat{B} \hat{C}} \,
(   \bar{Q}^{\la,\hat{C}}_{h_1+h_2-\frac{3}{2}-h} )_{m+r} \nonu \\
\!& + \!& \frac{1}{N}\, \delta^{\hat{A} \hat{B}}\,
(   \bar{Q}^{\la}_{h_1+h_2-\frac{3}{2}-h} )_{m+r} \Bigg) \
\nonu \\
\!&+& \!
\Bigg[  q^h\,  (-1)^h\,
q_{\mathrm{F}}^{h_1,h_2+\frac{1}{2}, h}(m,r,\la)
\, \Bigg( \frac{1}{2}\, d^{\hat{A} \hat{B} \hat{C}} \,
(   \bar{Q}^{\la,\hat{C}}_{h_1+h_2-\frac{3}{2}-h} )_{m+r} \nonu \\
\!& + \!& \frac{1}{N}\, \delta^{\hat{A} \hat{B}}\,
(   \bar{Q}^{\la}_{h_1+h_2-\frac{3}{2}-h} )_{m+r} \Bigg)\Bigg]_{h=h_1+h_2-2} \ ,
\nonu \\
\big[(W^{\la}_{\mathrm{B},h_1})_m,(\bar{Q}^{\la,\hat{A}}_{h_2+\frac{1}{2}})_r\big] 
\!&= &\!
\sum^{h_1+h_2-3}_{h=-1}\, q^h \,  (-1)^h \,
q_{\mathrm{B}}^{h_1,h_2+\frac{1}{2}, h}(m,r,\la) 
\, (\bar{Q}^{\la,\hat{A}}_{h_1+h_2-\frac{3}{2}-h})_{m+r}\
\nonu \\
\!&+ &\! \Bigg[  q^h \,  (-1)^h \,
  q_{\mathrm{B}}^{h_1,h_2+\frac{1}{2}, h}(m,r,\la) 
\, (\bar{Q}^{\la,\hat{A}}_{h_1+h_2-\frac{3}{2}-h})_{m+r}\Bigg]_{h=h_1+h_2-2},
\nonu\\
\big[(W^{\la,\hat{A}}_{\mathrm{B},h_1})_m,(\bar{Q}^{\la}_{h_2+\frac{1}{2}})_r\big] 
\!&= &\!
\sum^{h_1+h_2-3}_{h=-1}\, q^h \,  (-1)^h \,
q_{\mathrm{B}}^{h_1,h_2+\frac{1}{2}, h}(m,r,\la) 
\, (\bar{Q}^{\la,\hat{A}}_{h_1+h_2-\frac{3}{2}-h})_{m+r}\ 
\nonu \\
\!&+ &\! \Bigg[ q^h \,  (-1)^h \,
  q_{\mathrm{B}}^{h_1,h_2+\frac{1}{2}, h}(m,r,\la) 
\, (\bar{Q}^{\la,\hat{A}}_{h_1+h_2-\frac{3}{2}-h})_{m+r} \Bigg]_{h=h_1+h_2-2} \ , 
\nonu \\
\big[(W^{\la,\hat{A}}_{\mathrm{B},h_1})_m,
  (\bar{Q}^{\la,\hat{B}}_{h_2+\frac{1}{2}})_r\big] 
\!&=& \!
\sum^{h_1+h_2-3}_{h= -1} \, q^h\,  (-1)^h \,
q_{\mathrm{B}}^{h_1,h_2+\frac{1}{2}, h}(m,r,\la)
\, \frac{i}{2}\, f^{\hat{A} \hat{B}  \hat{C}} \,
(   \bar{Q}^{\la,\hat{C}}_{h_1+h_2-\frac{3}{2}-h} )_{m+r}
\nonu \\
\!&+& \! \Bigg[ q^h\,  (-1)^h \,
  q_{\mathrm{B}}^{h_1,h_2+\frac{1}{2}, h}(m,r,\la)
\, \frac{i}{2}\, f^{\hat{A} \hat{B}  \hat{C}} \,
(   \bar{Q}^{\la,\hat{C}}_{h_1+h_2-\frac{3}{2}-h} )_{m+r} \Bigg]_{h=h_1+h_2-2} \ 
\nonu \\
\!&+\!& \sum^{h_1+h_2-3}_{h= -1} \, q^h\,  (-1)^h \,
q_{\mathrm{B}}^{h_1,h_2+\frac{1}{2}, h}(m,r,\la)
\, \Bigg( \frac{1}{2}\, d^{\hat{A} \hat{B} \hat{C}} \,
(   \bar{Q}^{\la,\hat{C}}_{h_1+h_2-\frac{3}{2}-h} )_{m+r} \nonu \\
\!& +\! & \frac{1}{N}\, \delta^{\hat{A} \hat{B}}\,
(   \bar{Q}^{\la}_{h_1+h_2-\frac{3}{2}-h} )_{m+r} \Bigg) \ 
\nonu \\
\!&+\!& \Bigg[ q^h\,  (-1)^h \,q_{\mathrm{B}}^{h_1,h_2+\frac{1}{2}, h}(m,r,\la)
\, \Bigg( \frac{1}{2}\, d^{\hat{A} \hat{B} \hat{C}} \,
(   \bar{Q}^{\la,\hat{C}}_{h_1+h_2-\frac{3}{2}-h} )_{m+r} \nonu \\
\!& +\! & \frac{1}{N}\, \delta^{\hat{A} \hat{B}}\,
(   \bar{Q}^{\la}_{h_1+h_2-\frac{3}{2}-h} )_{m+r} \Bigg) \Bigg]_{h=h_1=h_2-2} \ ,
\nonu \\
\{(Q^{\la}_{h_1+\frac{1}{2}})_r,(\bar{Q}^{\la,\hat{A}}_{h_2+\frac{1}{2}})_s\} 
\!&=&\!
\sum^{h_1+h_2-1}_{h=0}\,
q^h  \,
\, o_{\mathrm{F}}^{h_1+\frac{1}{2},h_2+\frac{1}{2},h}(r,s,\la) \,
(W^{\la,\hat{A}}_{\mathrm{F},h_1+h_2-h})_{r+s}
\nonu \\
\!& + \!&
\sum^{h_1+h_2-2}_{h=0}\,
q^h  \,
o_{\mathrm{B}}^{h_1+\frac{1}{2},h_2+\frac{1}{2},h}(r,s,\la) \,
(W^{\la,\hat{A}}_{\mathrm{B},h_1+h_2-h})_{r+s}  \ 
\nonu \\
\!& + \!& \Bigg[q^h  \,
o_{\mathrm{B}}^{h_1+\frac{1}{2},h_2+\frac{1}{2},h}(r,s,\la) \,
(W^{\la,\hat{A}}_{\mathrm{B},h_1+h_2-h})_{r+s}  \Bigg]_{h=h_1+h_2-1} \ ,
\nonu \\
\{(Q^{\la,\hat{A}}_{h_1+\frac{1}{2}})_r,(\bar{Q}^{\la}_{h_2+\frac{1}{2}})_s\} 
\!&=&\!
\sum^{h_1+h_2-1}_{h=0} \,
q^h 
\, o_{\mathrm{F}}^{h_1+\frac{1}{2},h_2+\frac{1}{2},h}(r,s,\la) \,
(W^{\la,\hat{A}}_{\mathrm{F},h_1+h_2-h})_{r+s}
\nonu \\
\!& + \!& 
\,\sum^{h_1+h_2-2}_{h=0} \,
q^h 
o_{\mathrm{B}}^{h_1+\frac{1}{2},h_2+\frac{1}{2},h}(r,s,\la) \,
(W^{\la,\hat{A}}_{\mathrm{B},h_1+h_2-h})_{r+s}  \ 
\nonu \\
\!& + \!& \Bigg[ q^h 
o_{\mathrm{B}}^{h_1+\frac{1}{2},h_2+\frac{1}{2},h}(r,s,\la) \,
(W^{\la,\hat{A}}_{\mathrm{B},h_1+h_2-h})_{r+s} \Bigg]_{h=h_1+h_2-1} \ ,
\nonu \\
\{(Q^{\la,\hat{A}}_{h_1+\frac{1}{2}})_r,(\bar{Q}^{\la,\hat{B}}_{h_2+\frac{1}{2}})_s\} 
\!&=& \!
\sum^{h_1+h_2-1}_{h= 0} \, q^h \,o_{\mathrm{F}}^{h_1+\frac{1}{2},h_2+
\frac{1}{2}, h}(r,s,\la)
\, \frac{i}{2}\, f^{\hat{A} \hat{B}  \hat{C}} \,
(   W^{\la,\hat{C}}_{F,h_1+h_2-h} )_{r+s}
\nonu \\
\!&+\!& \sum^{h_1+h_2-1}_{h= 0} \, q^h \,o_{\mathrm{F}}^{h_1+\frac{1}{2},
h_2+\frac{1}{2}, h}(r,s,\la)
\, \Bigg( \frac{1}{2}\, d^{\hat{A} \hat{B} \hat{C}} \,
(   W^{\la,\hat{C}}_{F,h_1+h_2-h} )_{r+s} \nonu \\
\!& + \!& \frac{1}{N}\, \delta^{\hat{A} \hat{B}}\,
(   W^{\la}_{F,h_1+h_2-h} )_{r+s} \Bigg)
\nonu \\
\!&-\!& \sum^{h_1+h_2-2}_{h= 0} \, q^h \,
o_{\mathrm{B}}^{h_1+\frac{1}{2},h_2+\frac{1}{2}, h}(r,s,\la)
\, \frac{i}{2}\, f^{\hat{A} \hat{B}  \hat{C}} \,
(   W^{\la,\hat{C}}_{B,h_1+h_2-h} )_{r+s}
\nonu \\
\!& + \!& \Bigg[- q^h \,
o_{\mathrm{B}}^{h_1+\frac{1}{2},h_2+\frac{1}{2}, h}(r,s,\la)
\, \frac{i}{2}\, f^{\hat{A} \hat{B}  \hat{C}} \,
(   W^{\la,\hat{C}}_{B,h_1+h_2-h} )_{r+s} \Bigg]_{h=h_1+h_2-1} \
\nonu \\
\!&+\!& \sum^{h_1+h_2-2}_{h= 0} \, q^h \,
o_{\mathrm{B}}^{h_1+\frac{1}{2},h_2+\frac{1}{2}, h}(r,s,\la)
\, \Bigg( \frac{1}{2}\, d^{\hat{A} \hat{B} \hat{C}} \,
(   W^{\la,\hat{C}}_{B,h_1+h_2-h} )_{r+s} \nonu \\
\!& +\! & \frac{1}{N}\, \delta^{\hat{A} \hat{B}}\,
(   W^{\la}_{B,h_1+h_2-h} )_{r+s} \Bigg)
\nonu \\
\!&+\!& \Bigg[ q^h \,
o_{\mathrm{B}}^{h_1+\frac{1}{2},h_2+\frac{1}{2}, h}(r,s,\la)
\, \Bigg( \frac{1}{2}\, d^{\hat{A} \hat{B} \hat{C}} \,
(   W^{\la,\hat{C}}_{B,h_1+h_2-h} )_{r+s} \nonu \\
\!& +\! & \frac{1}{N}\, \delta^{\hat{A} \hat{B}}\,
(   W^{\la}_{B,h_1+h_2-h} )_{r+s} \Bigg) \Bigg]_{h=h_1+h_2-1}
\nonu \\
\!&+\!&  c_{Q\bar{Q}_{h_1+\frac{1}{2}}}(r,\la)  \, \delta^{\hat{A} \hat{B}}
\, \delta^{h_1 h_2} \, q^{2(h_1+\frac{1}{2}-1)}
\delta_{r+s} \ .
\label{Final}
  \eea
  As in the section $3$,
  we intentionally make the square brackets in the Appendix
  (\ref{Final}) in order to
  emphasize that the
  current $\bar{Q}^{\la}_{\frac{1}{2}}$ or the
  current $ W_{B,1}^{\la}$ occurs inside of the square brackets
    when we restrict to the operators in the left hand sides
    which do not have these weight-$\frac{1}{2}, 1$ currents. 

    \section{Some OPEs for $\la =\frac{1}{4}$}
    
    We present some OPEs for the particular value of $\la=\frac{1}{4}$
    for fixed $h_1$ and $h_2$ as follows.
    We keep the $\la$ dependence without substituting this value
    in order to see the factor $(1-4\la)$.
    It turns out that
    \bea
&&    V^{(4)+}_{\la}(\bar{z}) \,V^{(4)+}_{\la}(\bar{w})
     = \frac{1}{(\bar{z}-\bar{w})^8} \, \Bigg[-
      \frac{21}{5} (4 \lambda -1)
      (12 \lambda ^4-12 \lambda ^3-13 \lambda ^2+8 \lambda +3)\Bigg]
     \nonu \\
     && +
     \frac{1}{(\bar{z}-\bar{w})^6} \, \Bigg[
       \frac{56}{15} (\lambda -1) (\lambda +1) (2 \lambda -3)
       (2 \lambda +1) \, V^{(2)+}_{\la}
       \nonu \\
       && + \frac{8}{5} (\lambda +1) (2 \lambda -3) (4 \lambda -1)
     \, V^{(2)-}_{\la}  \Bigg](\bar{w})
     \nonu \\
      && +
     \frac{1}{(\bar{z}-\bar{w})^5} \, \Bigg[
     \frac{1}{2}\,  \frac{56}{15} (\lambda -1) (\lambda +1) (2 \lambda -3)
       (2 \lambda +1) \, \bar{\pa} \, V^{(2)+}_{\la}
       \nonu \\
       && +\frac{1}{2} \,
       \frac{8}{5} (\lambda +1) (2 \lambda -3) (4 \lambda -1)
     \, \bar{\pa} \, V^{(2)-}_{\la}  \Bigg](\bar{w})
     \nonu \\
      && +
     \frac{1}{(\bar{z}-\bar{w})^4} \, \Bigg[
     \frac{3}{20}\,  \frac{56}{15} (\lambda -1) (\lambda +1) (2 \lambda -3)
       (2 \lambda +1) \, \bar{\pa}^2 \, V^{(2)+}_{\la}
       \nonu \\
       && +\frac{3}{20} \,
       \frac{8}{5} (\lambda +1) (2 \lambda -3) (4 \lambda -1)
       \, \bar{\pa}^2 \, V^{(2)-}_{\la}
       - \frac{18}{35}  (12 \lambda ^2-6 \lambda -43) \,  V^{(4)+}_{\la}\nonu \\
       &&-
    \frac{6}{5}  (4 \lambda -1)\,  V^{(4)-}_{\la}   \Bigg](\bar{w})
     \nonu \\
      && +
     \frac{1}{(\bar{z}-\bar{w})^3} \, \Bigg[
     \frac{1}{30}\,  \frac{56}{15} (\lambda -1) (\lambda +1) (2 \lambda -3)
       (2 \lambda +1) \, \bar{\pa}^3 \, V^{(2)+}_{\la}
       \nonu \\
       && +\frac{1}{30} \,
       \frac{8}{5} (\lambda +1) (2 \lambda -3) (4 \lambda -1)
       \, \bar{\pa}^3 \, V^{(2)-}_{\la}
       - \frac{1}{2} \frac{18}{35}
       (12 \lambda ^2-6 \lambda -43) \, \bar{\pa}\, V^{(4)+}_{\la}\nonu \\
       &&- \frac{1}{2}\,
    \frac{6}{5}  (4 \lambda -1)\, \bar{\pa} \, V^{(4)-}_{\la}   \Bigg](\bar{w})
     \nonu \\
   && +
     \frac{1}{(\bar{z}-\bar{w})^2} \, \Bigg[
     \frac{1}{168}\,  \frac{56}{15} (\lambda -1) (\lambda +1) (2 \lambda -3)
       (2 \lambda +1) \, \bar{\pa}^4 \, V^{(2)+}_{\la}
       \nonu \\
       && +\frac{1}{168} \,
       \frac{8}{5} (\lambda +1) (2 \lambda -3) (4 \lambda -1)
       \, \bar{\pa}^4 \, V^{(2)-}_{\la}
       - \frac{5}{36} \frac{18}{35}
       (12 \lambda ^2-6 \lambda -43) \, \bar{\pa}^2\, V^{(4)+}_{\la}\nonu \\
       &&- \frac{5}{36}\,
       \frac{6}{5}  (4 \lambda -1)\, \bar{\pa}^2 \, V^{(4)-}_{\la}
       + 6 \,  V^{(6)+}_{\la} \Bigg](\bar{w})
     \nonu  \\
      && +
     \frac{1}{(\bar{z}-\bar{w})} \, \Bigg[
     \frac{1}{1120}\,  \frac{56}{15} (\lambda -1) (\lambda +1) (2 \lambda -3)
       (2 \lambda +1) \, \bar{\pa}^5 \, V^{(2)+}_{\la}
       \nonu \\
       && +\frac{1}{1120} \,
       \frac{8}{5} (\lambda +1) (2 \lambda -3) (4 \lambda -1)
       \, \bar{\pa}^5 \, V^{(2)-}_{\la}
       - \frac{1}{36} \frac{18}{35}
       (12 \lambda ^2-6 \lambda -43) \, \bar{\pa}^3\, V^{(4)+}_{\la}\nonu \\
       &&- \frac{1}{36}\,
       \frac{6}{5}  (4 \lambda -1)\, \bar{\pa}^3 \, V^{(4)-}_{\la}
       + \frac{1}{2} \, 6 \,  \bar{\pa}\, V^{(6)+}_{\la} \Bigg](\bar{w})
     + \cdots,
     \nonu  \\
     &&    V^{(4)+}_{\la}(\bar{z}) \,V^{(3)-}_{\la}(\bar{w})
      = \frac{1}{(\bar{z}-\bar{w})^6} \, \Bigg[
        \frac{2}{5} (\lambda -1) (\lambda +1) (2 \lambda -3)
        (2 \lambda +1) \, V^{(1)-}_{\la}    \Bigg](\bar{w})
      \nonu \\
      && +
  \frac{1}{(\bar{z}-\bar{w})^5} \, \Bigg[
        \frac{2}{5} (\lambda -1) (\lambda +1) (2 \lambda -3)
        (2 \lambda +1) \, \bar{\pa}\, V^{(1)-}_{\la}    \Bigg](\bar{w})
  \nonu \\
  && +
  \frac{1}{(\bar{z}-\bar{w})^4} \, \Bigg[
 \frac{1}{2}\,       \frac{2}{5} (\lambda -1) (\lambda +1) (2 \lambda -3)
 (2 \lambda +1) \, \bar{\pa}^2 \, V^{(1)-}_{\la}
 \nonu \\
 && + \frac{12}{125} (\lambda +1) (2 \lambda -3) (4 \lambda -1) \,
 V^{(3)+}_{\la}- \frac{6}{25} \, (12 \lambda ^2-6 \lambda -43)
 \,  V^{(3)-}_{\la} \Bigg](\bar{w})
  \nonu \\
    && +
  \frac{1}{(\bar{z}-\bar{w})^3} \, \Bigg[
 \frac{1}{6}\,       \frac{2}{5} (\lambda -1) (\lambda +1) (2 \lambda -3)
 (2 \lambda +1) \, \bar{\pa}^3 \, V^{(1)-}_{\la}
 \nonu \\
 && + \frac{2}{3} \,
 \frac{12}{125} (\lambda +1) (2 \lambda -3) (4 \lambda -1) \,
 \bar{\pa}\,
 V^{(3)+}_{\la}- \frac{2}{3} \, \frac{6}{25} \, (12 \lambda ^2-6 \lambda -43)
 \,  \bar{\pa} \, V^{(3)-}_{\la} \Bigg](\bar{w})
  \nonu \\
 && +
  \frac{1}{(\bar{z}-\bar{w})^2} \, \Bigg[
 \frac{1}{24}\,       \frac{2}{5} (\lambda -1) (\lambda +1) (2 \lambda -3)
 (2 \lambda +1) \, \bar{\pa}^4 \, V^{(1)-}_{\la}
 \nonu \\
 && + \frac{5}{21} \,
 \frac{12}{125} (\lambda +1) (2 \lambda -3) (4 \lambda -1) \,
 \bar{\pa}^2 \,
 V^{(3)+}_{\la}- \frac{5}{21} \, \frac{6}{25} \, (12 \lambda ^2-6 \lambda -43)
 \,  \bar{\pa}^2 \, V^{(3)-}_{\la} \nonu \\
 && -\frac{2}{9}  (4 \lambda -1)\,  V^{(5)+}_{\la}
 +5 \,  V^{(5)-}_{\la}\Bigg](\bar{w})
  \nonu  \\
   && +
  \frac{1}{(\bar{z}-\bar{w})} \, \Bigg[
 \frac{1}{120}\,       \frac{2}{5} (\lambda -1) (\lambda +1) (2 \lambda -3)
 (2 \lambda +1) \, \bar{\pa}^5 \, V^{(1)-}_{\la}
 \nonu \\
 && + \frac{5}{84} \,
 \frac{12}{125} (\lambda +1) (2 \lambda -3) (4 \lambda -1) \,
 \bar{\pa}^3 \,
 V^{(3)+}_{\la}- \frac{5}{84} \, \frac{6}{25} \, (12 \lambda ^2-6 \lambda -43)
 \,  \bar{\pa}^3 \, V^{(3)-}_{\la} \nonu \\
 && -\frac{3}{5} \, \frac{2}{9}  (4 \lambda -1)\,  \bar{\pa}\, V^{(5)+}_{\la}
 +\frac{3}{5} \, 5 \,  \bar{\pa} \, V^{(5)-}_{\la}\Bigg](\bar{w})
  + \cdots,
  \nonu    \\
    &&    V^{(4)+}_{\la}(\bar{z}) \,Q^{(4)+}_{\la}(\bar{w})
  = \frac{1}{(\bar{z}-\bar{w})^6} \, \Bigg[
    \frac{7}{5} (\lambda -1) (\lambda +1) (2 \lambda -3) (2 \lambda +1)
    \, Q^{(2)+}_{\la}    \nonu \\
   &&   +  \frac{1}{(\bar{z}-\bar{w})^5} \, \Bigg[
    \frac{2}{3}
    \, \frac{7}{5} (\lambda -1) (\lambda +1) (2 \lambda -3) (2 \lambda +1)
    \, \bar{\pa} \, Q^{(2)+}_{\la}\nonu \\
    && +
    \frac{8}{25} (\lambda +1) (2 \lambda -3) (4 \lambda -1)\,
     Q^{(3)-}_{\la}
    \Bigg](\bar{w})
    \nonu \\
   &&   +  \frac{1}{(\bar{z}-\bar{w})^4} \, \Bigg[
    \frac{1}{4}
    \, \frac{7}{5} (\lambda -1) (\lambda +1) (2 \lambda -3) (2 \lambda +1)
    \, \bar{\pa}^2 \, Q^{(2)+}_{\la}\nonu \\
    && +\frac{3}{5}\,
    \frac{8}{25} (\lambda +1) (2 \lambda -3) (4 \lambda -1)\,
    \bar{\pa}\, Q^{(3)-}_{\la} -
    \frac{9}{25}\, (12 \lambda ^2-6 \lambda -43)\,  Q^{(4)+}_{\la}
     \Bigg](\bar{w})
    \nonu  \\
      &&   +  \frac{1}{(\bar{z}-\bar{w})^3} \, \Bigg[
    \frac{1}{15}
    \, \frac{7}{5} (\lambda -1) (\lambda +1) (2 \lambda -3) (2 \lambda +1)
    \, \bar{\pa}^3 \, Q^{(2)+}_{\la}\nonu \\
    && +\frac{1}{5}\,
    \frac{8}{25} (\lambda +1) (2 \lambda -3) (4 \lambda -1)\,
    \bar{\pa}^2\, Q^{(3)-}_{\la} -
    \frac{4}{7}\,
    \frac{9}{25}\, (12 \lambda ^2-6 \lambda -43)\,  \bar{\p}\, Q^{(4)+}_{\la}
    \nonu \\
    && -\frac{2}{5}  (4 \lambda -1)\,  Q^{(5)-}_{\la}\Bigg](\bar{w})
    \nonu  \\
   &&   +  \frac{1}{(\bar{z}-\bar{w})^2} \, \Bigg[
    \frac{1}{72}
    \, \frac{7}{5} (\lambda -1) (\lambda +1) (2 \lambda -3) (2 \lambda +1)
    \, \bar{\pa}^4 \, Q^{(2)+}_{\la}\nonu \\
    && +\frac{1}{21}\,
    \frac{8}{25} (\lambda +1) (2 \lambda -3) (4 \lambda -1)\,
    \bar{\pa}^3\, Q^{(3)-}_{\la} -
    \frac{5}{28}\,
    \frac{9}{25}\, (12 \lambda ^2-6 \lambda -43)\,
    \bar{\pa}^2 \, Q^{(4)+}_{\la}
    \nonu \\
    && -\frac{5}{9} \,
    \frac{2}{5}  (4 \lambda -1)\,  \bar{\pa} \, Q^{(5)-}_{\la}
   +\frac{11}{2}\,  Q^{(6)+}_{\la} \Bigg](\bar{w})
    \nonu  \\
     &&   +  \frac{1}{(\bar{z}-\bar{w})} \, \Bigg[
    \frac{1}{420}
    \, \frac{7}{5} (\lambda -1) (\lambda +1) (2 \lambda -3) (2 \lambda +1)
    \, \bar{\pa}^5 \, Q^{(2)+}_{\la}\nonu \\
    && +\frac{1}{112}\,
    \frac{8}{25} (\lambda +1) (2 \lambda -3) (4 \lambda -1)\,
    \bar{\pa}^4\, Q^{(3)-}_{\la} -
    \frac{5}{126}\,
    \frac{9}{25}\, (12 \lambda ^2-6 \lambda -43)\,
    \bar{\pa}^3 \, Q^{(4)+}_{\la}
    \nonu \\
    && -\frac{1}{6} \,
    \frac{2}{5}  (4 \lambda -1)\,  \bar{\pa}^2 \, Q^{(5)-}_{\la}
    +\frac{6}{11}\, \frac{11}{2}\,  \bar{\pa}\,
    Q^{(6)+}_{\la} \Bigg](\bar{w})+ \cdots,
    \nonu  \\
 &&    V^{(4)+}_{\la}(\bar{z}) \,Q^{(3)-}_{\la}(\bar{w})
  = \frac{1}{(\bar{z}-\bar{w})^6} \, \Bigg[\frac{10}{3} (\lambda -1) \lambda  (2 \lambda -1) (2 \lambda +1)
    \, Q^{(1)-}_{\la}
    \nonu \\
    && + \frac{1}{(\bar{z}-\bar{w})^5} \,
    \Bigg[2\, \frac{10}{3} (\lambda -1) \lambda  (2 \lambda -1) (2 \lambda +1)
      \, \bar{\pa}\, Q^{(1)-}_{\la} +
      \frac{4}{5} (\lambda -1) (2 \lambda +1) (4 \lambda -1) \,
      Q^{(2)+}_{\la} \Bigg](\bar{w}) \nonu \\
    &&   + \frac{1}{(\bar{z}-\bar{w})^4} \,
    \Bigg[\frac{3}{2}\,
      \frac{10}{3} (\lambda -1) \lambda  (2 \lambda -1) (2 \lambda +1)
      \, \bar{\pa}^2\, Q^{(1)-}_{\la} +
      \frac{4}{5} (\lambda -1) (2 \lambda +1) (4 \lambda -1) \,\bar{\pa}\,
      Q^{(2)+}_{\la}\nonu \\
&& -\frac{21}{10} \,  (\lambda +1) (2 \lambda -3)\,   Q^{(3)-}_{\la}
      \Bigg](\bar{w}) \nonu \\
      &&   + \frac{1}{(\bar{z}-\bar{w})^3} \,
    \Bigg[\frac{2}{3}\,
      \frac{10}{3} (\lambda -1) \lambda  (2 \lambda -1) (2 \lambda +1)
      \, \bar{\pa}^3\, Q^{(1)-}_{\la} +
      \frac{1}{2}\,
      \frac{4}{5} (\lambda -1) (2 \lambda +1) (4 \lambda -1) \,\bar{\pa}^2\,
      Q^{(2)+}_{\la}\nonu \\
      && -\frac{4}{5}\, \frac{21}{10} \,  (\lambda +1) (2 \lambda -3)\,
      \bar{\pa} \, Q^{(3)-}_{\la}-\frac{8}{15} \, (4 \lambda -1)\,
       Q^{(4)+}_{\la}
       \Bigg](\bar{w}) \nonu \\
   &&   + \frac{1}{(\bar{z}-\bar{w})^2} \,
    \Bigg[\frac{5}{24}\,
      \frac{10}{3} (\lambda -1) \lambda  (2 \lambda -1) (2 \lambda +1)
      \, \bar{\pa}^4\, Q^{(1)-}_{\la} \nonu \\
      && +
      \frac{1}{6}\,
      \frac{4}{5} (\lambda -1) (2 \lambda +1) (4 \lambda -1) \,\bar{\pa}^3\,
      Q^{(2)+}_{\la}\nonu \\
      && -\frac{5}{12}\, \frac{21}{10} \,  (\lambda +1) (2 \lambda -3)\,
      \bar{\pa}^2 \, Q^{(3)-}_{\la}-\frac{5}{7}\,
      \frac{8}{15} \, (4 \lambda -1)\,
    \bar{\pa}\,    Q^{(4)+}_{\la}
    +\frac{9}{2}\,  Q^{(5)-}_{\la} \Bigg](\bar{w}) \nonu \\
     &&   + \frac{1}{(\bar{z}-\bar{w})} \,
    \Bigg[\frac{1}{20}\,
      \frac{10}{3} (\lambda -1) \lambda  (2 \lambda -1) (2 \lambda +1)
      \, \bar{\pa}^5\, Q^{(1)-}_{\la} \nonu \\
      && +
      \frac{1}{24}\,
      \frac{4}{5} (\lambda -1) (2 \lambda +1) (4 \lambda -1) \,\bar{\pa}^4\,
      Q^{(2)+}_{\la}\nonu \\
      && -\frac{5}{42}\, \frac{21}{10} \,  (\lambda +1) (2 \lambda -3)\,
      \bar{\pa}^3 \, Q^{(3)-}_{\la}-\frac{15}{56}\,
      \frac{8}{15} \, (4 \lambda -1)\,
    \bar{\pa}^2\,    Q^{(4)+}_{\la}
    +\frac{2}{3}\,
    \frac{9}{2}\,  \bar{\pa}\, Q^{(5)-}_{\la} \Bigg](\bar{w})+ \cdots.   
    \label{finalOPE}  
    \eea
    From Appendix (\ref{finalOPE}), we observe that
    the currents appearing in the right hand sides, $V_{\la}^{(2),-}$,
    $V_{\la}^{(4),-}$ (for nonzero $V_{\la}^{(h)+}$ with even
    $h$), $V_{\la}^{(3),+}$, $V_{\la}^{(5),+}$ (
    for nonzero $V_{\la}^{(h)-}$ with odd
    $h$),
    $Q_{\la}^{(3),-}$, $Q_{\la}^{(5),-}$ (for nonzero
    $Q_{\la}^{(h),\pm}$ with even $h$), $Q_{\la}^{(2),+}$
    and $Q_{\la}^{(4),+}$
 (for nonzero
    $Q_{\la}^{(h),\pm}$ with odd $h$)
    have the $(1-4\la)$ factor.
    Therefore, these currents are decoupled from the remaining
    subalgebra generated by (\ref{case1}) or (\ref{case2}).

    We can use the first and the third equations of (\ref{fourrelations})
    and calculate the commutator relation
    $\big[(V_{\la}^{(h_1),+})_m, (Q_{\la}^{(h_2+1),+})_r \big]$
    and focus on the coefficient function in front of
    $(Q_{\la}^{(h_1+h_2-2-h+1),-})_{m+r}$.
    Then we obtain the following result
    \bea
&& -\Bigg[\frac{1}{\frac{n_{W_{F,h_1}}}{q^{h_1-2}}\,
     \frac{(-1)^{h_1}}{\sum_{i=0}^{h_1-1}\, a^i( h_1, \frac{1}{2})}}
  \Bigg]\, \frac{1}{2}\,
\Bigg[\frac{1}{\frac{1}{2} \,\frac{n_{W_{Q,h_2+\frac{1}{2}}}}{q^{h_2-1}}
\, \frac{(-1)^{h_2+1}  \, h_2 }{
   \sum_{i=0}^{h_2-1} \, \beta^i( h_2+1, 0)}\,} \Bigg]
\nonu \\
&& \times \Bigg( \Bigg[ \frac{1}{2} \,\frac{n_{W_{Q,h_1+h_2-2-h+\frac{1}{2}}}}{q^{h_1+h_2-2-h-1}}
\, \frac{(-1)^{h_1+h_2-2-h+1}  \, (h_1+h_2-2-h) }{
   \sum_{i=0}^{h_1+h_2-2-h-1} \, \beta^i( h_1+h_2-2-h+1, 0)}\Bigg]
\, q^h \, q_F^{h_1,h_2+\frac{1}{2},h}(m,r,\la) \nonu \\
&& \times ((Q_{\la}^{(h_1+h_2-2-h+1),-})_{m+r}-(Q_{\la}^{(h_1+h_2-2-h+1),+})_{m+r}) 
\nonu \\
&& -
 \Bigg[ \frac{1}{2} \,\frac{n_{W_{Q,h_1+h_2-2-h+\frac{1}{2}}}}{q^{h_1+h_2-2-h-1}}
\, \frac{(-1)^{h_1+h_2-2-h+1}   }{
   \sum_{i=0}^{h_1+h_2-2-h} \, \al^i( h_1+h_2-2-h+1, 0)}\Bigg]
\, q^h \, q_F^{h_1,h_2+\frac{1}{2},h}(m,r,\la) \nonu \\
&& \times (-1)^h\, ( (Q_{\la}^{(h_1+h_2-2-h+1),-})_{m+r}+
(Q_{\la}^{(h_1+h_2-2-h+1),+})_{m+r}) \Bigg)
\nonu \\
&& -\Bigg[\frac{1}{\frac{n_{W_{B,h_1}}}{q^{h_1-2}}\,
     \frac{(-1)^{h_1}}{\sum_{i=0}^{h_1-1}\, a^i( h_1, 0)}}
  \Bigg]\, \frac{1}{2}\,
\Bigg[\frac{1}{\frac{1}{2} \,\frac{n_{W_{Q,h_2+\frac{1}{2}}}}{q^{h_2-1}}
\, \frac{(-1)^{h_2+1}   }{
   \sum_{i=0}^{h_2} \, \al^i( h_2+1, 0)}\,} \Bigg]
\nonu \\
&& \times \Bigg( \Bigg[ \frac{1}{2} \,\frac{n_{W_{Q,h_1+h_2-2-h+\frac{1}{2}}}}{q^{h_1+h_2-2-h-1}}
\, \frac{(-1)^{h_1+h_2-2-h+1}  \, (h_1+h_2-2-h) }{
   \sum_{i=0}^{h_1+h_2-2-h-1} \, \beta^i( h_1+h_2-2-h+1, 0)}\Bigg]
\, q^h \, q_B^{h_1,h_2+\frac{1}{2},h}(m,r,\la)
\nonu \\
&& \times ( (Q_{\la}^{(h_1+h_2-2-h+1),-})_{m+r}-(Q_{\la}^{(h_1+h_2-2-h+1),+})_{m+r}) 
\nonu \\
&& -
 \Bigg[ \frac{1}{2} \,\frac{n_{W_{Q,h_1+h_2-2-h+\frac{1}{2}}}}{q^{h_1+h_2-2-h-1}}
\, \frac{(-1)^{h_1+h_2-2-h+1}   }{
   \sum_{i=0}^{h_1+h_2-2-h} \, \al^i( h_1+h_2-2-h+1, 0)}\Bigg]
\, q^h \, q_F^{h_1,h_2+\frac{1}{2},h}(m,r,\la) \nonu \\
&& \times (-1)^h\, ( (Q_{\la}^{(h_1+h_2-2-h+1),-})_{m+r}+
(Q_{\la}^{(h_1+h_2-2-h+1),+})_{m+r}) \Bigg),
\label{finaleq}
\eea
which leads to zero at $\la =\frac{1}{4}$ for
the coefficient of $(Q_{\la}^{(h_1+h_2-2-h+1),-})_{m+r}$
with odd $(h_1+h_2-2-h+1)$. The property appearing in the footnote
\ref{equalcoeff} is used.

Similarly,
by using  the first and the fourth equations of (\ref{fourrelations})
    and calculate the commutator relation
    $\big[(V_{\la}^{(h_1),+})_m, (Q_{\la}^{(h_2+1),-})_r \big]$
    and focus on the coefficient function in front of
    $(Q_{\la}^{(h_1+h_2-2-h+1),+})_{m+r}$. We have checked that
    the coefficient function appearing in
    the mode  $(Q_{\la}^{(h_1+h_2-2-h+1),+})_{m+r}$
    with even $(h_1+h_2-2-h+1)$ vanishes at $\la =\frac{1}{4}$
    from similar equation of Appendix
    (\ref{finaleq}).


\end{document}